\documentclass{article}
\usepackage{amsfonts}
\usepackage[utf8]{inputenc} 
\usepackage[T1]{fontenc}    
\usepackage{textcomp}       
\usepackage{amsmath}
\usepackage{bm}
\usepackage{amssymb}
\usepackage{amsthm}
\usepackage{appendix}
\usepackage{array}
\usepackage{color}
\usepackage{comment}
\usepackage{dsfont}
\usepackage{enumerate}
\usepackage[mathscr]{euscript}
\usepackage{graphicx}
\usepackage{pdflscape}
\usepackage{afterpage}
\usepackage[utf8]{inputenc}
\usepackage{latexsym}
\usepackage{mathrsfs}
\usepackage{enumitem}
\usepackage{multicol}
\usepackage{mathtools}
\usepackage{multirow}
\usepackage{subfigure}
\usepackage[a4paper, top=3cm, bottom=3.5cm, left=3.5cm, right=3.5cm]{geometry}
\usepackage{tikz}
\usetikzlibrary{arrows.meta,calc}
\usepackage[font=small]{caption}
\usepackage{cite}
\RequirePackage[colorlinks,linkcolor=red,citecolor=blue,urlcolor=blue]{hyperref}
\hypersetup{pdftitle={Characterization of BSDE-induced dynamic risk measures via acceptance sets}, pdfauthor={M. Ferrari, R. J. A. Laeven, E. Rosazza Gianin, M. Zullino}}
\mathtoolsset{showonlyrefs}  
\usepackage[american]{babel}

\newcommand{\R}{\mathbb{R}}
\newcommand{\N}{\mathbb{N}}
\newcommand{\E}{\mathbb{E}}

\newcommand{\A}{\mathcal{A}}

\newcommand{\F}{\mathcal{F}}

\newcommand{\Q}{\mathbb{Q}}
\renewcommand{\P}{\mathbb{P}}
\newcommand{\1}{\mathds{1}}

\newcommand{\gr}[1]{\textcolor[gray]{0.45}{#1}}

\newcommand{\mE}{\mathcal{E}}

\newcommand{\be}{\begin{equation}}
\newcommand{\ee}{\end{equation}}
\newcommand{\bea}{\begin{eqnarray}}
\newcommand{\eea}{\end{eqnarray}}
\newcommand{\beas}{\begin{eqnarray*}}
\newcommand{\eeas}{\end{eqnarray*}}
\newcommand{\ds}{\displaystyle}
\newcommand{\itemref}[1]{\textup{(\ref{#1})}}

\newcommand{\smallbullet}{{\mathbin{\vcenter{\hbox{\scriptsize$\bullet$}}}}}
\newcommand\tinybullet{{\mathbin{\vcenter{\hbox{\tiny$\bullet$}}}}}

\makeatletter
\newcommand{\proplabel}[2]{\phantomsection\protected@edef\@currentlabel{#1}\label{#2}}
\makeatother

\DeclareRobustCommand{\axS}[1]{\ensuremath{\mathrm{S}_{\mathrm{#1}}}}
\DeclareRobustCommand{\axSp}[1]{\ensuremath{\mathrm{S}^\tinybullet_{\mathrm{#1}}}}

\def\a{{\alpha}}
\def\b{{\beta}}
\def\r{{\rho}}
\def\l{{\lambda}}
\def\O{{\Omega}}
\def\o{{\omega}}
\def\s{{\sigma}}

\def\t{{\tau}}
\def\g{{\gamma}}
\def\eps{{\varepsilon}}

\numberwithin{equation}{section}

\newtheorem{theorem}{Theorem}[section]

\newtheorem{corollary}[theorem]{Corollary}

\newtheorem{definition}[theorem]{Definition}

\newtheorem{lemma}[theorem]{Lemma}

\newtheorem{proposition}[theorem]{Proposition}

\renewcommand{\d}{\,\mathrm{d}}

\theoremstyle{remark}
\newtheorem{remark}[theorem]{Remark}
\newtheorem{example}[theorem]{Example}

\DeclareMathOperator*{\esssup}{ess\,sup}
\DeclareMathOperator*{\essinf}{ess\,inf}
\DeclareMathOperator*{\essmax}{ess\,max}

\title{Characterization of BSDE-induced dynamic risk measures via acceptance sets\thanks{
This research was funded in part by
the Netherlands Organization for Scientific Research (NWO) under grant NWO Vici 2020--2027 (Ferrari, Laeven, Zullino) and by an UvA-AC3R Midsize Grant (Ferrari, Laeven).
Emanuela Rosazza Gianin, Marco Zullino, and Matteo Ferrari are members of Gruppo Nazionale per l’Analisi Matematica, la Probabilità e le loro Applicazioni (GNAMPA)-INdAM, Italy.}} 
\author{
Matteo Ferrari \\
{\footnotesize Dept.~of Quantitative Economics}\\
{\footnotesize University of Amsterdam}\\
{\footnotesize and AC3R, The Netherlands}\\
{\footnotesize \texttt{m.ferrari@uva.nl}}\\
\and Roger J.~A.~Laeven\footnote{Corresponding author.} \\
{\footnotesize Dept.~of Quantitative Economics}\\
{\footnotesize University of Amsterdam, AC3R, CentER}\\
{\footnotesize and EURANDOM, The Netherlands}\\
{\footnotesize \texttt{R.J.A.Laeven@uva.nl}}\\
\and Emanuela Rosazza Gianin \\
{\footnotesize Dept.~of Statistics and Quantitative Methods}\\
{\footnotesize University of Milano-Bicocca, Italy}\\
{\footnotesize \texttt{emanuela.rosazza1@unimib.it}}\\
\and Marco Zullino \\
{\footnotesize Dept.~of Mathematics and Applications}\\
{\footnotesize University of Milano-Bicocca, Italy}\\
{\footnotesize \texttt{M.Zullino@campus.unimib.it}}\\
}

\date{November 26, 2026}

\begin{document}

\maketitle

\begin{abstract}
    We characterize dynamic risk measures induced by backward stochastic differential equations (BSDEs), in a Brownian filtration, directly through their acceptance sets.
    To include risk measures that are not cash-additive, we work with \emph{leveled acceptance families}, which collect the positions whose risk does not exceed a given level.
    We show that time-consistency is equivalent to a decomposition of these families across intermediate dates, without requiring conditional constancy.
    Our main result shows that a leveled acceptance family is induced by a BSDE with a Lipschitz driver if and only if it satisfies this decomposition and a set-valued counterpart of Peng's domination condition, together with a few standard axioms.
    The driver is then unique, and its properties can be read off the acceptance sets.
    In the cash-additive case, we also characterize, on $L^\infty$, the families induced by convex drivers with quadratic growth, where the growth condition amounts to a single inclusion of the entropic acceptance set over the whole time horizon.
    Applications include the infimal cocycle property of the penalty, the resilience rate, law-invariance, and the dynamic Value-at-Risk, whose only normalized, time-consistent, cash-additive majorant is the worst-case risk measure.
    Two examples show how an acceptance rule, given without a risk measure, is turned into a BSDE.
\end{abstract}



\newpage

\tableofcontents\bigskip 

\section{Introduction}
\label{sec:int}
 
Acceptance sets are the most primitive description of an attitude towards risk: they collect the positions that an agent, a regulator or a supervisor regards as acceptable.
The notion goes back to \cite{Yaari_1965_Convexity_theory_choice_under_risk}, which considers the set of gambles that a decision-maker accepts at a given wealth level and stresses that this set ``is observable, at least in principle, and its properties do not depend on any prior theoretical construction''.
In the axiomatic theory of risk measures initiated in \cite{Artzner+Delbaen+Eber+Heath_1999_Coherent_measures_risk}, an acceptance set $\A$ is the primitive object, and the associated risk measure
\[
    \r(X)=\inf\{m\in\R\ :\ X+m\in\A\}
\]
is the minimal amount of cash that makes the position $X$ acceptable; see also \cite[Chapter~4]{Follmer+Schied_2016_Stochastic_finance}.
Starting from acceptance sets rather than from risk measures has a clear rationale: one may know which positions are acceptable, because they are prescribed by a regulator or result from an internal assessment, without knowing the risk measure, which is then derived from them.
This viewpoint is taken, e.g., in \cite{Canna+Centrone+RosazzaGianin_2020_Capital_allocation_rules_acceptance_sets}, where capital allocation is formulated in terms of acceptance sets without necessarily involving a risk measure.
In the dynamic setting, the correspondence between conditional acceptance sets and cash-additive dynamic risk measures, including the characterization of time-consistency through a decomposition of the acceptance sets, is well established \cite{Detlefsen+Scandolo_2005_Conditional_dynamic_convex_risk_measures,Cheridito+Delbaen+Kupper_2006_Dynamic_monetary_risk_measures_bounded_discrete-time_processes,Follmer+Penner_2006_Convex_risk_measures_dynamics_their_penalty_functions,Bion-Nadal_2008_Dynamic_risk_measures_time_consistency__risk_measures_BMO_martingales,Acciaio+Penner_2011_Dynamic_risk_measures}, \cite[Chapter~11]{Follmer+Schied_2016_Stochastic_finance}.
Without cash-additivity, a single acceptance set no longer determines the risk measure, and one needs a family of acceptance sets indexed by the level of risk, as in the risk acceptance families of  \cite{Drapeau+Kupper_2013_Risk_preferences_robust_representation}, whose conditional versions for quasi-convex maps are given in \cite{Frittelli+Maggis_2011_Dual_representation_quasi-convex_conditional_maps,Bielecki+Cialenco+Drapeau+Karliczek_2016_Dynamic_assessment_indices}; the acceptance sets of \cite{Yaari_1965_Convexity_theory_choice_under_risk} are, in fact, already indexed by a level, namely the wealth of the decision-maker.
 
Dynamic risk measures induced by backward stochastic differential equations (BSDEs) form a central class of time-consistent dynamic risk measures in continuous time \cite{ElKaroui+Peng+Quenez_1997_Backward_stochastic_differential_equations_finance,Peng_1997_Backward_SDE_related_g-expectation,Frittelli+RosazzaGianin_2004_Dynamic_Convex_Risk_Measures,Peng_2004_Nonlinear_expectations_nonlinear_evaluations_risk_measures,RosazzaGianin_2006_Risk_measures_g-expectations,Barrieu+ElKaroui_2009_Pricing_hedging_optimally_designing_derivatives_minimization_risk_measures}; see \cite{Briand+Delyon+Hu+Pardoux+Stoica_2003_Lp_solutions_BSDEs} for terminal conditions in $L^p$ and \cite{Kobylanski_2000_Backward_stochastic_differential_equations_partial_differential_equations_quadratic_growth,Briand+Hu_2008_Quadratic_BSDEs_convex_generators_unbounded_terminal_conditions,Laeven+Stadje_2014_Robust_portfolio_choice_indifference_valuation} for drivers with quadratic growth.
Applications of BSDEs to problems such as valuation, portfolio choice, and optimal stopping are widespread; see, e.g., \cite{Lazrak+Quenez_2003_Generalized_stochastic_differential_utility,Mania+Schweizer_2005_Dynamic_exponential_utility_indifference_valuation,Kratschmer+Ladkau+Laeven+Schoenmakers+Stadje_2018_Optimal_stopping_under_uncertainty_drift_jump_intensity,Laeven+Schoenmakers+Schweizer+Stadje_2025_Robust_multiple_stopping_duality_approach}.
Given a driver $g$, the risk of a position $X$ at time $t$ evaluated at time $s\leq t$ is $\r^g_{s,t}(X):=Y_s$, where $(Y,Z)$ solves
\[
    Y_r=-X+\int_r^tg(u,Y_u,Z_u)\d u-\int_r^tZ_u\cdot\d W_u,\qquad r\in[0,t],
\]
and properties of $\r^g$, such as convexity or cash-additivity, can be read off from the driver \cite{Briand+Coquet+Hu+Memin+Peng_2000_Converse_comparison_theorem_BSDEs_related_properties_g-expectation,RosazzaGianin_2006_Risk_measures_g-expectations,Jiang_2008_Convexity_translation_invariance_subadditivity_g-expectations_related_risk_measures,Barrieu+ElKaroui_2009_Pricing_hedging_optimally_designing_derivatives_minimization_risk_measures}.
Conversely, a time-consistent dynamic risk measure is induced by a BSDE under a domination condition: a Lipschitz driver exists whenever the risk measure is dominated by the one induced by $\mu|z|$, as shown in \cite{Coquet+Hu+Memin+Peng_2002_Filtration-consistent_nonlinear_expectations_related_g-expectations} for translation-invariant filtration-consistent nonlinear expectations, or by the one induced by $\mu(|y|+|z|)$, as shown in \cite{Peng_2005_dynamically_consistent_nonlinear_evaluations} for general dynamically consistent nonlinear evaluations; see also \cite{Hu_2013_Dynamically_consistent_nonlinear_evaluations_generating_functions_Lp} for the $L^p$ case, \cite{Hu+Ma+Peng+Yao_2008_Representation_theorems_quadratic_F-consistent_nonlinear_expectations,Zheng_2024_g-expectations_filtration-consistent_nonlinear_expectations} for quadratic drivers, and \cite{Delbaen+Peng+RosazzaGianin_2010_Representation_penalty_term_dynamic_concave_utilities,Laeven+Stadje_2014_Robust_portfolio_choice_indifference_valuation} for the representation of the penalty.
These results take the risk measure as the starting point, and they require it to be well posed, monotone, time-consistent and dominated.
Moreover, apart from a few special cases, such as the conditional expectation and the entropic risk measure, BSDE-induced risk measures do not admit an explicit expression.
 
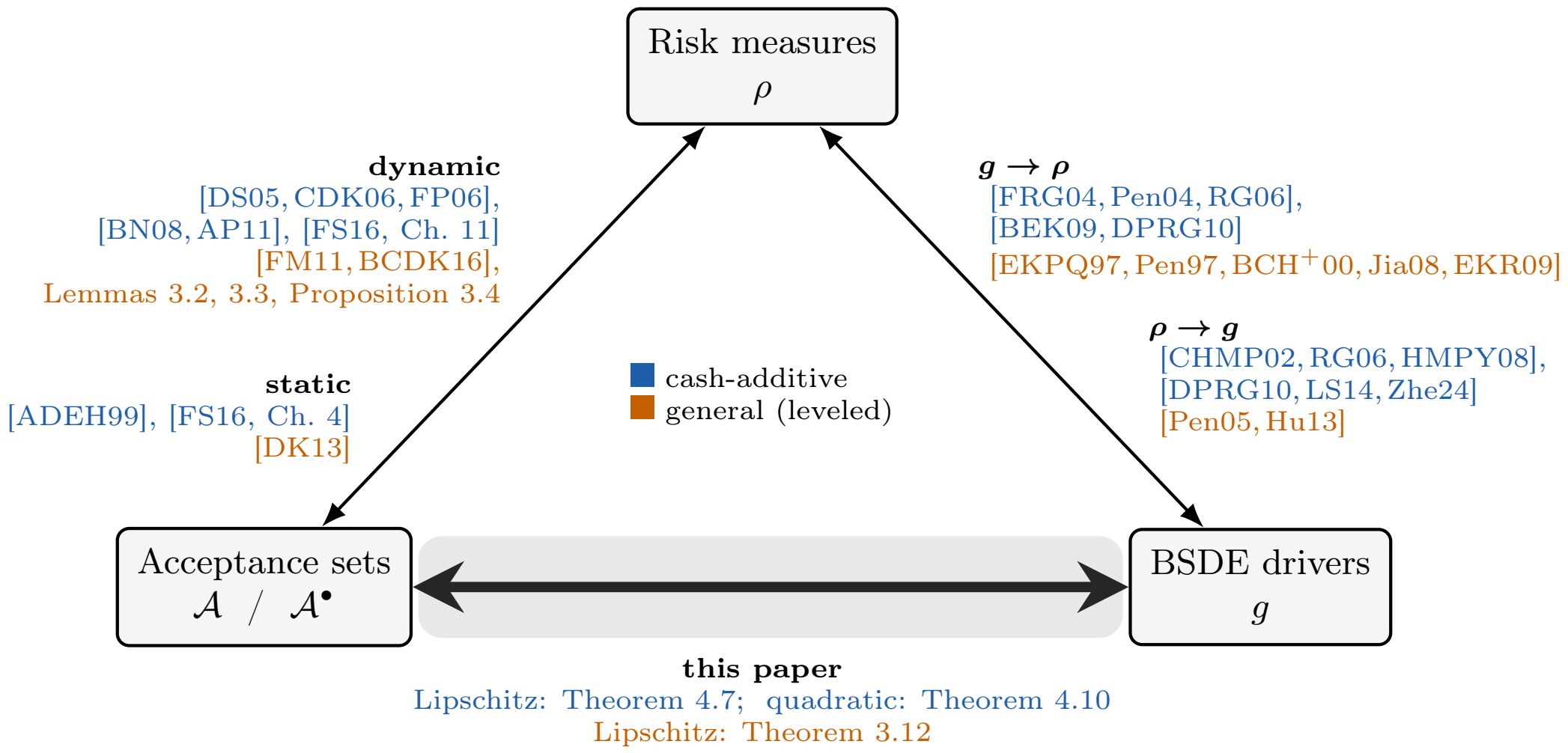
\begin{figure}[ht]
\centering
\definecolor{cacol}{RGB}{31,94,168}  
\definecolor{gencol}{RGB}{196,96,0}  
\newcommand{\cacite}[1]{{\hypersetup{citecolor=cacol,linkcolor=cacol}\color{cacol}#1}}
\newcommand{\gencite}[1]{{\hypersetup{citecolor=gencol,linkcolor=gencol}\color{gencol}#1}}
\begin{tikzpicture}[
    vertex/.style={draw, rounded corners=3pt, thick, fill=gray!8, align=center, inner sep=5pt, font=\small},
    edge/.style={{Latex[length=2.2mm]}-{Latex[length=2.2mm]}, thick},
    lab/.style={font=\scriptsize, inner sep=2pt},
]
    \node[vertex] (R) at (0,4.6)    {Risk measures\\ $\r$};
    \node[vertex] (A) at (-4.4,0)   {Acceptance sets\\ $\A$ \ / \ $\A^{\tinybullet}$};
    \node[vertex] (G) at (4.4,0)    {BSDE drivers\\ $g$};
 
    \fill[gray!18, rounded corners=6pt] ($(A.east)+(0.05,-0.45)$) rectangle ($(G.west)+(-0.05,0.45)$);
    \draw[{Stealth[length=4.5mm,width=4.5mm]}-{Stealth[length=4.5mm,width=4.5mm]}, line width=2.6pt, black!85] (A) -- (G);
    \draw[edge] (A) -- (R);
    \draw[edge] (R) -- (G);
    
    \node[lab, anchor=south east, align=right] at ($(A)!0.52!(R)+(-0.12,0)$) {%
        \textbf{dynamic}\\
        \cacite{\cite{Detlefsen+Scandolo_2005_Conditional_dynamic_convex_risk_measures,Cheridito+Delbaen+Kupper_2006_Dynamic_monetary_risk_measures_bounded_discrete-time_processes,Follmer+Penner_2006_Convex_risk_measures_dynamics_their_penalty_functions},}\\
        \cacite{\cite{Bion-Nadal_2008_Dynamic_risk_measures_time_consistency__risk_measures_BMO_martingales,Acciaio+Penner_2011_Dynamic_risk_measures}, \cite[Ch.~11]{Follmer+Schied_2016_Stochastic_finance}}\\
        \gencite{\cite{Frittelli+Maggis_2011_Dual_representation_quasi-convex_conditional_maps,Bielecki+Cialenco+Drapeau+Karliczek_2016_Dynamic_assessment_indices},}\\
        \gencite{Lemmas~\ref{LEM:N-from-A-to-rho}, \ref{LEM:N-from-rho-to-A}, Proposition~\ref{prop: suff cond-A3}}};
    \node[lab, anchor=south east, align=right] at ($(A)!0.22!(R)+(-0.12,0)$) {%
        \textbf{static}\\
        \cacite{\cite{Artzner+Delbaen+Eber+Heath_1999_Coherent_measures_risk}, \cite[Ch.~4]{Follmer+Schied_2016_Stochastic_finance}}\\
        \gencite{\cite{Drapeau+Kupper_2013_Risk_preferences_robust_representation}}};
        
    \node[lab, anchor=south west, align=left] at ($(R)!0.73!(G)+(0.12,0)$) {%
        $\bm{\r\to g}$\\
        \cacite{\cite{Coquet+Hu+Memin+Peng_2002_Filtration-consistent_nonlinear_expectations_related_g-expectations,RosazzaGianin_2006_Risk_measures_g-expectations,Hu+Ma+Peng+Yao_2008_Representation_theorems_quadratic_F-consistent_nonlinear_expectations},}\\
        \cacite{\cite{Delbaen+Peng+RosazzaGianin_2010_Representation_penalty_term_dynamic_concave_utilities,Laeven+Stadje_2014_Robust_portfolio_choice_indifference_valuation,Zheng_2024_g-expectations_filtration-consistent_nonlinear_expectations}}\\
        \gencite{\cite{Peng_2005_dynamically_consistent_nonlinear_evaluations,Hu_2013_Dynamically_consistent_nonlinear_evaluations_generating_functions_Lp}}};
    \node[lab, anchor=south west, align=left] at ($(R)!0.45!(G)+(-0.15,0.1)$) {%
        $\bm{g\to\r}$\\
        \cacite{\cite{Frittelli+RosazzaGianin_2004_Dynamic_Convex_Risk_Measures,Peng_2004_Nonlinear_expectations_nonlinear_evaluations_risk_measures,RosazzaGianin_2006_Risk_measures_g-expectations},}\\
        \cacite{\cite{Barrieu+ElKaroui_2009_Pricing_hedging_optimally_designing_derivatives_minimization_risk_measures,Delbaen+Peng+RosazzaGianin_2010_Representation_penalty_term_dynamic_concave_utilities}}\\
        \gencite{\cite{ElKaroui+Peng+Quenez_1997_Backward_stochastic_differential_equations_finance,Peng_1997_Backward_SDE_related_g-expectation,Briand+Coquet+Hu+Memin+Peng_2000_Converse_comparison_theorem_BSDEs_related_properties_g-expectation, Jiang_2008_Convexity_translation_invariance_subadditivity_g-expectations_related_risk_measures, ElKaroui+Ravanelli_2009_Cash_sub-additive_risk_measures_interest_rate_ambiguity}}};
        
    \node[lab, anchor=north, align=center] at ($(A)!0.5!(G)+(0,-0.55)$) {%
        \textbf{this paper}\\
        \cacite{Lipschitz: Theorem~\ref{THM:Lp-additive}; \ quadratic: Theorem~\ref{THM:BSDE_quad}}\\
        \gencite{Lipschitz: Theorem~\ref{THM:Lp-nonadditive}}};
        
    \node[lab, align=left, anchor=center] at (0,1.7) {%
        \textcolor{cacol}{\rule{6pt}{6pt}} cash-additive\\
        \textcolor{gencol}{\rule{6pt}{6pt}} general (leveled)};
\end{tikzpicture}
\caption{Correspondences between acceptance sets, dynamic risk measures and BSDE drivers, in the cash-additive (blue) and general (orange) case.}
\label{FIG:triangle}
\end{figure}

In this paper, we bring the two viewpoints together and characterize BSDE-induced dynamic risk measures through their acceptance sets.
We start from a family of acceptance sets, define the risk measure as the essential infimum of the levels at which a position is acceptable, and give axioms on the acceptance sets under which this risk measure is induced by a BSDE, together with an expression of the driver in terms of the acceptance sets.
This has two advantages.
First, the axioms are conditions on the positions deemed acceptable, which can be checked without knowing the risk measure.
Second, it provides a formula for the BSDE-induced risk measure, the essential infimum of the acceptable levels, which is useful precisely because BSDE-induced risk measures rarely admit an explicit expression.
Our axioms are not stronger than the assumptions of \cite{Coquet+Hu+Memin+Peng_2002_Filtration-consistent_nonlinear_expectations_related_g-expectations,Peng_2005_dynamically_consistent_nonlinear_evaluations}, but are their counterpart on acceptance sets: well-posedness, monotonicity and time-consistency of the risk measure are not assumed, but follow from the axioms on the acceptance sets.
Figure~\ref{FIG:triangle} summarizes how our results relate to the literature: the correspondences between acceptance sets and risk measures, and between risk measures and BSDE drivers, are known, whereas the direct link between acceptance sets and drivers is established here.
 
Our first set of results concerns leveled acceptance families $\A^\tinybullet=(\A^m_{s,t})$, the dynamic version of the risk acceptance families of \cite{Drapeau+Kupper_2013_Risk_preferences_robust_representation}.
For every $0\leq s\leq t\leq T$ and every $\F_s$-measurable level $m$, the set $\A^m_{s,t}$ collects the $\F_t$-measurable positions whose risk at time $s$ does not exceed $m$, and the induced dynamic risk functional is
\[
    \r_{s,t}(X)=\essinf\{m\ :\ X\in\A^m_{s,t}\}.
\]
Leveled acceptance families describe general, possibly cash-nonadditive, dynamic risk functionals, while monetary acceptance families $\A=(\A_{s,t})$, with $\A^m_{s,t}=\A_{s,t}-m$, correspond to the cash-additive ones.
We first establish the correspondence between axioms on $\A^\tinybullet$ and properties of $\r$ (Lemmas~\ref{LEM:N-from-A-to-rho} and~\ref{LEM:N-from-rho-to-A}), which extends the conditional one of \cite{Frittelli+Maggis_2011_Dual_representation_quasi-convex_conditional_maps,Bielecki+Cialenco+Drapeau+Karliczek_2016_Dynamic_assessment_indices} beyond the quasi-convex case, and we show that a local dynamic risk measure is time-consistent if and only if its leveled acceptance family satisfies the level decomposition
\begin{equation}
\label{EQ:dec_bull_intro}
    \A^m_{s,t}=\bigcup_{n\in\A^m_{s,u}}\A^{-n}_{u,t},\qquad 0\leq s\leq u\leq t\leq T
    \tag{S$_{\mathrm{dec}}^\tinybullet$}
\end{equation}
(Proposition~\ref{prop: suff cond-A3}).
This extends the decomposition $\A_{s,t}=\A_{s,u}+\A_{u,t}$ of the cash-additive case and, unlike the characterizations of time-consistency available in \cite[Lemma~11.11]{Follmer+Schied_2016_Stochastic_finance} and \cite[Proposition~8]{Mastrogiacomo+RosazzaGianin_2015_Time-consistency_cash-subadditive_risk_measures}, does not require conditional constancy, which fails, e.g., for the discounted conditional expectation (Example~\ref{EX:CS-discount}).
Our main result in the leveled case (Theorem~\ref{THM:Lp-nonadditive}) states that, for $p\in(1,+\infty]$, a leveled acceptance family in $L^p$ is induced by a $\mu$-Lipschitz driver $g(t,y,z)$ if and only if it satisfies level right-continuity, normalization, locality, the level decomposition \eqref{EQ:dec_bull_intro} and the inclusion
\begin{equation}
\label{EQ:intro_inclusion_mu}
    \A^m_{s,t}+\A^{n,\mu}_{s,t}\subseteq\A^{m+n}_{s,t},
\tag{S$_{\mathrm{Lip}}^\tinybullet$}
\end{equation}
where $\A^{\tinybullet,\mu}$ is the leveled acceptance family induced by the driver $\mu(|y|+|z|)$.
This inclusion is the set-valued counterpart of the domination condition \cite[(A5)]{Peng_2005_dynamically_consistent_nonlinear_evaluations}.
The driver is unique and is recovered from the acceptance sets through the representation formula of \cite[Lemma~2.1]{Jiang_2008_Convexity_translation_invariance_subadditivity_g-expectations_related_risk_measures}, and the main properties of $\A^\tinybullet$ are equivalent to corresponding properties of $g$.
The core of the proof of Theorem~\ref{THM:Lp-nonadditive} is the case $p=2$: the inclusion \eqref{EQ:intro_inclusion_mu} yields well-posedness, monotonicity and domination of $\r$, and the level decomposition \eqref{EQ:dec_bull_intro} yields its time-consistency (Proposition~\ref{prop: suff cond-A3}), so that \cite[Theorem~3.1]{Peng_2005_dynamically_consistent_nonlinear_evaluations} applies and provides the driver.
The other values of $p$ are reduced to $p=2$ through domination (Lemma~\ref{LEM:N-reduction}): for $p\in(1,2)$, $\r$ is restricted to $L^2$, as in \cite{Hu_2013_Dynamically_consistent_nonlinear_evaluations_generating_functions_Lp}, and, for $p\in(2,+\infty]$, it is extended to $L^2$ by density.
 
Our second set of results concerns the cash-additive case.
For Lipschitz drivers, the characterization follows from the leveled one: by Theorem~\ref{THM:Lp-additive}, for $p<+\infty$, a monetary acceptance family is induced by a $\mu$-Lipschitz driver $g(t,z)$ if and only if it satisfies normalization, closedness, locality, the decomposition property and the inclusion 
\begin{equation}
\label{EQ:intro_inclusion_mu_CA}
    \A_{s,t}+\A^\mu_{s,t}\subseteq\A_{s,t}.
\tag{S$_{\mathrm{Lip}}$}
\end{equation}
\noeqref{EQ:intro_inclusion_mu_CA}
Quadratic drivers require a different approach: the representation theorems for quadratic filtration-consistent nonlinear expectations \cite{Hu+Ma+Peng+Yao_2008_Representation_theorems_quadratic_F-consistent_nonlinear_expectations,Zheng_2024_g-expectations_filtration-consistent_nonlinear_expectations} rely on domination conditions together with an independent increments assumption, and therefore only produce deterministic drivers, whereas we allow for random drivers and replace these domination conditions by a single inclusion at $(0,T)$; see \eqref{EQ:intro_inclusion_exp} below.
We first show that, on $L^\infty$, under standard axioms on a monetary acceptance family (Proposition~\ref{prop: charact-accept-f}), the minimal penalty of the induced risk measure has an integral form:
\[
    c^\A_{s,t}(\Q)=\E_\Q\bigg[\int_s^tf(u,q_u)\d u\,\bigg|\,\F_s\bigg],
\]
where $f$ is a convex function and $q$ is the Girsanov kernel of $\Q$.
To derive this representation from \cite{Delbaen+Peng+RosazzaGianin_2010_Representation_penalty_term_dynamic_concave_utilities}, whose assumptions are formulated for families indexed by stopping times, we show that the induced risk measure admits a càdlàg extension to stopping times (Lemma~\ref{LEM:extension}).
We then prove that a single inclusion at $(0,T)$, namely 
\begin{equation}
\label{EQ:intro_inclusion_exp}
    \A^\g_{0,T}\subseteq\A_{0,T},
\tag{S$_{\mathrm{quad}}$}
\end{equation}
\noeqref{EQ:intro_inclusion_exp} where $\A^\g$ is the acceptance family of the entropic risk measure,  characterizes the monetary acceptance families induced by convex drivers with $0\leq g\leq\frac\g2|z|^2$; the driver is recovered as the convex conjugate of $f$ (Theorem~\ref{THM:BSDE_quad}).
 
Section~\ref{SEC:applications} illustrates the results with four applications and two examples.
First, for leveled acceptance families that are convex and level cash-subadditive, the level decomposition \eqref{EQ:dec_bull_intro} is equivalent to the infimal cocycle property of the induced minimal penalty \cite{Ferrari+Laeven+RosazzaGianin+Zullino_2026_Time-consistency_cash-subadditive_infimal_cocycle} (Corollary~\ref{COR:N-dec-IC}).
Second, the resilience rate and evaluation of \cite{Laeven+Ferrari+Rosazza+Zullino_2025_measuringfinancialresilienceusing, Ferrari+Laeven+Rosazza+Zullino_2026_financialresilienceevaluation}  are well defined for every acceptance family satisfying our axioms, so that their well-posedness can be guaranteed directly in terms of the acceptable positions, and the resilience-acceptance sets of \cite{Laeven+Ferrari+Rosazza+Zullino_2025_measuringfinancialresilienceusing} are described through the acceptance family and the driver (Section~\ref{SUBSEC:resilience}).
Third, for monetary acceptance families induced by Lipschitz or quadratic drivers, law-invariance of $\A_{0,T}$ is equivalent to conditional law-invariance of every $\A_{s,t}$, and it holds only for the conditional expectation  and the entropic risk measures (Corollary~\ref{COR:LI}).
This is a continuous-time version of the result of \cite{Kupper+Schachermayer_2009_Representation_results_law_invariant_time_consistent_functions}, formulated in terms of acceptance sets and derived from \cite{Bensaid+Dumitrescu+Matoussi+Sabbagh_2026_Law-invariant_BSDEs_dynamic_risk_measures}.
Fourth, we determine which of our axioms are satisfied or violated by the acceptance families of the dynamic Value-at-Risk and Conditional Value-at-Risk; in particular, neither of them is induced by a BSDE (Proposition~\ref{CVaR}). 
A related result is that the only normalized, time-consistent, cash-additive dynamic risk measure on $L^\infty$ dominating either of them is the worst-case risk measure (Theorem~\ref{TH:VaR_majorant}).
Finally, two examples, namely stress tests with an interest-rate corridor and capital requirements relative to approved benchmarks (Examples~\ref{EX:corridor} and~\ref{EX:benchmarks}), show how the characterizations are used: an acceptance rule is described through the positions it accepts, the axioms are checked on the sets, and the theorems yield the driver.
 
The paper is organized as follows.
Section~\ref{sec:prel} introduces the setting, the properties of dynamic risk measures, the axioms on acceptance families and the BSDE preliminaries.
Section~\ref{SEC:N} studies leveled acceptance families: the correspondence with risk measures, time-consistency (Section~\ref{SUBSEC:N-TC}) and the characterization through Lipschitz BSDEs (Section~\ref{SUBSEC:N-Lipschitz}).
Section~\ref{SEC:CA} specializes to the cash-additive case: penalty representation (Section~\ref{SUBSEC:CA-penalty}), Lipschitz and quadratic BSDEs (Sections~\ref{SUBSEC:CA-Lipschitz} and~\ref{SUBSEC:CA-quadratic}).
Section~\ref{SEC:applications} contains the applications: infimal cocycle property (Section~\ref{SUBSEC:IC}), resilience (Section~\ref{SUBSEC:resilience}), law-invariance (Section~\ref{SUBSEC:LI}), VaR and CVaR (Section~\ref{SEC:VaR_CVaR}), and a few examples (Section~\ref{SUBSEC:examples}).
Appendix~\ref{APP:correspondence_proofs} collects the proofs of the correspondence results, Appendix~\ref{APP:extension} proves the extension to stopping times used in Section~\ref{SUBSEC:CA-penalty} (Lemma~\ref{LEM:extension}), and Appendix~\ref{APP:examples} contains the verifications for the examples of Section~\ref{SUBSEC:examples}.

\section{Preliminaries}
\label{sec:prel}
In this section we fix the setting and collect the notions that are used throughout the paper: properties of dynamic risk measures (Section~\ref{SUBSEC:P-rho}), acceptance families and their axioms (Section~\ref{SUBSEC:P-acceptance}), and risk measures induced by BSDEs (Section~\ref{SUBSEC:P-BSDE}).

\subsection{Setting and notation}
\label{SUBSEC:P-setting}
 
Fix a finite time horizon $T>0$ and a probability space $(\Omega,\F,\P)$ supporting a $d$-dimensional Brownian motion $W=(W_t)_{t\in[0,T]}$, and let $\mathbb F=(\F_t)_{t\in[0,T]}$ be the augmented filtration generated by $W$, with $\F=\F_T$.
Then $\mathbb F$ satisfies the usual conditions and $\F_0$ is trivial.
For $t\in[0,T]$ and $p\in[1,+\infty]$, we write $L^p(\F_t):=L^p(\Omega,\F_t,\P)$, and $L^p_+(\F_t)$ and $L^p_-(\F_t)$ for the sets of its nonnegative and nonpositive elements, respectively; $L^0\big(\F_t;[-\infty,+\infty]\big)$ denotes the set of $\F_t$-measurable random variables with values in $[-\infty,+\infty]$.
When no argument is specified, $L^p:=L^p(\Omega,\F,\P)$.
Random variables that coincide $\P$-a.s.\ are identified, and equalities and inequalities between random variables are understood $\P$-a.s., unless otherwise stated.
We denote by $\ell$ the Lebesgue measure on Borel sets of $[0,T]$, and by $\mathcal P$ the predictable $\sigma$-algebra on $\Omega\times[0,T]$.
Essential infima and suprema \cite[Section~A.5]{Follmer+Schied_2016_Stochastic_finance} are taken with the conventions $\essinf\varnothing=+\infty$ and $\esssup\varnothing=-\infty$.

For $0\leq s\leq t\leq T$, we denote by $\mathcal Q^{\mathrm{ac}}_{s,t}$ the set of probability measures $\Q$ on $(\Omega,\F_t)$ such that $\Q\ll\P$ on $\F_t$ and $\Q=\P$ on $\F_s$, and by $\mathcal Q_{s,t}$ the set of $\Q\in\mathcal Q^{\mathrm{ac}}_{s,t}$ that are equivalent to $\P$ on $\F_t$.
For $\Q\in\mathcal Q^{\mathrm{ac}}_{s,t}$, since $\Q=\P$ on $\F_s$, Bayes’ formula gives $\E_\Q[Y|\F_s]=\E[(\!\d \Q/ \!\d\P)Y|\F_s]$ for every $\F_t$-measurable $Y$ with $(\!\d \Q/ \!\d\P)Y\in L^1$.
 
\subsection{Dynamic risk functionals}
\label{SUBSEC:P-rho}
 
A \emph{(dynamic) risk functional on $L^p$} is a family $\r=(\r_{s,t})_{0\leq s\leq t\leq T}$ of maps $\r_{s,t}:L^p(\F_t)\to L^0\big(\F_s;[-\infty,+\infty]\big)$, where $\r_{s,t}(X)$ is interpreted as the capital requirement at time $s$ for a financial position $X$ with maturity $t$.
We use the conventions $0\cdot(\pm\infty)=0$ and $+\infty-\infty=+\infty$.
In the following list, each property is required for all $0\leq s\leq t\leq T$ and $X,Y\in L^p(\F_t)$.
We say that a risk functional $\r$ on $L^p$ is
\begin{itemize}[nosep, leftmargin=*, label=$\smallbullet$]
    \item \proplabel{WP}{RM:WP}
        \emph{well posed in $L^p$} if $\r_{s,t}\big(L^p(\F_t)\big)\subseteq L^p$;
    \item \proplabel{M}{RM:M}
        \emph{monotone} if $X\leq Y$ implies $\r_{s,t}(X)\geq\r_{s,t}(Y)$;
    \item \proplabel{CA}{RM:CA}
        \emph{cash-additive} if $\r_{s,t}(X+m)=\r_{s,t}(X)-m$, for all $m \in L^p(\F_s)$;
    \item \proplabel{CC}{RM:CC}
        \emph{conditionally constant} if $\r_{s,t}(\eta)=-\eta$ for all $\eta\in L^p(\F_s)$;
    \item \proplabel{N}{RM:N}
        \emph{normalized} if $\r_{s,t}(0)=0$;
    \item \proplabel{LP}{RM:LP}
        \emph{local} if $\r_{s,t}(\1_AX+\1_{A^c}Y)=\1_A\r_{s,t}(X)+\1_{A^c}\r_{s,t}(Y)$ for all $A\in\F_s$;
    \item \proplabel{Cx}{RM:Cx}
        \emph{convex} if $\r_{s,t}\big(\l X+(1-\l)Y\big)\leq\l\r_{s,t}(X)+(1-\l)\r_{s,t}(Y)$ for all $\l\in[0,1]$;
    \item \proplabel{CCx}{RM:CCx}
        \emph{conditionally convex} if $\r_{s,t}\big(\Lambda X+(1-\Lambda)Y\big)\leq\Lambda\r_{s,t}(X)+(1-\Lambda)\r_{s,t}(Y)$ for all $\Lambda\in L^\infty(\F_s)$ with $0\leq\Lambda\leq1$;
    \item \proplabel{PH}{RM:PH}
        \emph{positively homogeneous} if $\r_{s,t}(\l X)=\l\r_{s,t}(X)$ for all $\l>0$;
    \item \proplabel{CPH}{RM:CPH}
        \emph{conditionally positively homogeneous} if $\r_{s,t}(\Lambda X)=\Lambda\r_{s,t}(X)$ for all $\Lambda\in L^\infty_+(\F_s)$;
    \item \proplabel{SS}{RM:SS}
        \emph{star-shaped} if $\r_{s,t}(\l X)\leq\l\r_{s,t}(X)$ for all $\l\in(0,1]$;
    \item \proplabel{CS}{RM:CS}
        \emph{cash-subadditive} if $\r_{s,t}(X-k)\leq\r_{s,t}(X)+k$ for all $k\in L^p_+(\F_s)$;
    \item \proplabel{C$\downarrow$}{RM:Cabove}
        \emph{continuous from above} if $X_n\downarrow X$ $\P$-a.s., with $X_n\in L^p(\F_t)$, implies $\r_{s,t}(X_n)\uparrow\r_{s,t}(X)$ $\P$-a.s.;
    \item \proplabel{C$\uparrow$}{RM:Cbelow}
        \emph{continuous from below} if $X_n\uparrow X$ $\P$-a.s., with $X_n\in L^p(\F_t)$, implies $\r_{s,t}(X_n)\downarrow\r_{s,t}(X)$ $\P$-a.s.;
    \item \proplabel{TC}{RM:TC}
        \emph{time-consistent} if it is well posed in $L^p$ and $\r_{s,t}(X)=\r_{s,u}\big(-\r_{u,t}(X)\big)$ for all $u\in[s,t]$.
\end{itemize}
A \emph{(dynamic) risk measure on $L^p$} is a risk functional that is well posed in $L^p$ and monotone.
These notions are all standard; see, e.g., \cite{Artzner+Delbaen+Eber+Heath_1999_Coherent_measures_risk} and \cite[Chapter~4]{Follmer+Schied_2016_Stochastic_finance} in the static case, \cite{Detlefsen+Scandolo_2005_Conditional_dynamic_convex_risk_measures,Cheridito+Delbaen+Kupper_2006_Dynamic_monetary_risk_measures_bounded_discrete-time_processes,Bion-Nadal_2008_Dynamic_risk_measures_time_consistency__risk_measures_BMO_martingales} and \cite[Chapter~11]{Follmer+Schied_2016_Stochastic_finance} in the dynamic case, \cite{ElKaroui+Ravanelli_2009_Cash_sub-additive_risk_measures_interest_rate_ambiguity} for cash-subadditivity and \cite{Laeven+RosazzaGianin+Zullino_2023_Dynamic_return_star-shaped_risk_measures_BSDEs} for star-shapedness.
Time-consistency is also known as recursiveness; see, e.g., \cite[Theorem~12]{Delbaen_2006_Structure_m-stable_sets_particular_set_risk_neutral_measures}, \cite[Lemma 11.11]{Follmer+Schied_2016_Stochastic_finance}, \cite{Bion-Nadal_2008_Dynamic_risk_measures_time_consistency__risk_measures_BMO_martingales}.

\subsection{Acceptance families}
\label{SUBSEC:P-acceptance}

For $p\in[1,+\infty]$, a \emph{leveled acceptance family in $L^p$} is a collection $\A^\tinybullet\!=\!(\A^m_{s,t})_{0\leq s\leq t\leq T,\,m\in L^p(\F_s)}$ of sets $\A^m_{s,t}\subseteq L^p(\F_t)$, where $\A^m_{s,t}$ is interpreted as the set of $\F_t$-measurable positions whose risk, assessed at time $s$, does not exceed the level $m$.
This is the dynamic version of the risk acceptance families of \cite{Drapeau+Kupper_2013_Risk_preferences_robust_representation}.
The risk functional induced by $\A^\tinybullet$ is given by
\begin{equation}
\label{eq: rho given A-m}
    \rho_{s,t}^{\mathcal{A}^\tinybullet}(X):=\essinf\{m \in L^{p}(\F_s) : X \in \mathcal{A}^{m}_{s,t}\}, \qquad X \in L^{p}(\mathcal{F}_t).
\end{equation}
Vice versa, every dynamic risk functional $\r$ as in Section~\ref{SUBSEC:P-rho} induces the leveled acceptance family $\A^{\tinybullet,\r}$ given by
\begin{equation}
\label{eq: A-m given rho}
    \mathcal{A}^{m, \rho}_{s,t}:=\{ X \in L^{p}(\mathcal{F}_t) : \rho_{s,t}(X) \leq m\}, \qquad m\in L^p(\F_s).
\end{equation}

A \emph{monetary acceptance family in $L^p$} is a collection $\A=(\A_{s,t})_{0\leq s\leq t\leq T}$ of sets $\A_{s,t}\subseteq L^p(\F_t)$, where $\A_{s,t}$ is interpreted as the set of $\F_t$-measurable positions that are acceptable at time $s$.
It induces the cash-additive risk functional
\begin{equation}
\label{eq: rho given A}
    \rho^{\mathcal{A}}_{s,t}(X):=\essinf\{m \in L^{p}(\F_s) : m+X \in \mathcal{A}_{s,t}\}, \qquad X \in L^{p}(\mathcal{F}_t),
\end{equation}
and, vice versa, a risk functional $\r$ induces the monetary acceptance family
\begin{equation}
\label{eq: A given rho}
    \mathcal{A}^{\rho}_{s,t}:=\mathcal{A}^{0,\rho}_{s,t}=\big\{ X \in L^{p}(\mathcal{F}_t) : \rho_{s,t}(X) \leq 0\big\};
\end{equation}
see, e.g., \cite{Detlefsen+Scandolo_2005_Conditional_dynamic_convex_risk_measures}, \cite[Section~3]{Cheridito+Delbaen+Kupper_2006_Dynamic_monetary_risk_measures_bounded_discrete-time_processes} and \cite[Chapter~11]{Follmer+Schied_2016_Stochastic_finance}.
Every monetary acceptance family $\A$ induces the leveled acceptance family $\A^m_{s,t}:=\A_{s,t}-m$, and \eqref{eq: rho given A-m} and \eqref{eq: rho given A} give $\r^{\A^\tinybullet}=\r^\A$, which is cash-additive by direct inspection.
Conversely, if $\r$ is cash-additive, then $\A^{m,\r}_{s,t}=\A^\r_{s,t}-m$.
Thus, monetary acceptance families describe cash-additive risk functionals, whereas leveled acceptance families describe the general ones.
When no ambiguity arises, we write $\r$ instead of $\r^{\A^\tinybullet}$ or $\r^\A$, and we drop the dependence of $\A^\r$ and $\A^{\tinybullet,\r}$ on $\r$.

\afterpage{%
\begin{landscape}
\begin{table}[p]
\centering
\renewcommand{\arraystretch}{1.2}
\setlength{\tabcolsep}{3pt}
\begin{tabular}{@{}l|l>{$\textstyle}p{7.7cm}<{$}|l>{$\textstyle}p{7.9cm}<{$}@{}}
    & \multicolumn{2}{l|}{\textbf{Leveled family} $\A^\tinybullet$} & \multicolumn{2}{l}{\textbf{Monetary family} $\A$}\\
    \hline
    \multicolumn{5}{c}{\emph{Axioms on the levels}}\\
    level monotonicity
        & (\axSp{lmon})\proplabel{\axSp{lmon}}{IT:L-monotone} & m\leq n\ \Rightarrow\ \A^m_{s,t}\subseteq\A^n_{s,t}
        & & \text{\gr{implied by \itemref{IT:S-monotone}}}\\
    level right-continuity
        & (\axSp{lrc})\proplabel{\axSp{lrc}}{IT:L-rightcont} & \ds\bigcap_{k}\A^{n_k}_{s,t}\subseteq\A^n_{s,t}\ \text{ if } n_k\downarrow n \text{ a.s., } n_k\in L^p(\F_s)
        & & \text{\gr{implied by \itemref{IT:S-closed}}}\\
    level cash-subadditivity
        & (\axSp{csa})\proplabel{\axSp{csa}}{IT:N-cashsub} & \A^m_{s,t}\subseteq\A^{m+k}_{s,t}+k,\ \ k\in L^p_+(\F_s)
        & & \text{\gr{automatic}}\\
    level cash-additivity
        & (\axSp{ca})\proplabel{\axSp{ca}}{IT:N-cashadd} & \A^0_{s,t}=\A^m_{s,t}+m
        & & \text{\gr{automatic}}\\
    \hline
    \multicolumn{5}{c}{\emph{Axioms on the sets}}\\
    upper boundedness
        & (\axSp{ub})\proplabel{\axSp{ub}}{IT:N-wellposed} & L^p(\F_t)=\ds\bigcup_{m\in L^p(\F_s)}\A^m_{s,t}
        & (\axS{ub})\proplabel{\axS{ub}}{IT:S-upperbound} & L^p(\F_t)=\A_{s,t}+L^p(\F_s)\\
    lower boundedness
        & (\axSp{lb})\proplabel{\axSp{lb}}{IT:N-lowerbound} & \forall X\in L^p(\F_t)\ \exists \, k\in L^p(\F_s):\ X\in\A^m_{s,t}\Rightarrow m\geq k
        & (\axS{lb})\proplabel{\axS{lb}}{IT:S-lowerbound} & \forall X\in L^p(\F_t)\ \exists \, k\in L^p(\F_s):\ X+m\in\A_{s,t}\Rightarrow m\geq k\\
    monotonicity
        & (\axSp{mon})\proplabel{\axSp{mon}}{IT:N-monotone} & \A^m_{s,t}+L^p_+(\F_t)\subseteq\A^m_{s,t}
        & (\axS{mon})\proplabel{\axS{mon}}{IT:S-monotone} & \A_{s,t}+L^p_+(\F_t)\subseteq\A_{s,t}\\
    convexity
        & (\axSp{cx})\proplabel{\axSp{cx}}{IT:N-convex} & \l\A^m_{s,t}+(1-\l)\A^n_{s,t}\subseteq\A^{\l m+(1-\l)n}_{s,t},\ \ \l\in[0,1]
        & (\axS{cx})\proplabel{\axS{cx}}{IT:S-convex} & \A_{s,t}\text{ is convex}\\
    conicity
        & (\axSp{cone})\proplabel{\axSp{cone}}{IT:N-conic} & \l\A^m_{s,t}\subseteq\A^{\l m}_{s,t},\ \ \l>0
        & (\axS{cone})\proplabel{\axS{cone}}{IT:S-conic} & \A_{s,t}\text{ is a cone}\\
    star-shapedness
        & (\axSp{star})\proplabel{\axSp{star}}{IT:N-star} & \l\A^m_{s,t}\subseteq\A^{\l m}_{s,t},\ \ \l\in(0,1]
        & (\axS{star})\proplabel{\axS{star}}{IT:S-star} & \A_{s,t}\text{ is star-shaped}\\
    closedness
        & (\axSp{cl})\proplabel{\axSp{cl}}{IT:N-closed} & \A^m_{s,t}\text{ is closed }\begin{cases}\text{in } L^p & \text{if } p<\infty,\\ \text{in } \sigma(L^\infty,L^1) & \text{if } p=\infty\end{cases}
        & (\axS{cl})\proplabel{\axS{cl}}{IT:S-closed} & \A_{s,t}\text{ is closed }\begin{cases}\text{in } L^p & \text{if } p<\infty,\\ \text{in } \sigma(L^\infty,L^1) & \text{if } p=\infty\end{cases}\\
    conditional constancy
        & (\axSp{cc})\proplabel{\axSp{cc}}{IT:N-constancy} & \A^m_{s,t}\cap L^p(\F_s)=\{\eta\in L^p(\F_s):\eta\geq-m\}
        & (\axS{cc})\proplabel{\axS{cc}}{IT:S-constancy} & \A_{s,t}\cap L^p(\F_s)=L^p_+(\F_s)\\
    normalization
        & (\axSp{nor})\proplabel{\axSp{nor}}{IT:N-normalized} & \essinf\{m\in L^p(\F_s): 0\in\A^m_{s,t}\}=0
        & (\axS{nor})\proplabel{\axS{nor}}{IT:S-normalized} & \essinf\big(\A_{s,t}\cap L^p(\F_s)\big)=0\\
    locality
        & (\axSp{loc})\proplabel{\axSp{loc}}{IT:N-pasting} & \1_A\A^m_{s,t}+\1_{A^c}\A^n_{s,t}\subseteq\A^{\1_Am+\1_{A^c}n}_{s,t},\ \ A\in\F_s
        & (\axS{loc})\proplabel{\axS{loc}}{IT:S-pasting} & \1_A\A_{s,t}+\1_{A^c}\A_{s,t}\subseteq\A_{s,t},\ \ A\in\F_s\\
    expectation-boundedness
        & (\axSp{exp})\proplabel{\axSp{exp}}{IT:N-expectation} & \E[X|\F_s]\geq-m\ \text{ for } X\in\A^m_{s,t}
        & (\axS{exp})\proplabel{\axS{exp}}{IT:S-expectation} & \E[X|\F_s]\geq0\ \text{ for } X\in\A_{s,T}\\
    \hline
    \multicolumn{5}{c}{\emph{Axioms across time} (for all $u\in[s,t]$)}\\
    decomposition
        & (\axSp{dec})\proplabel{\axSp{dec}}{IT:N-decomposition} & \A^m_{s,t}=\ds\bigcup_{n\in\A^m_{s,u}}\A^{-n}_{u,t}
        & (\axS{dec})\proplabel{\axS{dec}}{IT:S-decomposition} & \A_{s,t}=\A_{s,u}+\A_{u,t}\\
    \hline
    \multicolumn{5}{c}{\emph{Axioms for BSDEs} (for some $\mu,\g>0$)}\\
    Lipschitz domination
        & (\axSp{Lip})\proplabel{\axSp{Lip}}{IT:N-SL} & \A^m_{s,t}+\A^{n,\mu}_{s,t}\subseteq\A^{m+n}_{s,t}
        & (\axS{Lip})\proplabel{\axS{Lip}}{IT:S-SL} & \A_{s,t}+\A^\mu_{s,t}\subseteq\A_{s,t}\\
    entropic bound
        & & \text{\gr{not considered}}
        & (\axS{quad})\proplabel{\axS{quad}}{IT:S-SQ} & \A^\g_{0,T}\subseteq\A_{0,T}\\
    \hline
\end{tabular}
\vspace{-4.3pt}
\caption{Axioms on acceptance families in $L^p$, $p\in[1,+\infty]$, required for all $0\leq s\leq u\leq t\leq T$, all $m,n\in L^p(\F_s)$, and some $\mu,\gamma>0$.
}
\label{TAB:axioms}
\end{table}
\end{landscape}
}

\paragraph{Axioms.}
Table~\ref{TAB:axioms} lists the axioms we consider for a leveled or monetary acceptance family.
Since $\A^m_{s,t}$ plays the role of the sublevel set $\{\r_{s,t}\leq m\}$ of the induced risk functional $\r=\r^{\A^\tinybullet}$, each axiom is best read as a property of sublevel sets.
The axioms do not play the same role in our results: a few of them are the basic assumptions of our characterizations, namely regularity (Definition~\ref{DEF:regular}) and the domination axioms for BSDEs, while the others either follow from these or characterize properties of the induced risk functional.
\begin{itemize}[nosep, leftmargin=*, label=\scriptsize$\bullet$]
    \item
        \emph{The axioms on the levels} describe how $\A^m_{s,t}$ changes with $m$.
        Under level monotonicity \itemref{IT:L-monotone}, a higher level accepts more positions.
        Level right-continuity \itemref{IT:L-rightcont} says that a position accepted at every level slightly above $m$ is accepted at $m$; together with level monotonicity and locality, it ensures that the essential infimum in \eqref{eq: rho given A-m} is itself an acceptable level, so that $\A^m_{s,t}$ is exactly the sublevel set $\{\r_{s,t}\leq m\}$ (Lemma~\ref{LEM:N-from-A-to-rho}).
        Level cash-additivity \itemref{IT:N-cashadd} and cash-subadditivity \itemref{IT:N-cashsub} say that adding $k$ units of cash to a position lowers its level by exactly $k$, respectively by at most $k$.
    \item
        \emph{Upper and lower boundedness} make the risk finite: under \itemref{IT:N-wellposed} every position is accepted at some level, so that $\r<+\infty$, and under \itemref{IT:N-lowerbound} no position is accepted at all levels, so that $\r>-\infty$.
    \item
        \emph{The other axioms on the sets} are the set-valued versions of the usual properties of risk functionals.
        The correspondence is discussed in Lemmas \ref{LEM:N-from-A-to-rho}, \ref{LEM:N-from-rho-to-A}, \ref{LEM:from_A_to_rho} and \ref{LEM:from_rho_to_A}.
    \item
        \emph{The decomposition axioms} link different times.
        \itemref{IT:N-decomposition} will be motivated in Section~\ref{SUBSEC:N-TC}: $X$ is accepted at time $s$ at level $m$ if and only if, at the intermediate time $u$, it is accepted at a level $n$ such that the position $-n$ is itself accepted at time $s$ at level $m$.
        In the monetary case it simplifies to \itemref{IT:S-decomposition}: a position is acceptable over $[s,t]$ if and only if it is the sum of a position acceptable over $[s,u]$ and one acceptable over $[u,t]$.
    \item 
        \emph{The axioms for BSDEs} involve the acceptance families $\A^{\tinybullet,\mu},\A^\mu,\A^\g$ of the reference drivers $\mu(|y|+|z|)$, $\mu|z|$ and $\frac\g2|z|^2$, introduced in Sections~\ref{SUBSEC:N-Lipschitz}, \ref{SUBSEC:CA-Lipschitz} and \ref{SUBSEC:CA-quadratic}, respectively, where they are interpreted.
\end{itemize}
The monetary axioms are the leveled ones for the embedding $\A^m_{s,t}:=\A_{s,t}-m$ (with the exception of \itemref{IT:S-expectation}, which is only required for $t=T$); for this embedding, the axioms on the levels are automatic or implied by the monetary ones, see Remark~\ref{REM:P-embedding}.

\begin{remark}[Literature]
\label{REM:P-axioms-literature}
    The domination axioms \itemref{IT:N-SL}, \itemref{IT:S-SL} and \itemref{IT:S-SQ} appear here for the first time.
    The other axioms coincide with, or are adaptations of, axioms in the literature:
    \begin{itemize}[nosep, leftmargin=*, label=\scriptsize$\bullet$]
        \item \emph{Leveled axioms.}
            In the static case, \itemref{IT:L-monotone}, \itemref{IT:N-monotone} and \itemref{IT:L-rightcont} are part of the definition of a risk acceptance family \cite[Definition~3]{Drapeau+Kupper_2013_Risk_preferences_robust_representation}, where \itemref{IT:L-rightcont} reads $\A^m=\bigcap_{n>m}\A^n$.
            Moreover, \itemref{IT:N-convex} and \itemref{IT:N-conic} are in \cite[Proposition~1]{Drapeau+Kupper_2013_Risk_preferences_robust_representation}, \itemref{IT:N-cashadd} and \itemref{IT:N-cashsub} in \cite[Propositions~2 and~3]{Drapeau+Kupper_2013_Risk_preferences_robust_representation}, and \itemref{IT:N-closed} in \cite[Remark~6]{Drapeau+Kupper_2013_Risk_preferences_robust_representation}.
            Unlike \cite{Drapeau+Kupper_2013_Risk_preferences_robust_representation}, we do not require the sets $\A^m_{s,t}$ to be convex.
        \item \emph{Monetary axioms.}
            \itemref{IT:S-monotone}, \itemref{IT:S-normalized}, \itemref{IT:S-pasting} and conditional versions of \itemref{IT:S-convex}, \itemref{IT:S-conic} can be found in \cite[Proposition~3.6]{Cheridito+Delbaen+Kupper_2006_Dynamic_monetary_risk_measures_bounded_discrete-time_processes}.
            \itemref{IT:S-decomposition} characterizes time-consistency: \cite[Theorem~4.6]{Cheridito+Delbaen+Kupper_2006_Dynamic_monetary_risk_measures_bounded_discrete-time_processes}, \cite[Theorem~2.5]{Bion-Nadal_2008_Dynamic_risk_measures_time_consistency__risk_measures_BMO_martingales}, \cite[Proposition~11.15]{Follmer+Schied_2016_Stochastic_finance}.
            \itemref{IT:S-expectation} is assumption~(h) of \cite[Section~3]{Delbaen+Peng+RosazzaGianin_2010_Representation_penalty_term_dynamic_concave_utilities}.
            The static versions of  \itemref{IT:S-convex}, \itemref{IT:S-conic}, \itemref{IT:S-closed} and \itemref{IT:S-star} are in \cite[Propositions~4.6, 4.7 and Theorem~4.33]{Follmer+Schied_2016_Stochastic_finance} and  \cite{Castagnoli+Cattelan+Maccheroni+Tebaldi+Wang_2022_Star-shaped_risk_measures}, respectively.
    \end{itemize}
\end{remark}

\begin{remark}[Basic relations between the axioms]
\label{REM:P-embedding}
\label{REM:not_empty}
    Fix a leveled acceptance family $\A^\tinybullet$ in $L^p$.
    \begin{multicols}{2}
    \begin{itemize}[nosep, leftmargin=*, label=\scriptsize$\bullet$]
        \item \itemref{IT:N-cashadd} $\Rightarrow$ \itemref{IT:N-cashsub}
        \item \itemref{IT:N-constancy} $\Rightarrow$ \itemref{IT:N-normalized}
        \item \itemref{IT:N-conic} $\Rightarrow$ \itemref{IT:N-star}
        \item \itemref{IT:N-convex} $+$ $\big(0\in\A^0_{s,t}\big)$ $\Rightarrow$ \itemref{IT:N-star}
    \end{itemize}
    \end{multicols}
    \noindent Moreover, if $\A^\tinybullet$ satisfies \itemref{IT:N-cashadd}, then $\A_{s,t}:=\A^0_{s,t}=\A_{s,t}^m+m$ is a monetary acceptance family and $\r^{\A^\tinybullet}=\r^{\A}$ is cash-additive.
    
    Conversely, if $\A$ is a monetary acceptance family in $L^p$, then $\A^m_{s,t}:=\A_{s,t}-m$ defines a leveled acceptance family that automatically satisfies \itemref{IT:N-cashadd}, hence \itemref{IT:N-cashsub}, and $\r^{\A^\tinybullet}=\r^\A$ is cash-additive.
    Moreover:
    \begin{itemize}[nosep, leftmargin=*, label=\scriptsize$\bullet$]
        \item 
            (S$_j$) $\Leftrightarrow$ (S$^\tinybullet_j$) for $j\in\{\mathrm{ub},\mathrm{lb},\mathrm{mon},\mathrm{cx},\mathrm{cone},\mathrm{star},\mathrm{cl},\mathrm{cc},\mathrm{nor},\mathrm{loc},\mathrm{dec}\}$;
    \end{itemize}
    {\setlength{\multicolsep}{0pt}
    \begin{multicols}{2}
        \begin{itemize}[nosep, leftmargin=*, label=\scriptsize$\bullet$]
        \item 
            \itemref{IT:S-monotone} $\Rightarrow$ \itemref{IT:L-monotone};
        \item 
            \itemref{IT:N-expectation}  $\Rightarrow$  \itemref{IT:S-expectation};
        \end{itemize}
    \end{multicols}}
    \begin{itemize}[nosep, leftmargin=*, label=\scriptsize$\bullet$]
        \item 
            \itemref{IT:S-closed} $\Rightarrow$ $\A_{s,t}$ is closed under $\P$-a.s.\ decreasing limits in $L^p$ $\Rightarrow$ \itemref{IT:L-rightcont};
        \item 
            \itemref{IT:S-normalized} $\Rightarrow$ $\A_{s,t}\cap L^p(\F_s)\neq\varnothing$, since $\essinf\varnothing=+\infty$.
    \end{itemize}
    
    Most of these implications 
    follow directly from the definitions.
    If \itemref{IT:S-closed} holds, $X_n\in\A_{s,t}$, $X_n\downarrow X$ $\P$-a.s.\ and $X\in L^p$, then $X_n\to X$ in $L^p$ if $p<\infty$, and $\E[ZX_n]\to\E[ZX]$ for all $Z\in L^1$ if $p=\infty$, by dominated convergence; hence $X\in\A_{s,t}$.
    Applied to $X+n_k\downarrow X+n$, this gives \itemref{IT:L-rightcont}.

    The domination axioms \itemref{IT:N-SL}, \itemref{IT:S-SL} and \itemref{IT:S-SQ} will be discussed in Remarks~\ref{REM:N-SL} and~\ref{REM:SL_implies_S1} and in Lemma~\ref{LEM:SQ_f}, respectively.
\end{remark}

 \begin{definition}
\label{DEF:regular}
    A leveled acceptance family is \emph{regular} if it satisfies normalization \itemref{IT:N-normalized}, locality \itemref{IT:N-pasting}, decomposition \itemref{IT:N-decomposition} and level right-continuity \itemref{IT:L-rightcont}.
    A monetary acceptance family is \emph{regular} if it satisfies normalization \itemref{IT:S-normalized}, locality \itemref{IT:S-pasting}, decomposition \itemref{IT:S-decomposition} and closedness \itemref{IT:S-closed}.
\end{definition}
By Remark~\ref{REM:P-embedding}, if a monetary acceptance family $\A$ is regular, then so is the induced leveled family $\A^m_{s,t}:=\A_{s,t}-m$.

\begin{example}
\label{EX:P-AE}
    For $p\in[1,+\infty]$, let
    \[
        \A^{\E}_{s,t}:=\big\{X\in L^p(\F_t)\ :\ \E[X|\F_s]\geq0\big\},\qquad 0\leq s\leq t\leq T.
    \]
    Then $\r^{\A^\E}_{s,t}(X)=\E[-X|\F_s]$, and $\A^\E$ satisfies all the monetary axioms on the sets and across time of Table~\ref{TAB:axioms}.
    They follow directly from the properties of the conditional expectation, except possibly \itemref{IT:S-closed} with $p=\infty$ and \itemref{IT:S-decomposition}.
    For \itemref{IT:S-closed} with $p=\infty$, note that $\A^\E_{s,t}=\bigcap_{B\in\F_s}\{X\in L^\infty(\F_t) : \E[X\1_B]\geq0\}$ is an intersection of $\sigma(L^\infty,L^1)$-closed half-spaces.
    For \itemref{IT:S-decomposition}, every $X\in\A^\E_{s,t}$ can be written as $X=\E[X|\F_u]+(X-\E[X|\F_u])$, with $\E[X|\F_u]\in\A^\E_{s,u}$ and $X-\E[X|\F_u]\in\A^\E_{u,t}$, while $\A^\E_{s,u}+\A^\E_{u,t}\subseteq\A^\E_{s,t}$ follows from the tower property.
\end{example}

\subsection{BSDE-induced risk measures and acceptance families}
\label{SUBSEC:P-BSDE}

Whenever BSDEs are considered, $p\in(1,+\infty]$; see Remark~\ref{REM:N-p} for $p=1$.
A \emph{driver} is a $\mathcal P\otimes\mathcal B(\R)\otimes\mathcal B(\R^d)$-measurable function $g:\Omega\times[0,T]\times\R\times\R^d\to\R:(\o,u,y,z)\mapsto g(\o,u,y,z)$.
Drivers that coincide outside a $\P\otimes\ell$-null predictable set are identified; thus, a property holds for a driver if it is satisfied by at least one representative.
A driver is \emph{monetary} if it does not depend on $y$ (we then write $g(\o,u,z)$), \emph{convex} if $g(\o,u,\cdot,\cdot)$ is convex, and \emph{star-shaped} if $g(\o,u,\l y,\l z)\leq\l g(\o,u,y,z)$ for all $\l\in(0,1]$ and $(y,z)$.
For $t\in[0,T]$ and $X\in L^0(\F_t)$, a \emph{solution} of the BSDE with parameters $(g,t,X)$ is a pair $(Y,Z)$, with $Y$ continuous adapted and $Z$ predictable, such that $\int_0^t\big(|Z_u|^2+|g(u,Y_u,Z_u)|\big)\d u<\infty$ a.s.\ and
\begin{equation}
\label{EQ:N-BSDE}
    Y_r=-X+\int_r^tg(u,Y_u,Z_u)\d u-\int_r^tZ_u\cdot\d W_u,\qquad r\in[0,t].
\end{equation}
The BSDEs with driver $g$ are \emph{well posed in $L^p$} if, for all $t$ and $X\in L^p(\F_t)$, \eqref{EQ:N-BSDE} has a unique solution with $\sup_{r\leq t}|Y_r|\in L^p$.
In this case, we set $\r^g_{s,t}(X):=Y_s$, $0\leq s\leq t\leq T$, and denote the leveled acceptance family induced by $\r^g$ by $\A^{\tinybullet,g}:=\A^{\tinybullet,\r^g}$.
Analogously, when $\r^g$ is cash-additive, we simply write $\A^g:=\A^{\r^g}$.
We say that $\A^\tinybullet$, $\A$ or $\rho$ is induced by a driver $g$ if $\A^\tinybullet=\A^{\tinybullet,g}$, $\A=\A^g$, or $\r=\r^g$ on $L^p$, respectively.
 
\begin{definition}
\label{DEF:P-driver}
    Let $\mu,\gamma>0$.
    A \emph{$\mu$-Lipschitz driver} is a driver $g$ such that $g(\,\cdot\,,0,0)=0$ and $|g(\,\cdot\,,y,z)-g(\,\cdot\,,y',z')|\leq\mu\big(|y-y'|+|z-z'|\big)$ for all $(y,z),(y',z')\in\R\times\R^d$.
    A \emph{quadratic convex driver with constant $\g$} is a convex monetary driver with $0\leq g(\cdot,z)\leq\frac\g2|z|^2$.
\end{definition}

If $g$ is a $\mu$-Lipschitz driver, the BSDEs with driver $g$ are well posed in $L^p$ for every $p\in(1,+\infty]$ \cite[Theorems~4.3.1 and~4.4.4]{Zhang_2017_Backward_stochastic_differential_equations}, \cite[Theorem~4.2]{Briand+Delyon+Hu+Pardoux+Stoica_2003_Lp_solutions_BSDEs}, and the following standard facts hold for all $0\le s\le t\le T$ and $X,X'\in L^p(\F_t)$ \cite{ElKaroui+Peng+Quenez_1997_Backward_stochastic_differential_equations_finance,Peng_1997_Backward_SDE_related_g-expectation,RosazzaGianin_2006_Risk_measures_g-expectations,Zhang_2017_Backward_stochastic_differential_equations}:
\begin{enumerate}[label=(F\arabic*),ref=F\arabic*, nosep, leftmargin=*]
    \item \label{IT:B-consistency}
        $\r^g_{s,t}(X)$ does not depend on $p$;
    \item \label{IT:B-stability}
        $\|\r^g_{s,t}(X)-\r^g_{s,t}(X')\|_{L^p}\leq C\|X-X'\|_{L^p}$, $C=C(p,\mu,T)$;
    \item \label{IT:B-comparison}
        if $h$ is a $\mu$-Lipschitz driver, $g\leq h$ and $X\geq X'$, then $\r^g_{s,t}(X)\leq\r^{h}_{s,t}(X')$;
    \item \label{IT:B-flow}
        $\r^g$ is time-consistent and local, and conditionally constant if $g(\,\cdot\,,y,0)=0$;
    \item \label{IT:B-convex}\label{IT:B-cash}
        $g$ convex $\Rightarrow$ $\r^g$ convex \cite[Theorem~6.7]{Barrieu+ElKaroui_2009_Pricing_hedging_optimally_designing_derivatives_minimization_risk_measures}; $g$ monetary $\Rightarrow$ $\r^g$ cash-additive \cite[Lemma~1.1(iv)]{Jiang_2008_Convexity_translation_invariance_subadditivity_g-expectations_related_risk_measures}.
\end{enumerate}
In particular, $\r^g$ is a normalized dynamic risk measure on $L^p$.
 
\smallskip
For $\Q\in\mathcal{Q}_{s,t}$, the \emph{Girsanov kernel} of $\Q$ is the $\P\otimes\ell$-a.e.\ unique predictable process $q:\Omega\times[0,t]\to\R^d$ such that $\int_0^t|q_u|^2\d u<+\infty$ $\P$-a.s., $q=0$ on $[0,s]$ and $\d\Q/\d\P=\mathcal E_t\big(\int  q\cdot\d W\big)$, where $\mathcal E$ is the stochastic exponential.
For $h:\Omega\times[0,T]\times\R^d\to(-\infty,+\infty]$, 
\[
    h^\ast(\o,u,q):=\sup_{z\in\R^d}\{q\cdot z-h(\o,u,z)\}
\]
is its convex conjugate in the last variable.
If $h$ is $\mathcal P\otimes\mathcal B(\R^d)$-measurable and $h(\o,u,\cdot)$ is finite and convex for every $(\o,u)$, then $h(\o,u,\cdot)$ is continuous, so that the supremum can be taken over vectors $z$ of rationals, ensuring $\mathcal P\otimes\mathcal B(\R^d)$-measurability of $h^\ast$.
If moreover $h\geq0$ and $h(\cdot,0)=0$, then $h^\ast\geq0=h^\ast(\cdot,0)$.

The next lemma is a special case of \cite[Theorem~3.2(i)]{Fan+Hu+Tang_2025_Unbounded_dynamic_concave_utilities_BSDEs}; see also \cite{Kobylanski_2000_Backward_stochastic_differential_equations_partial_differential_equations_quadratic_growth,Briand+Hu_2008_Quadratic_BSDEs_convex_generators_unbounded_terminal_conditions} for well-posedness and \cite[Theorem~7.4]{Barrieu+ElKaroui_2009_Pricing_hedging_optimally_designing_derivatives_minimization_risk_measures}, \cite{Delbaen+Hu+Bao_2011_Backward_SDEs_superquadratic_growth,Delbaen+Peng+RosazzaGianin_2010_Representation_penalty_term_dynamic_concave_utilities} for the dual representation.
 
\begin{lemma}
\label{LEM:quad_BSDE}
    Let $g$ be a quadratic convex driver with constant $\g$, $t\in[0,T]$ and $X\in L^\infty(\F_t)$.
    The BSDE with parameters $(g,t,X)$ has a unique solution $(Y,Z)$ with $Y$ bounded; moreover $|Y_s|\leq\|X\|_{L^\infty}$ and
    \begin{equation}
    \label{EQ:dual_quad}
        Y_s=\essmax_{\Q\in\mathcal Q_{s,t}}\E_\Q\bigg[-X-\int_s^tg^\ast(u,q_u)\d u\,\bigg|\,\F_s\bigg],\qquad s\in[0,t].
    \end{equation}
    In particular, the BSDEs with driver $g$ are well posed in $L^\infty$.
\end{lemma}

\begin{proof}
    Let $f:=g^\ast$.
    Then $f(\o,u,\cdot)$ is convex and lower semicontinuous, $f(\cdot,0)=0$, $f(\cdot,q)\geq|q|^2/(2\g)$ and $f^\ast=g$, so \cite[Theorem~3.2(i)]{Fan+Hu+Tang_2025_Unbounded_dynamic_concave_utilities_BSDEs}, applied on $[0,t]$ with the sign change between concave utilities and risk measures, gives a unique solution $(Y,Z)$ with $\sup_{u\leq t}|Y_u|$ exponentially integrable, and \eqref{EQ:dual_quad} with the essential supremum over the $\Q\in\mathcal Q_{s,t}$ with $\E_\Q[\int_s^tf(u,q_u)\d u]<\infty$, attained at some $\hat\Q\in\mathcal Q_{s,t}$ whose kernel satisfies  $\hat q_u\in\partial g(u,Z_u)$.
    Every other $\Q\in\mathcal Q_{s,t}$ can be added: for $A_k:=\big\{\E_\Q\big[\int_s^tf(u,q_u)\d u\,\big|\,\F_s\big]\leq k\big\}\in\F_s$, the measure $\1_{A_k}\Q+\1_{A_k^c}\P$ belongs to $\mathcal Q_{s,t}$ (since $\Q=\P$ on $\F_s$), has integrable penalty and coincides with $\Q$ on $A_k$, while on $\bigcap_kA_k^c$ the right-hand side is $-\infty$.
    Finally, since $f\geq0$, the right-hand side of \eqref{EQ:dual_quad} is at most $\|X\|_{L^\infty}$, and by choosing $\Q=\P$, it is at least $-\|X\|_{L^\infty}$; hence $|Y_s|\leq\|X\|_{L^\infty}$, so $Y$ is bounded.
    Since every solution with bounded first component belongs to the uniqueness class of \cite[Theorem~3.2(i)]{Fan+Hu+Tang_2025_Unbounded_dynamic_concave_utilities_BSDEs}, it coincides with $(Y,Z)$.
\end{proof}

\section{Leveled acceptance families}
\label{SEC:N}

Throughout, we work with a leveled acceptance family  $\A^\tinybullet$ in $L^p$, for some $p\in[1,+\infty]$.
We characterize time-consistency of  $\r^{\A^\tinybullet}$ through \itemref{IT:N-decomposition} in Section~\ref{SUBSEC:N-TC}, and the families induced by Lipschitz BSDEs through \itemref{IT:N-SL} in Section~\ref{SUBSEC:N-Lipschitz}.
Only Section~\ref{SUBSEC:N-Lipschitz} uses the Brownian structure; all the other results hold in any filtration satisfying the usual conditions.

We start with a technical lemma: under \itemref{IT:N-pasting}, the essential infimum that defines $\r^{\A^\tinybullet}$ in \eqref{eq: rho given A-m}  is the limit of a decreasing sequence of acceptable levels.

\begin{lemma}
\label{LEM:N-directedness}
    Assume that $\A^\tinybullet$ satisfies \itemref{IT:N-pasting}.
    For any $X\in L^p(\F_t)$ such that $\r^{\A^\tinybullet}_{s,t}(X)<+\infty$, there exists $(m_k)_{k\in\N}\subset L^p(\F_s)$ such that $X\in\A^{m_k}_{s,t}$ for every $k\in\N$ and $m_k\downarrow\r^{\A^\tinybullet}_{s,t}(X)$ $\P$-a.s.
\end{lemma}

\begin{proof}
    For $X\in L^p(\F_t)$, set $D^X_{s,t}:=\{m\in L^p(\F_s)\ :\ X\in\A^m_{s,t}\}$, which is nonempty because $\r^{\A^\tinybullet}_{s,t}(X)<+\infty$, and $\r^{\A^\tinybullet}_{s,t}(X)=\essinf D^X_{s,t}$ by \eqref{eq: rho given A-m}.
    For $m,n\in D^X_{s,t}$, setting $B:=\{m\leq n\}\in\F_s$, we have $X=\1_BX+\1_{B^c}X\in\A^{\1_Bm+\1_{B^c}n}_{s,t}=\A^{m\wedge n}_{s,t}$ by \itemref{IT:N-pasting}, i.e., $m\wedge n\in D^X_{s,t}$.
    Hence $D^X_{s,t}$ is directed downward, and \cite[Theorem~A.37(b) and Definition~A.38]{Follmer+Schied_2016_Stochastic_finance} yield a decreasing sequence $(m_k)_{k\in\N}\subset D^X_{s,t}$ with $m_k\downarrow\essinf D^X_{s,t}$ $\P$-a.s.
\end{proof}

The next two lemmas, together with their cash-additive versions Lemmas~\ref{LEM:from_A_to_rho} and~\ref{LEM:from_rho_to_A}, are a dictionary between the axioms of Table~\ref{TAB:axioms} and the properties of Section~\ref{SUBSEC:P-rho}: they translate statements about acceptance families into statements about risk functionals, and back.
They extend to the dynamic setting part of the static correspondence of \cite[Theorem~1]{Drapeau+Kupper_2013_Risk_preferences_robust_representation}; conditional versions for quasi-convex maps, with levels in $L^0$, are given in \cite[Theorem~2.4]{Bielecki+Cialenco+Drapeau+Karliczek_2016_Dynamic_assessment_indices} and \cite{Frittelli+Maggis_2011_Dual_representation_quasi-convex_conditional_maps}, whereas here no convexity is assumed.
The proofs are routine and deferred to Appendix~\ref{APP:N}.

\begin{lemma}[From $\A^\tinybullet$ to $\r^{\A^\tinybullet}$]
\label{LEM:N-from-A-to-rho}
    Let $\A^\tinybullet$ be a leveled acceptance family in $L^p$, and write $\r:=\r^{\A^\tinybullet}$.
    Then $\A^m_{s,t}\subseteq\A^{m,\r}_{s,t}$ and
    \begin{enumerate}[label=(\alph*)]
    \item 
        \begin{multicols}{2}
        \begin{itemize}[nosep, leftmargin=*, label=\scriptsize$\bullet$]
            \item \itemref{IT:N-normalized} $\Leftrightarrow$ $\r$ is normalized
            \item \itemref{IT:N-expectation} $\Leftrightarrow$ $\r_{s,t}(\,\cdot\,)+\E[\,\cdot\,|\F_s]\geq 0$
            \item \itemref{IT:L-monotone} $+$ \itemref{IT:L-rightcont} $+$ \itemref{IT:N-pasting} $\Rightarrow$ $\A^m_{s,t}=\A^{m,\r}_{s,t}$
            \item \itemref{IT:N-wellposed} $+$ \itemref{IT:N-lowerbound} $\Leftrightarrow$ $\r$ is well posed in $L^p$
            \item \itemref{IT:N-wellposed} $+$ \itemref{IT:N-convex} $+$ \itemref{IT:N-normalized} $\Rightarrow$ \itemref{IT:N-lowerbound}
            \item \itemref{IT:N-wellposed} $+$ \itemref{IT:N-convex} $+$ \itemref{IT:N-constancy} $\Rightarrow$ $\r$ cash-additive
        \end{itemize}
        \end{multicols}
    \item
        \begin{multicols}{2}
        \begin{itemize}[nosep, leftmargin=*, label=\scriptsize$\bullet$]
            \item \itemref{IT:N-monotone} $\Rightarrow$ $\r$ is monotone
            \item \itemref{IT:N-constancy} $\Rightarrow$ $\r$ is conditionally constant
            \item \itemref{IT:N-conic} $\Rightarrow$ $\r$ is positively homogeneous
            \item \itemref{IT:N-star} $\Rightarrow$ $\r$ is star-shaped
            \item \itemref{IT:N-cashsub} $\Rightarrow$ $\r$ is cash-subadditive
            \item \itemref{IT:N-cashadd} $\Rightarrow$ $\r$ is cash-additive
            \item \itemref{IT:N-convex} $\Rightarrow$ $\r$ is convex
            \item \itemref{IT:N-wellposed} $+$ \itemref{IT:N-pasting} $\Rightarrow$ $\r$ is local
        \end{itemize}
        \end{multicols}
    \item
        If $p=\infty$, \itemref{IT:N-monotone} $+$ \itemref{IT:N-constancy} $\Rightarrow$ $\r$ is well posed in $L^\infty$ and ${\|\r_{s,t}(X)\|_{L^\infty}\leq\|X\|_{L^\infty}}$.
    \end{enumerate}
\end{lemma}

\begin{lemma}[From $\r$ to $\A^{\tinybullet,\r}$]
\label{LEM:N-from-rho-to-A}
    Let $\r$ be a dynamic risk functional well posed in $L^p$.
    Then $\A^{\tinybullet,\r}$ satisfies \itemref{IT:L-monotone}, \itemref{IT:L-rightcont}, \itemref{IT:N-wellposed}, \itemref{IT:N-lowerbound}, $\r^{\A^{\tinybullet,\r}}=\r$, and
    all the implications in Lemma~\ref{LEM:N-from-A-to-rho}~(b) are equivalences.
\end{lemma}

In particular, if $\A^\tinybullet$ satisfies \itemref{IT:N-wellposed}, \itemref{IT:N-convex}, \itemref{IT:N-constancy}, \itemref{IT:L-monotone}, \itemref{IT:L-rightcont} and \itemref{IT:N-pasting}, then $\A^\tinybullet=\A^{\tinybullet,\r}$ with $\r$ cash-additive, so that $\A^\tinybullet$ satisfies \itemref{IT:N-cashadd} and is induced by the monetary family $\A^0$; this case is studied in Section~\ref{SEC:CA}.

\subsection{Time-consistency}
\label{SUBSEC:N-TC}
 
In the cash-additive case, time-consistency of $\r^\A$ is equivalent to the decomposition property \itemref{IT:S-decomposition} of $\A$; see Section~\ref{SUBSEC:P-acceptance} and Lemma~\ref{LEM:from_rho_to_A}.
In this subsection, we prove the leveled counterpart of this equivalence, with \itemref{IT:S-decomposition} replaced by \itemref{IT:N-decomposition}: $\A^m_{s,t}=\bigcup_{n\in\A^m_{s,u}}\A^{-n}_{u,t}$ for all $0\leq s\leq u\leq t\leq T$ and $m\in L^p(\F_s)$.

First, notice that if $\A$ is a monetary acceptance family in $L^p$ and $\A^m_{s,t}:=\A_{s,t}-m$, then $\A^\tinybullet$ satisfies \itemref{IT:N-decomposition} if and only if $\A$ satisfies \itemref{IT:S-decomposition}, since $\bigcup_{n\in\A_{s,u}-m}(\A_{u,t}+n)=\A_{s,u}+\A_{u,t}-m$.

\begin{proposition}
\label{prop: suff cond-A3}
    Let $\A^\tinybullet$ be a leveled acceptance family in $L^p$ satisfying \itemref{IT:L-monotone}, \itemref{IT:L-rightcont}, \itemref{IT:N-wellposed}, \itemref{IT:N-lowerbound} and \itemref{IT:N-pasting}, and write $\r:=\r^{\A^\tinybullet}$.
    Then the following are equivalent:
    \begin{enumerate}[label=(\alph*)]
        \item
            $\r$ is time-consistent;
        \item
            for all $0\leq s\leq u\leq t\leq T$, $m\in L^p(\F_s)$ and $X\in L^p(\F_t)$:
            \begin{equation}
            \label{eq: cond-A-tc}
                X\in\A^m_{s,t}\ \Longleftrightarrow\ -\r_{u,t}(X)\in\A^m_{s,u}.
            \end{equation}
    \end{enumerate}
    If, moreover, $\A^\tinybullet$ satisfies \itemref{IT:N-monotone}, then (a) and (b) are also equivalent to
    \begin{enumerate}[label=(\alph*), resume]
        \item
            $\A^\tinybullet$ satisfies \itemref{IT:N-decomposition}.
    \end{enumerate}
    In particular, a local dynamic risk measure $\r$ on $L^p$ is time-consistent if and only if $\A^{\tinybullet,\r}$ satisfies \itemref{IT:N-decomposition}.
\end{proposition}
 
\begin{proof}
    By Lemma~\ref{LEM:N-from-A-to-rho}~(a), $\r$ is well posed in $L^p$ and
    \begin{equation}
    \label{EQ:N-A-rho-A}
        \A^m_{a,b}=\A^{m,\r}_{a,b},\qquad 0\leq a\leq b\leq T,\ m\in L^p(\F_a).
    \end{equation}
    Fix $0\leq s\leq u\leq t\leq T$, $m\in L^p(\F_s)$ and $X\in L^p(\F_t)$; note that $-\r_{u,t}(X)\in L^p(\F_u)$.
    \smallskip
 
    \emph{(a)$\Rightarrow$(b).}
        By \eqref{EQ:N-A-rho-A} at $(s,t)$ and $(s,u)$ and by time-consistency,
        \[
            X\in\A^m_{s,t}
            \ \Longleftrightarrow\ \r_{s,t}(X)\leq m
            \ \Longleftrightarrow\ \r_{s,u}\big(-\r_{u,t}(X)\big)\leq m
            \ \Longleftrightarrow\ -\r_{u,t}(X)\in\A^m_{s,u}.
        \]
 
    \emph{(b)$\Rightarrow$(a).}
        By \eqref{eq: rho given A-m} and \eqref{eq: cond-A-tc},
        \begin{align}
            \r_{s,t}(X)
            &=\essinf\{m\in L^p(\F_s)\ :\ X\in\A^m_{s,t}\}\\
            &=\essinf\{m\in L^p(\F_s)\ :\ -\r_{u,t}(X)\in\A^m_{s,u}\}\\
            &=\r_{s,u}\big(-\r_{u,t}(X)\big).
        \end{align}
 
    \emph{(c)$\Rightarrow$(b), under \itemref{IT:N-monotone}.}
        Let $X\in\A^m_{s,t}$.
        By \itemref{IT:N-decomposition}, there is $n\in\A^m_{s,u}$ such that $X\in\A^{-n}_{u,t}$, hence $\r_{u,t}(X)\leq-n$ by \eqref{eq: rho given A-m}.
        Since $-\r_{u,t}(X)\geq n$, \itemref{IT:N-monotone} yields $-\r_{u,t}(X)\in\A^m_{s,u}$.
        Conversely, let $n:=-\r_{u,t}(X)\in\A^m_{s,u}$.
        By \eqref{EQ:N-A-rho-A}, $X\in\A^{-n}_{u,t}$, hence $X\in\A^m_{s,t}$ by \itemref{IT:N-decomposition}.
 
    \emph{(b)$\Rightarrow$(c), under \itemref{IT:N-monotone}.}
        Let $X\in\A^m_{s,t}$.
        By \eqref{eq: cond-A-tc}, $n:=-\r_{u,t}(X)\in\A^m_{s,u}$, and $X\in\A^{-n}_{u,t}$ by \eqref{EQ:N-A-rho-A}.
        Conversely, let $n\in\A^m_{s,u}$ and $X\in\A^{-n}_{u,t}$.
        Then $\r_{u,t}(X)\leq-n$, so that $-\r_{u,t}(X)\geq n$, and \itemref{IT:N-monotone} gives $-\r_{u,t}(X)\in\A^m_{s,u}$.
        Hence $X\in\A^m_{s,t}$ by \eqref{eq: cond-A-tc}.
    \smallskip

    The last statement follows from (a)$\Leftrightarrow$(c)  and Lemma~\ref{LEM:N-from-rho-to-A}.
\end{proof}

\begin{example}[Discounted expectation]
\label{EX:CS-discount}
    Let $r>0$ and $\r_{s,t}(X):=-e^{-r(t-s)}\E[X|\F_s]$ for $X\in L^p(\F_t)$, $p\in[1,+\infty]$.
    Then $\r$ is monotone, local, convex, cash-subadditive, time-consistent, normalized, and continuous from above and from below.
    Its leveled acceptance family is $\A^m_{s,t}=\{X\in L^p(\F_t)\ :\ \E[X|\F_s]\geq-me^{r(t-s)}\}$, which satisfies \itemref{IT:N-decomposition} by Proposition~\ref{prop: suff cond-A3} and Lemma~\ref{LEM:N-from-rho-to-A}.
\end{example}
 
\begin{remark}
\label{REM:N-res-discount}
    Other notions of time-consistency are used in the literature; for $0\leq s\leq u\leq t\leq T$, they include the order version $\r_{u,t}(X)\leq\r_{u,t}(Y)\Rightarrow\r_{s,t}(X)\leq\r_{s,t}(Y)$ for $X,Y\in L^p(\F_t)$ (called (strong) time-consistency in \cite[Definition~11.10]{Follmer+Schied_2016_Stochastic_finance} and weak time-consistency in \cite{Mastrogiacomo+RosazzaGianin_2015_Time-consistency_cash-subadditive_risk_measures}) and the restriction property $\r_{s,t}(\eta)=\r_{s,u}(\eta)$ for $\eta\in L^p(\F_u)$ \cite{DiNunno+RosazzaGianin_2024_Fully_dynamic_risk_measures_horizon_risk_time-consistency_relations_BSDEs_BSVIEs}.
    For a conditionally constant dynamic risk measure, time-consistency is equivalent to the two of them together; cf.\ \cite[Lemma~11.11]{Follmer+Schied_2016_Stochastic_finance} and \cite[Proposition~8]{Mastrogiacomo+RosazzaGianin_2015_Time-consistency_cash-subadditive_risk_measures}.
    Without conditional constancy, time-consistency does not imply the restriction property, as the discounted expectation of Example~\ref{EX:CS-discount} shows.
    For this reason, in the leveled case we characterize time-consistency through \itemref{IT:N-decomposition}, which does not involve conditional constancy.
\end{remark}

\subsection{Lipschitz BSDEs}
\label{SUBSEC:N-Lipschitz}

Throughout this subsection, we fix $p\in(1,+\infty]$.
For $\mu>0$, we use the reference $\mu$-Lipschitz driver $g^{\tinybullet,\mu}(y,z):=\mu(|y|+|z|)$, $(y,z)\in\R\times\R^d$, with induced risk measure $\r^{\tinybullet,\mu}:=\r^{g^{\tinybullet,\mu}}$ and leveled acceptance family $\A^{\tinybullet,\mu}:=\A^{\tinybullet,g^{\tinybullet,\mu}}$.
We consider the following domination property of a risk functional $\r$ on $L^p$.
\begin{itemize}[leftmargin=4em, labelwidth=3.4em, labelsep=0.6em, align=left]
    \item[(Dom)] \proplabel{Dom}{RM:Dom}
        There exists $\mu>0$ such that, for all $0\leq s\leq t\leq T$,
        \begin{equation}
        \label{eq: cond rho-dom-m}
            \r_{s,t}(X)\leq\r_{s,t}(X')+\r^{\tinybullet,\mu}_{s,t}(X-X'),
            \qquad X,X'\in L^p(\F_t).
        \end{equation}
        We then say that \itemref{RM:Dom} holds with constant $\mu$.
\end{itemize}

\begin{remark}
\label{REM:P-Dom}
    Let $\r$ be a normalized risk functional on $L^p$ satisfying \itemref{RM:Dom} with constant $\mu$.
    Then $\r$ is a dynamic risk measure, $\r_{t,t}(X)=-X$ for all $t\in[0,T]$ and $X\in L^p(\F_t)$, and
    \begin{equation}
    \label{EQ:N-two-sided}
        -\r^{\tinybullet,\mu}_{s,t}(X'-X)\leq\r_{s,t}(X)-\r_{s,t}(X')\leq\r^{\tinybullet,\mu}_{s,t}(X-X'),
        \qquad X,X'\in L^p(\F_t).
    \end{equation}
    Indeed, \eqref{eq: cond rho-dom-m}
    gives $-\r^{\tinybullet,\mu}_{s,t}(-X)\leq\r_{s,t}(X)\leq\r^{\tinybullet,\mu}_{s,t}(X)$.
    Since $\r^{\tinybullet,\mu}_{s,t}(\pm X)\in L^p(\F_s)$ by \itemref{IT:B-stability}, $\r$ is well posed in $L^p$; since $\r^{\tinybullet,\mu}_{t,t}(X)=-X$, the case $s=t$ gives $\r_{t,t}(X)=-X$.
    Well-posedness allows us to rearrange \eqref{eq: cond rho-dom-m}, and exchanging $X$ and $X'$ yields \eqref{EQ:N-two-sided}.
    Finally, if $X\geq X'$, then $\r^{\tinybullet,\mu}_{s,t}(X-X')\leq\r^{\tinybullet,\mu}_{s,t}(0)=0$ by \itemref{IT:B-comparison}, so that $\r_{s,t}(X)\leq\r_{s,t}(X')$.
\end{remark}
 
\begin{remark}
\label{REM:P-Peng}
    Let $\r$ be a well-posed risk functional on $L^2$.
    Then the assumptions of \cite{Peng_2005_dynamically_consistent_nonlinear_evaluations} translate as follows: (A1) is monotonicity; (A2) is $\r_{t,t}(X)=-X$ for $X\in L^2(\F_t)$; (A2$'$) is conditional constancy; (A3) is time-consistency; (A4) is locality, by \cite[Proposition~2.3]{Peng_2005_dynamically_consistent_nonlinear_evaluations}; (A4$_0$) is normalization; and the domination (A5) of \cite[Section~3]{Peng_2005_dynamically_consistent_nonlinear_evaluations} is \itemref{RM:Dom}.
\end{remark}

We recall the axiom \itemref{IT:N-SL} of Table~\ref{TAB:axioms}: there exists $\mu>0$ such that $\A^m_{s,t}+\A^{n,\mu}_{s,t}\subseteq\A^{m+n}_{s,t}$ for all $0\leq s\leq t\leq T$ and $m,n\in L^p(\F_s)$; we then say that \itemref{IT:N-SL} holds with constant $\mu$.
Lipschitz domination says that adding a position $Y$ never raises the risk by more than $\r^{\tinybullet,\mu}_{s,t}(Y)$, the worst expected discounted loss of $Y$ over all scenarios in which the short rate stays in $[-\mu,\mu]$ and the Girsanov kernel is bounded by $\mu$ (Example~\ref{EX:corridor} with $r=-\mu$ and $R=\kappa=\mu$).
 
\begin{remark}
\label{REM:N-SL}
    For a leveled acceptance family $\A^\tinybullet$ in $L^p$, the following implications hold:
    \begin{multicols}{2}
    \begin{itemize}[nosep, leftmargin=*, label=\scriptsize$\bullet$]
        \item \itemref{IT:N-SL} $\Rightarrow$ \itemref{IT:L-monotone} $+$ \itemref{IT:N-monotone}
        \item \itemref{IT:N-SL} $+$ \itemref{IT:L-rightcont} $+$ \itemref{IT:N-normalized} $+$ \itemref{IT:N-pasting} $\Rightarrow$ \itemref{IT:N-wellposed}
    \end{itemize}
    \end{multicols}
    Indeed, $\r^{\tinybullet,\mu}_{s,t}(U)\leq\r^{\tinybullet,\mu}_{s,t}(0)=0$ for $U\in L^p_+(\F_t)$; hence $0\in\A^{n,\mu}_{s,t}$ for $n\geq0$ and $L^p_+(\F_t)\subseteq\A^{0,\mu}_{s,t}$, and \itemref{IT:N-SL} gives the first implication.
    For the second, $\r^{\A^\tinybullet}_{s,t}(0)=0$ by Lemma~\ref{LEM:N-from-A-to-rho}~(a), so $0\in\A^{0,\r^{\A^\tinybullet}}_{s,t}=\A^0_{s,t}$ by Lemma~\ref{LEM:N-from-A-to-rho}~(a), and \itemref{IT:N-SL} gives $X=0+X\in\A^{\r^{\tinybullet,\mu}_{s,t}(X)}_{s,t}$ for all $X\in L^p(\F_t)$.
\end{remark}
 
\begin{lemma}
\label{lem: A-m-cond-dom}
    Let $\A^\tinybullet$ be a leveled acceptance family in $L^p$, let $\mu>0$, and write $\r:=\r^{\A^\tinybullet}$.
    \begin{enumerate}[label=(\alph*)]
        \item
            If \itemref{IT:N-SL} holds with constant $\mu$, then $\r$ satisfies \itemref{RM:Dom} with constant $\mu$.
            In particular, if $\r$ is also normalized, Remark~\ref{REM:P-Dom} applies.
        \item
            If $\A^\tinybullet$ satisfies \itemref{IT:L-monotone}, \itemref{IT:L-rightcont} and \itemref{IT:N-pasting}, then \itemref{IT:N-SL} with constant $\mu$ is equivalent to \itemref{RM:Dom} with constant $\mu$.
    \end{enumerate}
\end{lemma}
\begin{proof}
    \emph{(a)}
    Fix $0\leq s\leq t\leq T$ and $X,X'\in L^p(\F_t)$, and set $b:=\r^{\tinybullet,\mu}_{s,t}(X-X')$, which belongs to $L^p(\F_s)$ by \itemref{IT:B-stability}, so that $X-X'\in\A^{b,\mu}_{s,t}$.
    If $m\in L^p(\F_s)$ is such that $X'\in\A^m_{s,t}$, then \itemref{IT:N-SL} yields $X=X'+(X-X')\in\A^{m+b}_{s,t}$, hence $\r_{s,t}(X)\leq m+b$.
    Taking the essential infimum over all such $m$ gives \eqref{eq: cond rho-dom-m}.

    \smallskip
 
    \emph{(b)}
    By \textit{(a)}, \itemref{IT:N-SL} implies \itemref{RM:Dom}.
    Conversely, let $X'\in\A^m_{s,t}$ and $Y\in\A^{n,\mu}_{s,t}$, so that $\r_{s,t}(X')\leq m$ and $\r^{\tinybullet,\mu}_{s,t}(Y)\leq n$.
    By \eqref{eq: cond rho-dom-m} with $X=X'+Y$, we get $\r_{s,t}(X'+Y)\leq m+n$, hence $X'+Y\in\A^{m+n}_{s,t}$ by Lemma~\ref{LEM:N-from-A-to-rho}~(a).
\end{proof}

\begin{remark}
\label{REM:N-p}
    For $p=1$, the $L^1$-theory of Lipschitz BSDEs in \cite[Section~6, (H7)]{Briand+Delyon+Hu+Pardoux+Stoica_2003_Lp_solutions_BSDEs} requires a sublinear growth of the driver in $z$, which fails for $g^{\tinybullet,\mu}$, so that \itemref{IT:N-SL}
    is not available.
    For this reason, when dealing with Lipschitz BSDEs, we always assume $p>1$.
\end{remark}

We now characterize the leveled acceptance families induced by Lipschitz BSDEs.
The proof is given in Section~\ref{SUBSEC:Lp}; its core is the case $p=2$, where the characterization \cite[Theorem~3.1]{Peng_2005_dynamically_consistent_nonlinear_evaluations} of time-consistent dynamic risk functionals dominated by $\r^{\tinybullet,\mu}$ is available.

\begin{theorem}[Lipschitz BSDEs, leveled case]
\label{THM:Lp-nonadditive}
    Let $p\in(1,+\infty]$ and $\mu>0$.
    \begin{enumerate}[label=(\roman*)]
        \item
            A leveled acceptance family $\A^\tinybullet$ in $L^p$ is induced by a $\mu$-Lipschitz driver $g$ if and only if it is regular and satisfies Lipschitz domination \itemref{IT:N-SL} with constant $\mu$.
        \item
            In this case, $g$ is  unique, $\r^{\A^\tinybullet}=\r^g$ on $L^p$, and $\A^\tinybullet$ also satisfies \itemref{IT:L-monotone}, \itemref{IT:N-wellposed}, \itemref{IT:N-lowerbound} and \itemref{IT:N-monotone}.
            If $p<\infty$ or if $g$ is convex, then $\A^\tinybullet$ satisfies \itemref{IT:N-closed}.
            Moreover, if $p<\infty$, then, for every $(y,z)\in\R\times\R^d$, every $q\in[1,2)$ and $\ell$-a.e.\ $t\in[0,T)$, with $\Delta_\eps W_t:=W_{t+\eps}-W_t$,  
            \begin{equation}
            \label{EQ:N-g-from-A}
                g(t,y,z)
                =L^q\text{-}\lim_{\eps\downarrow0}\frac{\r^{\A^\tinybullet}_{t,t+\eps}(-y-z\cdot\Delta_\eps W_t)-y}\eps.
            \end{equation}
        \item
            If $g$ is a $\mu$-Lipschitz driver, then $\A^{\tinybullet,g}$ satisfies
            \begin{multicols}{2}
            \begin{itemize}[nosep, leftmargin=*, label=\scriptsize$\bullet$]
                \item \itemref{IT:N-convex} $\Leftrightarrow$ $g$ convex,
                \item \itemref{IT:N-star} $\Leftrightarrow$ $g$  star-shaped,
                \item \itemref{IT:N-constancy} $\Leftrightarrow$ ,$g(\,\cdot\,,y,0)=0$ for all $y\in\R$
                \item \itemref{IT:N-expectation} $\Leftrightarrow$ $g\geq0$,
                \item \itemref{IT:N-cashsub} $\Leftrightarrow$ $g$ nonincreasing in $y$,
                \item \itemref{IT:N-cashadd} $\Leftrightarrow$ $g$ monetary,
                \item \itemref{IT:N-convex} $+$ \itemref{IT:N-constancy} $\Leftrightarrow$ $g$ convex monetary.
            \end{itemize}
            \end{multicols}
    \end{enumerate}
\end{theorem}

For $p=\infty$, \eqref{EQ:N-g-from-A} does not make sense for $\r^{\A^{\tinybullet}}$, since $z\cdot\Delta_\eps W_t\notin L^\infty$ in general; it holds with $\r^{\A^{\tinybullet}}$ replaced by its extension $\bar\r$ to $L^2$ of Lemma~\ref{LEM:N-reduction}~(b) below.

Level right-continuity, which is part of Definition~\ref{DEF:regular} of a regular family, is only needed in the previous theorem to identify the sets.
In the next corollary we show that, without \itemref{IT:L-rightcont}, we still obtain $\r^{\A^\tinybullet}=\r^g$, but not $\A^\tinybullet=\A^{\tinybullet,g}$.

\begin{corollary}
\label{COR:N-no-lrc}
    Let $p\in(1,+\infty]$, $\mu>0$, and let $\A^\tinybullet$ be a leveled acceptance family in $L^p$ satisfying \itemref{IT:N-normalized}, \itemref{IT:N-pasting}, \itemref{IT:N-decomposition} and \itemref{IT:N-SL} with constant $\mu$.
    Then $\r^{\A^\tinybullet}$ is induced by a unique $\mu$-Lipschitz driver $g$, and $\A^\tinybullet\subseteq\A^{\tinybullet,g}$, with equality if and only if \itemref{IT:L-rightcont} holds.
\end{corollary}
\begin{proof}
    Let $\r:=\r^{\A^\tinybullet}$.
    The statement follows from applying Theorem~\ref{THM:Lp-nonadditive} to the family $\A^{\tinybullet,\r}$ of sublevel sets of $\r$, which satisfies \itemref{IT:L-rightcont} by construction and contains $\A^\tinybullet$, with equality if and only if $\A^\tinybullet$ satisfies \itemref{IT:L-rightcont} (Lemma~\ref{LEM:N-from-A-to-rho}~(a)).
    Normalization, locality and \itemref{IT:N-SL} pass from $\A^\tinybullet$ to $\r$ and back to $\A^{\tinybullet,\r}$ by the correspondence lemmas, and \itemref{IT:N-decomposition} for $\A^{\tinybullet,\r}$ is equivalent to the time-consistency of $\r$ (Proposition~\ref{prop: suff cond-A3}); the only point to prove is therefore that $\r$ is time-consistent, with \itemref{IT:N-SL} playing the role of \itemref{IT:L-rightcont}.
    
    By \itemref{IT:N-normalized}, $\r$ is normalized; hence, by Lemma~\ref{lem: A-m-cond-dom}~(a) and Remark~\ref{REM:P-Dom}, $\r$ is a normalized dynamic risk measure on $L^p$ satisfying \itemref{RM:Dom} with constant $\mu$.
    In particular, $\r$ is finite and \itemref{IT:N-wellposed} holds.

    Let $0\leq s\leq u\leq t\leq T$, $X\in L^p(\F_t)$ and $m\in L^p(\F_s)$.
    If $X\in\A^m_{s,t}$, then $X\in\A^{-n}_{u,t}$ for some $n\in\A^m_{s,u}$ by \itemref{IT:N-decomposition}, so $\r_{u,t}(X)\leq-n$, and $-\r_{u,t}(X)\in\A^m_{s,u}$ since $\A^\tinybullet$ satisfies \itemref{IT:N-monotone} by Remark~\ref{REM:N-SL}; hence $\r_{s,u}(-\r_{u,t}(X))\leq\r_{s,t}(X)$.

    Conversely, let $-\r_{u,t}(X)\in\A^m_{s,u}$.
    Without \itemref{IT:L-rightcont}, $\r_{u,t}(X)$ need not be an acceptable level, so we approximate it from above and bound the error with \itemref{IT:N-SL}.
    By Lemma~\ref{LEM:N-directedness}, $X\in\A^{m_k}_{u,t}$ for some $m_k\in L^p(\F_u)$ with $m_k\downarrow\r_{u,t}(X)$.
    With $\delta_k:=m_k-\r_{u,t}(X)\in L^p_+(\F_u)$ and $e_k:=\r^{\tinybullet,\mu}_{s,u}(-\delta_k)$, \itemref{IT:N-SL} gives $-m_k=-\r_{u,t}(X)-\delta_k\in\A^{m+e_k}_{s,u}$, and \itemref{IT:N-decomposition} gives $X\in\A^{m+e_k}_{s,t}$, i.e., $\r_{s,t}(X)\leq m+e_k$.
    Since $0\leq\delta_k\leq\delta_1$ and $\delta_k\to0$ a.s., dominated convergence gives $\delta_k\to0$ in $L^p$ if $p<\infty$ and in $L^2$ if $p=\infty$; hence \itemref{IT:B-consistency} and \itemref{IT:B-stability}, applied with the corresponding exponent, give $e_k\to0$ a.s.\ along a subsequence, so $\r_{s,t}(X)\leq m$ and $\r_{s,t}(X)\leq\r_{s,u}(-\r_{u,t}(X))$.
\end{proof}

\subsubsection{Proof of Theorem~\ref{THM:Lp-nonadditive}}
\label{SUBSEC:Lp}

The proof reduces the case $p\neq2$ to $p=2$ through \itemref{RM:Dom}, which allows one to transfer a risk functional from $L^p$ to $L^2$ and back (Lemma~\ref{LEM:N-reduction}).
For $p\in(1,2)$, this transfer is a simple restriction to $L^2$, and the corresponding $L^p$-version of \cite[Theorem~3.1]{Peng_2005_dynamically_consistent_nonlinear_evaluations} is \cite[Theorem~3.1]{Hu_2013_Dynamically_consistent_nonlinear_evaluations_generating_functions_Lp};
for $p\in(2,+\infty]$, $\r$ has to be extended from $L^p$ to $L^2$ by density.
Throughout, $p\in(1,+\infty]$; the facts \itemref{IT:B-consistency}--\itemref{IT:B-cash} are used both on $L^p$ and on $L^2$.

\begin{lemma}
\label{LEM:N-reduction}
    Let $\mu>0$, and let $\r$ be a normalized, local and time-consistent risk functional on $L^p$ satisfying \itemref{RM:Dom} with constant $\mu$.
    Then:
    \begin{enumerate}[label=(\alph*)]
        \item
            for every $p'\in(1,+\infty]$, there exists $C=C(p',\mu,T)>0$ such that for $0\leq s\leq t\leq T$ and $X,X'\in L^p(\F_t)$ with $X-X'\in L^{p'}(\F_t)$,
            \begin{equation}
            \label{EQ:N-Lq-cont}
                \big\|\r_{s,t}(X)-\r_{s,t}(X')\big\|_{L^{p'}}\leq 2C \,\|X-X'\|_{L^{p'}};
            \end{equation}
        \item
            there exists a unique risk functional $\bar\r=(\bar\r_{s,t})_{0\leq s\leq t\leq T}$ on $L^2$ such that each $\bar\r_{s,t}:L^2(\F_t)\to L^2(\F_s)$ is $L^2$-continuous and coincides with $\r_{s,t}$ on $L^p(\F_t)\cap L^2(\F_t)$;
            it is monotone, normalized, local and time-consistent, it satisfies $\bar\r_{t,t}(X)=-X$ for $X\in L^2(\F_t)$ and \itemref{RM:Dom} with constant $\mu$ on $L^2$, and it is convex if $\r$ is;
        \item
            if $\bar\r=\r^g$ on $L^2$ for some $\mu$-Lipschitz driver $g$, then $\r=\r^g$ on $L^p$.
    \end{enumerate}
\end{lemma}
\begin{proof}
    \emph{(a)}
    By \eqref{EQ:N-two-sided} (Remark~\ref{REM:P-Dom}), $|\r_{s,t}(X)-\r_{s,t}(X')|\leq|\r^{\tinybullet,\mu}_{s,t}(X-X')|+|\r^{\tinybullet,\mu}_{s,t}(X'-X)|$, and \eqref{EQ:N-Lq-cont} follows from \itemref{IT:B-consistency}, \itemref{IT:B-stability} with exponent $p'$ and $\r^{\tinybullet,\mu}_{s,t}(0)=0$.
    \smallskip
 
    \emph{(b)}
    Since $\r_{s,t}(0)=0$, \eqref{EQ:N-Lq-cont} with $p'=2$ shows that $\r_{s,t}$ maps $L^p(\F_t)\cap L^2(\F_t)$ into $L^2(\F_s)$ and is Lipschitz continuous for the $L^2$-norm.
    Since $L^p(\F_t)\cap L^2(\F_t)\supseteq L^\infty(\F_t)$ is dense in $L^2(\F_t)$, $\r_{s,t}$ has a unique $L^2$-continuous extension $\bar\r_{s,t}:L^2(\F_t)\to L^2(\F_s)$.
    For $p\in(1,2]$, $\bar\r_{s,t}$ is simply the restriction of $\r_{s,t}$ to $L^2(\F_t)$.
 
    For $k\in\N$, let $T_k(x):=(x\wedge k)\vee(-k)$, $x\in\R$.
    For $X\in L^2(\F_t)$, we have $T_k(X)\in L^\infty(\F_t)$ and $T_k(X)\to X$ in $L^2$, hence $\bar\r_{s,t}(X)=L^2\text{-}\lim_k\r_{s,t}(T_k(X))$, and the convergence holds $\P$-a.s.\ along a subsequence.
    Since $X\leq X'$ implies $T_k(X)\leq T_k(X')$, monotonicity passes to the limit.
    For $\eta\in L^2(\F_t)$, $\r_{t,t}(T_k(\eta))=-T_k(\eta)\to-\eta$ by Remark~\ref{REM:P-Dom}, which gives $\bar\r_{t,t}(\eta)=-\eta$; normalization is immediate.
    For $A\in\F_s$, $T_k(\1_AX+\1_{A^c}Y)=\1_AT_k(X)+\1_{A^c}T_k(Y)$, which gives locality.
    \eqref{eq: cond rho-dom-m} passes to the limit by \itemref{IT:B-stability} for $\r^{\tinybullet,\mu}$ with exponent $2$, and convexity passes to the limit along the bounded approximations $\l T_k(X)+(1-\l)T_k(Y)$.
    For time-consistency, let $0\leq s\leq u\leq t\leq T$ and $X\in L^2(\F_t)$.
    By time-consistency of $\r$, $\r_{s,t}(T_k(X))=\r_{s,u}(-\r_{u,t}(T_k(X)))=\bar\r_{s,u}(-\r_{u,t}(T_k(X)))$, where $\r_{u,t}(T_k(X))\in L^p(\F_u)\cap L^2(\F_u)$.
    As $k\to\infty$, the left-hand side converges to $\bar\r_{s,t}(X)$ and $\r_{u,t}(T_k(X))\to\bar\r_{u,t}(X)$ in $L^2$, so that the $L^2$-continuity of $\bar\r_{s,u}$ yields $\bar\r_{s,t}(X)=\bar\r_{s,u}(-\bar\r_{u,t}(X))$.
    \smallskip
 
    \emph{(c)}
    If $p\in[2,+\infty]$, then $L^p\subseteq L^2$ and $\r=\bar\r=\r^g$ on $L^p$ by \itemref{IT:B-consistency}.
    If $p\in(1,2)$, then $\r$ and $\r^g$ coincide on $L^2(\F_t)$, which is dense in $L^p(\F_t)$, and both are continuous on $L^p(\F_t)$ by \eqref{EQ:N-Lq-cont} with $p'=p$ and by \itemref{IT:B-stability}; hence they coincide on $L^p(\F_t)$.
\end{proof}
 
For risk measures induced by a $\mu$-Lipschitz driver $g$, the properties that depend on $g$ can be checked on $L^2$.
 
\begin{lemma}
\label{LEM:Lp-transfer}
    Let $g$ be a $\mu$-Lipschitz driver.
    Each of the following properties holds for $\r^g$ on $L^p$ if and only if it holds for $\r^g$ on $L^2$:
    {\setlength{\multicolsep}{2pt}%
    \begin{multicols}{2}
    \begin{itemize}[nosep, leftmargin=*, label=\scriptsize$\bullet$]
        \item convexity
        \item star-shapedness
        \item conditional constancy
        \item cash-subadditivity
        \item cash-additivity
        \item $\r^g_{s,t}(\,\cdot\,) + \E[\,\cdot\,|\F_s]\geq 0$ for all $0\leq s\leq t\leq T$
    \end{itemize}
    \end{multicols}}%
\end{lemma}
\begin{proof}
    We prove the claim for conditional constancy; the other properties are analogous.
    Assume that $\r^g$ is conditionally constant on $L^2$, and let $\eta\in L^p(\F_s)$.
    For $n\in\N$, the truncation $\eta_n:=(\eta\wedge n)\vee(-n)$ belongs to $L^\infty(\F_s)\subseteq L^2(\F_s)\cap L^p(\F_s)$, so that, by \itemref{IT:B-consistency}, $\r^g_{s,t}(\eta_n)=-\eta_n$ also for $\r^g$ on $L^p$.
    If $p=\infty$, then $\eta=\eta_n$ for $n\geq\|\eta\|_{L^\infty}$.
    If $p<\infty$, then $\eta_n\to\eta$ in $L^p$, hence $\r^g_{s,t}(\eta_n)\to\r^g_{s,t}(\eta)$ in $L^p$ by \itemref{IT:B-stability}, and $\r^g_{s,t}(\eta)=-\eta$.
    The converse is obtained by exchanging the roles of $L^p$ and $L^2$.
\end{proof}
 
\begin{proof}[\textbf{Proof of Theorem~\ref{THM:Lp-nonadditive}}]
    For a $\mu$-Lipschitz driver $g$, write $\mE^g_{s,t}[\xi]:=\r^g_{s,t}(-\xi)$.
    \smallskip

\noindent\textbf{\textit{(i) and (ii).}}
    \emph{Step 1: properties of $\A^\tinybullet$ and $\r:=\r^{\A^\tinybullet}$.}
    Let $\A^\tinybullet$ satisfy the axioms in (i).
    By Remark~\ref{REM:N-SL}, $\A^\tinybullet$ satisfies \itemref{IT:L-monotone}, \itemref{IT:N-wellposed} and \itemref{IT:N-monotone}.
    Hence Lemma~\ref{LEM:N-from-A-to-rho}~(a) gives $\A^\tinybullet=\A^{\tinybullet,\r}$.
    By Lemma~\ref{LEM:N-from-A-to-rho}~(a), $\r$ is normalized, so Lemma~\ref{lem: A-m-cond-dom}~(a) and Remark~\ref{REM:P-Dom} show that $\r$ is well posed in $L^p$ and satisfies \itemref{RM:Dom} and $\r_{t,t}(X)=-X$ for $X\in L^p(\F_t)$; in particular, \itemref{IT:N-lowerbound} holds.
    By Lemma~\ref{LEM:N-from-A-to-rho}~(b), $\r$ is monotone and local, and, by Proposition~\ref{prop: suff cond-A3}, (c)$\Rightarrow$(a), it is time-consistent.
    \smallskip

    \emph{Step 2: the case $p=2$, via \cite[Theorem~3.1]{Peng_2005_dynamically_consistent_nonlinear_evaluations}.}
    Let $p=2$ and set $\mE_{s,t}[\xi]:=\r_{s,t}(-\xi)$, $\xi\in L^2(\F_t)$.
    By Step~1 and Remark~\ref{REM:P-Peng}, $\mE$ satisfies (A1)--(A4) and (A4$_0$), and (A5) is \itemref{RM:Dom}.
    Hence there is a driver $g$, $\mu$-Lipschitz in $(y,z)$ with $g(\cdot,0,0)=0$, such that $\r=\r^g$ on $L^2$; taking a predictable version of $g(\cdot,y,z)$ for $(y,z)$ in a countable dense set (every progressively measurable process coincides $\P\otimes\ell$-a.e.\ with a predictable one) and extending by continuity, we may assume that $g$ is $\mathcal P\otimes\mathcal B(\R)\otimes\mathcal B(\R^d)$-measurable and $\mu$-Lipschitz for every $(\omega,t)$, i.e., a $\mu$-Lipschitz driver.
    By \eqref{eq: A-m given rho}, $\A^m_{s,t}=\A^{m,\r}_{s,t}=\A^{m,g}_{s,t}$, i.e., $\A^\tinybullet=\A^{\tinybullet,g}$.
    By \cite[Lemma~2.1]{Jiang_2008_Convexity_translation_invariance_subadditivity_g-expectations_related_risk_measures}, for every $(y,z)$, $q\in[1,2)$ and a.e.\ $t$,
    \begin{equation}
    \label{EQ:N-Jiang}
        g(t,y,z)
        =L^q\text{-}\lim_{\eps\downarrow0}\frac1\eps\big(\mE^{g}_{t,t+\eps}[y+z\cdot\Delta_\eps W_t]-y\big),
    \end{equation}
    which is \eqref{EQ:N-g-from-A}, since $\mE^g_{t,t+\eps}[y+z\cdot\Delta_\eps W_t]=\r_{t,t+\eps}(-y-z\cdot\Delta_\eps W_t)$.
    If $g'$ is another $\mu$-Lipschitz driver with $\A^{\tinybullet,g'}=\A^\tinybullet$, then $\r^{g'}=\r^{\A^{\tinybullet,g'}}=\r=\r^g$ on $L^2$ by Lemma~\ref{LEM:N-from-rho-to-A}.
    Hence \eqref{EQ:N-Jiang} applied to $g$ and to $g'$ gives $g'(\cdot,y,z)=g(\cdot,y,z)$ $\P\otimes\ell$-a.e.\ for $(y,z)$ in a countable dense set, and thus, by continuity in $(y,z)$, $g'=g$ $\P\otimes\ell$-a.e.
    \smallskip

    \emph{Step 3: reduction of the case $p\neq2$ to $p=2$.}
    Let $\bar\r$ be the extension of $\r$ to $L^2$ given by Lemma~\ref{LEM:N-reduction}~(b).
    By Lemmas~\ref{LEM:N-from-rho-to-A} and~\ref{lem: A-m-cond-dom}~(b) and Proposition~\ref{prop: suff cond-A3}, $\A^{\tinybullet,\bar\r}$ satisfies the axioms in (i) in $L^2$, and Step~2 yields a $\mu$-Lipschitz driver $g$ with $\bar\r=\r^{\A^{\tinybullet,\bar\r}}=\r^g$ on $L^2$.
    By Lemma~\ref{LEM:N-reduction}~(c), $\r=\r^g$ on $L^p$, hence $\A^\tinybullet=\A^{\tinybullet,\r}=\A^{\tinybullet,g}$.
    If $p<\infty$, then $y+z\cdot\Delta_\eps W_t\in L^p\cap L^2$, where $\r$ and $\r^g$ on $L^2$ coincide by \itemref{IT:B-consistency}, so that \eqref{EQ:N-g-from-A} follows from \eqref{EQ:N-Jiang}.
    Uniqueness holds because two $\mu$-Lipschitz drivers inducing the same risk functional on $L^\infty$ induce the same risk functional on $L^2$, by \itemref{IT:B-consistency}, \itemref{IT:B-stability} and density, and Step~2 applies.
    \smallskip

    \emph{Step 4: the ``only if'' part of (i) and the axioms in (ii).}
    By \itemref{IT:B-stability}--\itemref{IT:B-flow}, $\r^g$ is a local and time-consistent dynamic risk measure on $L^p$, and it is normalized since $g(\cdot,0,0)=0$.
    By Lemmas~\ref{LEM:N-from-rho-to-A} and~\ref{LEM:N-from-A-to-rho}~(a),(b), $\A^{\tinybullet,g}=\A^{\tinybullet,\r^g}$ satisfies \itemref{IT:L-monotone}, \itemref{IT:L-rightcont}, \itemref{IT:N-wellposed}, \itemref{IT:N-lowerbound}, \itemref{IT:N-monotone}, \itemref{IT:N-normalized} and \itemref{IT:N-pasting}, and $\r^{\A^{\tinybullet,g}}=\r^g$.
    \itemref{IT:N-decomposition} then follows from Proposition~\ref{prop: suff cond-A3}, (a)$\Rightarrow$(c).
 
    For \itemref{IT:N-SL}, let $X^1,X^2\in L^p(\F_t)$ and let $(Y^i,Z^i)$ be the solution of \eqref{EQ:N-BSDE} with terminal condition $-X^i$, $i=1,2$.
    Then $(Y^1-Y^2,Z^1-Z^2)$ is the solution of \eqref{EQ:N-BSDE} with terminal condition $-(X^1-X^2)$ and driver
    \[
        f(r,y,z):=g(r,Y^2_r+y,Z^2_r+z)-g(r,Y^2_r,Z^2_r)\leq g^{\tinybullet,\mu}(y,z),
    \]
    which is $\mu$-Lipschitz with $f(\cdot,0,0)=0$.
    Hence \itemref{IT:B-comparison} gives $\r^g_{s,t}(X^1)-\r^g_{s,t}(X^2)=\r^f_{s,t}(X^1-X^2)\leq\r^{\tinybullet,\mu}_{s,t}(X^1-X^2)$, and \itemref{IT:N-SL} follows from Lemma~\ref{lem: A-m-cond-dom}~(b).
 
    For \itemref{IT:N-closed} with $p<\infty$, let $X_k\in\A^{m,g}_{s,t}$ with $X_k\to X$ in $L^p$; by \itemref{IT:B-stability}, $\r^g_{s,t}(X_k)\to\r^g_{s,t}(X)$ in $L^p$, hence a.s.\ along a subsequence, and $\r^g_{s,t}(X)\leq m$.
    For $p=\infty$ and $g$ convex, $\A^{m,g}_{s,t}$ is convex, and its intersection with each ball $\{\|X\|_{L^\infty}\leq k\}$ is closed in $L^1$, since on this ball $\|X_j-X\|_{L^2}^2\leq2k\|X_j-X\|_{L^1}$ and \itemref{IT:B-stability} holds with exponent $2$.
    By the Krein--\v{S}mulian theorem \cite[Theorem~A.67]{Follmer+Schied_2016_Stochastic_finance}, $\A^{m,g}_{s,t}$ is $\sigma(L^\infty,L^1)$-closed.
    \medskip

\noindent\textbf{\textit{(iii).}}
    By Lemmas~\ref{LEM:N-from-A-to-rho}~(a),(b) and~\ref{LEM:N-from-rho-to-A}, each axiom in (iii) holds for $\A^{\tinybullet,g}$ in $L^p$ if and only if the corresponding property holds for $\r^g$ on $L^p$, hence, by Lemma~\ref{LEM:Lp-transfer}, for $\r^g$ on $L^2$; it therefore suffices to consider $p=2$.
    If $g$ is convex, then $\r^g$ is convex by \itemref{IT:B-convex}, and \itemref{IT:N-convex} follows from Lemma~\ref{LEM:N-from-rho-to-A}.
    
    Conversely, if $\A^{\tinybullet,g}$ satisfies \itemref{IT:N-convex}, then $\r^g=\r^{\A^{\tinybullet,g}}$ is convex by Lemma~\ref{LEM:N-from-A-to-rho}~(b).
    Hence $\xi\mapsto\mE^g_{t,t+\eps}[\xi]$ is convex on $L^2(\F_{t+\eps})$.
    For $(y_i,z_i)\in\R\times\R^d$, $i=1,2$, and $\l\in[0,1]$, set $\xi_i:=y_i+z_i\cdot\Delta_\eps W_t$; then
    \[
        \frac1\eps\Big(\mE^g_{t,t+\eps}\big[\l\xi_1+(1-\l)\xi_2\big]-\l y_1-(1-\l)y_2\Big)
        \leq\sum_{i=1}^2\l_i\,\frac1\eps\Big(\mE^g_{t,t+\eps}[\xi_i]-y_i\Big),
    \]
    with $\l_1:=\l$ and $\l_2:=1-\l$.
    Passing to the limit in \eqref{EQ:N-Jiang} along an a.s.\ convergent subsequence gives the convexity inequality for $g(t,\cdot,\cdot)$, $\P\otimes\ell$-a.e., for rational $\l$ and parameters in a countable dense set; by continuity, this holds for all $\l$ and $(y_i,z_i)$ outside a $\P\otimes\ell$-null set $N\in\mathcal P$, i.e., $g$ is convex $\P\otimes\ell$-a.e.; see also \cite[Theorem~3.2]{Jiang_2008_Convexity_translation_invariance_subadditivity_g-expectations_related_risk_measures}.

    If $g$ is star-shaped, fix $\l\in(0,1]$ and set $g_\l(r,y,z):=\l g(r,y/\l,z/\l)$, which is a $\mu$-Lipschitz driver with $g\leq g_\l$.
    If $(Y,Z)$ solves the BSDE with parameters $(g,t,X)$, then $(\l Y,\l Z)$ solves the BSDE with parameters $(g_\l,t,\l X)$, so that \itemref{IT:B-comparison} gives $\r^g_{s,t}(\l X)\leq\r^{g_\l}_{s,t}(\l X)=\l\r^g_{s,t}(X)$.
    Hence $\r^g$ is star-shaped, and \itemref{IT:N-star} follows from Lemma~\ref{LEM:N-from-rho-to-A}.
 
    Conversely, if $\A^{\tinybullet,g}$ satisfies \itemref{IT:N-star}, then $\r^g$ is star-shaped by Lemma~\ref{LEM:N-from-A-to-rho}~(b), that is, $\mE^g_{t,t+\eps}[\l\xi]\leq\l\mE^g_{t,t+\eps}[\xi]$ for all $\l\in(0,1]$.
    By \eqref{EQ:N-Jiang} with $\xi:=y+z\cdot\Delta_\eps W_t$, for every $(y,z)$, $\l\in(0,1]$ and a.e.\ $t$,
    \[
        g(t,\l y,\l z)
        =L^q\text{-}\lim_{\eps\downarrow0}\frac1\eps\big(\mE^g_{t,t+\eps}[\l\xi]-\l y\big)
        \leq\l\,L^q\text{-}\lim_{\eps\downarrow0}\frac1\eps\big(\mE^g_{t,t+\eps}[\xi]-y\big)
        =\l g(t,y,z),
    \]
    since the inequality passes to the limit along an a.s.\ convergent subsequence.
    Taking rational $\l$ and $(y,z)$ in a countable dense set, by continuity this holds for all $\l\in(0,1]$ and $(y,z)$ outside a $\P\otimes\ell$-null set $N\in\mathcal P$, i.e., $g$ is star-shaped $\P\otimes\ell$-a.e.
    
    For \itemref{IT:N-constancy}: if $g(\cdot,y,0)=0$ for all $y$, then $\r^g$ is conditionally constant by \itemref{IT:B-flow}, and \itemref{IT:N-constancy} follows from Lemma~\ref{LEM:N-from-rho-to-A}.
    Conversely, if \itemref{IT:N-constancy} holds, then $\mE^g_{t,t+\eps}[y]=\r^g_{t,t+\eps}(-y)=y$, and \eqref{EQ:N-Jiang} with $z=0$ gives $g(t,y,0)=0$ $\P\otimes\ell$-a.e.\ for every $y$.
 
    By Lemma~\ref{LEM:N-from-A-to-rho}~(a), \itemref{IT:N-expectation} holds if and only if $\r^g_{s,t}(X)\geq\E[-X|\F_s]=\r^0_{s,t}(X)$ for all $0\leq s\leq t\leq T$ and $X\in L^2(\F_t)$, where $\r^0$ is induced by the zero driver.
    If $g\geq0$, this follows from \itemref{IT:B-comparison}.
    Conversely, applying the inequality with $(s,t)$ replaced by $(t,t+\eps)$ to $X=-y-z\cdot\Delta_\eps W_t$ in \eqref{EQ:N-Jiang} gives $g(t,y,z)\geq L^q\text{-}\lim_{\eps\downarrow0}\frac1\eps\big(\E[y+z\cdot\Delta_\eps W_t|\F_t]-y\big)=0$, $\P\otimes\ell$-a.e., for $(y,z)$ in a countable dense set, hence, by continuity, for all $(y,z)$ outside a null set.
 
    For \itemref{IT:N-cashsub}, by Lemmas~\ref{LEM:N-from-A-to-rho}~(b) and~\ref{LEM:N-from-rho-to-A} it suffices to show that $\r^g_{s,t}(X-n)\leq\r^g_{s,t}(X)+n$ for all $X\in L^2(\F_t)$ and $n\in L^2_+(\F_s)$ if and only if $g$ is nonincreasing in $y$.
    If $g$ is nonincreasing in $y$, let $(Y,Z)$ solve \eqref{EQ:N-BSDE} with terminal condition $-X$ and set $\bar Y_r:=Y_r+n$, $r\in[s,t]$.
    Then $(\bar Y,Z)$ solves \eqref{EQ:N-BSDE} on $[s,t]$ with terminal condition $-(X-n)$ and driver $g(\cdot,y-n,z)\geq g(\cdot,y,z)$, so the comparison theorem \cite[Theorem~2.2]{ElKaroui+Peng+Quenez_1997_Backward_stochastic_differential_equations_finance} gives $\r^g_{s,t}(X-n)\leq\bar Y_s=\r^g_{s,t}(X)+n$ (it applies, since $|g(\cdot,-n,0)|\leq\mu n\in L^2(\F_s)$).
    Conversely, for $y,z$ and a constant $n\geq0$, applying the inequality to $X=-y-z\cdot\Delta_\eps W_t$ with $(s,t)=(t,t+\eps)$ yields $\mE^g_{t,t+\eps}[y+n+z\cdot\Delta_\eps W_t]\leq\mE^g_{t,t+\eps}[y+z\cdot\Delta_\eps W_t]+n$, and \eqref{EQ:N-Jiang} gives $g(t,y+n,z)\leq g(t,y,z)$ $\P\otimes\ell$-a.e.; by continuity, this holds for all $y,z$ and $n\geq0$ outside a null set.

    For \itemref{IT:N-cashadd}, by Lemmas~\ref{LEM:N-from-A-to-rho}~(b) and~\ref{LEM:N-from-rho-to-A} it suffices to show that $\r^g$ is cash-additive if and only if $g$ is monetary.
    If $g$ is monetary, this is \itemref{IT:B-cash}.
    Conversely, if $\r^g$ is cash-additive, then $\mE^g_{t,t+\eps}[y+z\cdot\Delta_\eps W_t]-y=\mE^g_{t,t+\eps}[z\cdot\Delta_\eps W_t]$, and \eqref{EQ:N-Jiang} gives $g(t,y,z)=g(t,0,z)$ $\P\otimes\ell$-a.e.\ for $(y,z)$ in a countable dense set, hence, by continuity, for all $(y,z)$ outside a null set, i.e., $g$ is monetary.

    For \itemref{IT:N-convex} $+$ \itemref{IT:N-constancy}, by the items for \itemref{IT:N-convex} and \itemref{IT:N-constancy} it suffices to show that a convex $g$ with $g(\cdot,y,0)=0$ for all $y$ is monetary, the converse being clear.
    Writing $(y,z)$ as the midpoint of $(0,2z)$ and $(2y,0)$ gives $g(\cdot,y,z)\leq\frac12g(\cdot,0,2z)$ for all $y$, so the convex function $\R\ni y \mapsto g(\cdot,y,z)$ is bounded from above, hence constant.
\end{proof}

\section{Cash-additive case}
\label{SEC:CA}

Throughout, $p\in[1,+\infty]$, $\A$ is a monetary acceptance family in $L^p$ and $\r^\A$ is given by \eqref{eq: rho given A}, unless otherwise stated.
Many results of Section~\ref{SEC:N} specialize to this case through Remark~\ref{REM:P-embedding}.
We first collect them, together with  the dual representation of $\r^\A$, which is specific to the cash-additive case (Lemmas~\ref{LEM:penalty}, \ref{LEM:from_A_to_rho} and~\ref{LEM:from_rho_to_A}); they hold in any filtration satisfying the usual conditions.
In Section~\ref{SUBSEC:CA-penalty}, we represent the minimal penalty of $\r^\A$ in terms of $\A$ and of Girsanov kernels.
Sections~\ref{SUBSEC:CA-Lipschitz} and~\ref{SUBSEC:CA-quadratic} characterize the acceptance families induced by Lipschitz and by quadratic BSDEs, respectively.

For $0\leq s\leq t\leq T$, we set
\[
    c^\A_{s,t}(\Q):=\esssup_{X\in\A_{s,t}}\E_\Q[-X|\F_s],
    \qquad \Q\in\mathcal Q^{\mathrm{ac}}_{s,t}\text{ with }\frac{\d\Q}{\d\P}\in L^q(\F_t),
\]
where $q\in[1,+\infty]$ is the conjugate exponent of $p$.
By direct inspection, $c^\A_{s,t}(\Q)$ is an $\F_s$-measurable $[-\infty,+\infty]$-valued random variable; under further assumptions, it is the minimal penalty of $\r^\A$ (Lemmas~\ref{LEM:penalty} and~\ref{LEM:from_A_to_rho}), with $\r^\A$ defined as in \eqref{eq: rho given A}.
The next lemma expresses $c^\A$ as the convex conjugate of $\r^\A$; for $p=+\infty$, it is \cite[Remark~9]{Detlefsen+Scandolo_2005_Conditional_dynamic_convex_risk_measures}. 
Since its proof is elementary, we defer it to Appendix~\ref{APP:CA}.
 
\begin{lemma}
\label{LEM:penalty}
    Let $\A$ be a monetary acceptance family in $L^p$, let $0\leq s\leq t\leq T$, and let $\Q\in\mathcal Q^{\mathrm{ac}}_{s,t}$ with $\d\Q/\d\P\in L^q(\F_t)$, where $q$ is the conjugate exponent of $p$.
    Then
    \begin{equation}
    \label{EQ:c^A_dual}
        c^\A_{s,t}(\Q)=\esssup_{X\in L^p(\F_t)}\big\{\E_\Q[-X|\F_s]-\r^\A_{s,t}(X)\big\} \in L^0\big(\F_s;[-\infty,+\infty]\big).
    \end{equation}
    Moreover:
    \begin{itemize}[nosep, leftmargin=*, label=\scriptsize$\bullet$]
        \item \itemref{IT:S-normalized} $\Rightarrow$ $c^\A_{s,t}(\Q)\in[0,+\infty]$
        \item \itemref{IT:S-normalized} $+$ \itemref{IT:S-expectation} $\Rightarrow$ $c^\A_{s,T}(\P)=0$
        \item \itemref{IT:S-normalized} $+$ \itemref{IT:S-expectation} $+$ \itemref{IT:S-decomposition} $\Rightarrow$ $0\leq c^\A_{s,t}(\P)\leq\E[Y|\F_s]$ for all $Y\in\A_{t,T}\cap L^p(\F_t)$.
    \end{itemize}
\end{lemma}
 
The next two lemmas are the cash-additive versions of Lemmas~\ref{LEM:N-from-A-to-rho} and~\ref{LEM:N-from-rho-to-A}, from which they follow through the embedding $\A^m_{s,t}:=\A_{s,t}-m$, except for part~(c) of Lemma~\ref{LEM:from_A_to_rho}, which concerns $p=+\infty$ and the dual representation; see \cite[Proposition~3.6]{Cheridito+Delbaen+Kupper_2006_Dynamic_monetary_risk_measures_bounded_discrete-time_processes}, \cite[Section~11.1]{Follmer+Schied_2016_Stochastic_finance} and \cite{Detlefsen+Scandolo_2005_Conditional_dynamic_convex_risk_measures} for the correspondence, and \cite[Theorem~4.6]{Cheridito+Delbaen+Kupper_2006_Dynamic_monetary_risk_measures_bounded_discrete-time_processes}, \cite[Theorem~2.5]{Bion-Nadal_2008_Dynamic_risk_measures_time_consistency__risk_measures_BMO_martingales} and \cite[Proposition~11.15]{Follmer+Schied_2016_Stochastic_finance} for time-consistency. We only state them in the form used below, and defer their proofs to Appendix~\ref{APP:CA}.
 
\begin{lemma}[From $\A$ to $\r^\A$]
\label{LEM:from_A_to_rho}
    Let $\A$ be a monetary acceptance family in $L^p$, $p\in[1,+\infty]$.
    Then $\r^\A$ is cash-additive, $\A\subseteq\A^{\r^\A}$ and 
    \begin{enumerate}[label=(\alph*)]
    \item
        \begin{multicols}{2}
        \begin{itemize}[nosep, leftmargin=*, label=\scriptsize$\bullet$]
            
            \item \itemref{IT:S-monotone} $+$ \itemref{IT:S-closed} $+$ \itemref{IT:S-pasting} $\Rightarrow$ $\A=\A^{\r^\A}$
            \item \itemref{IT:S-normalized} $\Leftrightarrow$ $\r^\A$ is normalized
            \item \itemref{IT:S-upperbound} $+$ \itemref{IT:S-lowerbound} $\Leftrightarrow$ $\r^\A$ is well posed in $L^p$
            \item $\A\subseteq\A^\E$ $\Rightarrow$ \itemref{IT:S-lowerbound}
        \end{itemize}
        \end{multicols}
    \item 
        \begin{multicols}{2}
        \begin{itemize}[nosep, leftmargin=*, label=\scriptsize$\bullet$]
            \item \itemref{IT:S-monotone} $\Rightarrow$ $\r^\A$ is monotone
            \item \itemref{IT:S-conic} $\Rightarrow$ $\r^\A$ is positively homogeneous
            \item \itemref{IT:S-star} $\Rightarrow$ $\r^\A$ is star-shaped
            \item \itemref{IT:S-convex} $\Rightarrow$ $\r^\A$ is convex
            \item \itemref{IT:S-upperbound} $+$ \itemref{IT:S-pasting} $\Rightarrow$ $\r^\A$ is local
        \end{itemize}
        \end{multicols}
    \item
        If $p=+\infty$ and $\A$ satisfies \itemref{IT:S-monotone} and \itemref{IT:S-normalized}, then $\r^\A$ is a normalized and local cash-additive dynamic risk measure on $L^\infty$ with $\|\r^\A_{s,t}(X)\|_{L^\infty}\leq\|X\|_{L^\infty}$.
        Moreover:
        \begin{enumerate}[label=(\roman*), ref=(c)(\roman*), nosep]
            \item
                \itemref{IT:S-convex} $\Rightarrow$ $\r^\A$ is conditionally convex;
            \item
                \itemref{IT:S-conic} $\Rightarrow$ $\r^\A$ is conditionally positively homogeneous;
            \item
            \label{item:cont_above}
                \itemref{IT:S-closed} $+$ \itemref{IT:S-pasting} $\Rightarrow$ $\r^\A$ is continuous from above;
            \item
            \label{item:dual_eq}
                \itemref{IT:S-closed} $+$ \itemref{IT:S-pasting} $+$ \itemref{IT:S-convex} $+$ $\big(c^\A_{s,t}(\P)<+\infty\big)$ $\Rightarrow$ $c^\A_{s,t}$ restricted to $\mathcal Q_{s,t}$ is the minimal penalty of $\r^\A_{s,t}$, namely
                \begin{equation}
                \label{EQ:dual_ac}
                    \r^\A_{s,t}(X)
                    =
                    \esssup_{\Q\in\mathcal Q_{s,t}}
                    \big\{
                    \E_\Q[-X|\F_s]-c^\A_{s,t}(\Q)
                    \big\},\qquad X\in L^\infty(\F_t),
                \end{equation}
                and  $c^\A_{s,t}(\Q)\leq\a(\Q)$, $\P$-a.s., for all $\Q\in\mathcal Q_{s,t}$ and all $\a:\mathcal Q_{s,t}\to L^0(\F_s;[0,+\infty])$ such that \eqref{EQ:dual_ac} holds with $\a$ in place of $c^\A_{s,t}$.
        \end{enumerate}
    \end{enumerate}
\end{lemma}

\begin{lemma}[From $\r$ to $\A^\r$]
\label{LEM:from_rho_to_A}
    Let $\r$ be a cash-additive dynamic risk functional well posed in $L^p$.
    Then $\A^\r$ satisfies \itemref{IT:S-upperbound} and \itemref{IT:S-lowerbound}, $\r^{\A^\r}=\r$, and all the implications in Lemma~\ref{LEM:from_A_to_rho}~(b) are equivalences.
    Moreover, if $\r$ is monotone and local, then $\r$ is time-consistent if and only if $\A^\r$ satisfies \itemref{IT:S-decomposition}.
\end{lemma}
 
\subsection{Penalty representation}
\label{SUBSEC:CA-penalty}
 
In this subsection, we give axioms on a monetary family in $L^\infty$ under which the minimal penalty of the induced risk measure is an integral functional of the Girsanov kernel.
Throughout, $p=+\infty$ and $q$ denotes the Girsanov kernel of $\Q\in\mathcal Q_{s,t}$.
 
\begin{proposition}
\label{prop: charact-accept-f}
    Let $\A$ be a regular monetary acceptance family in $L^\infty$ satisfying  monotonicity \itemref{IT:S-monotone}, convexity \itemref{IT:S-convex} and expectation-boundedness \itemref{IT:S-expectation}.
    Then $\r^\A$ is a normalized, conditionally convex, local and time-consistent cash-additive dynamic risk measure on $L^\infty$, which is continuous from above.
    Moreover, there exists a $\mathcal P\otimes\mathcal B(\R^d)$-measurable function $f:\Omega\times[0,T]\times\R^d\to[0,+\infty]$ such that, for every $(\o,u)\in\Omega\times[0,T]$, $f(\o,u,\cdot)$ is proper, convex and lower semicontinuous with $f(\o,u,0)=0$, and, for all $0\leq s\leq t\leq T$,
    \begin{equation}
    \label{eq: rho-A-f}
        \rho_{s,t}^{\mathcal{A}}(X)=\esssup_{\Q \in \mathcal{Q}_{s,t}} \, \E_{\Q}\bigg[-X -\int_s^t f(u,q_u) \d u\,\bigg\vert\, \mathcal{F}_s \bigg], \qquad X \in L^{\infty}(\mathcal{F}_t).
    \end{equation}
    Furthermore, the restriction of $c^\A_{s,t}$ to $\mathcal Q_{s,t}$ is the minimal penalty of $\r^\A_{s,t}$ and
    \begin{equation}
    \label{EQ:c^A_f}
        c^\A_{s,t}(\Q)=\E_\Q\bigg[\int_s^tf(u,q_u)\d u\,\bigg|\,\F_s\bigg], \qquad \Q\in\mathcal Q_{s,t}.
    \end{equation}
    Finally, for all $0\leq s\leq t\leq T$,
    \begin{equation}
    \label{eq: A-f}
        \mathcal{A}_{s,t}= \A^{\r^\A}_{s,t}=\bigg\{X \in L^{\infty}(\mathcal{F}_t)\ :\ \E_{\Q}[-X| \mathcal{F}_s] \leq \E_{\Q}\bigg[\int_s^t f(u,q_u) \d u\,\bigg\vert\, \mathcal{F}_s \bigg] \text{ for all } \Q \in \mathcal{Q}_{s,t}\bigg\}.
    \end{equation}
\end{proposition}
 
The integral representation for the penalty is deduced from \cite[Theorem~3.2]{Delbaen+Peng+RosazzaGianin_2010_Representation_penalty_term_dynamic_concave_utilities}, whose assumptions (a)--(g) are formulated for a family $(u_{\s,\t})$ indexed by pairs of \emph{stopping times}, whereas $\r^\A$ is only defined at deterministic times.
Lemma~\ref{LEM:extension} below, proved in Appendix~\ref{APP:extension}, shows that this is not an additional restriction: under the properties of $\r^\A$ established in Lemma~\ref{LEM:from_A_to_rho}, an extension to stopping times satisfying all the assumptions of \cite{Delbaen+Peng+RosazzaGianin_2010_Representation_penalty_term_dynamic_concave_utilities} is automatically available.

We denote by $\Theta$ the set of $\mathbb F$-stopping times with values in $[0,T]$.

\begin{lemma}
\label{LEM:extension}
    Let $\r$ be a normalized, conditionally convex, time-consistent, cash-additive dynamic risk measure on $L^\infty$ which is continuous from above and satisfies $\r_{t,T}(X)\geq\E[-X|\F_t]$ for all $t\in[0,T]$ and $X\in L^\infty(\F_T)$.
    Then, for every $X\in L^\infty(\F_T)$, the process $(\r_{t,T}(X))_{t\in[0,T]}$ admits a càdlàg modification $(\bar\r_t(X))_{t\in[0,T]}$, and
    \begin{equation}
    \label{EQ:extension}
        \hat\r_{\s,\t}(X):=\bar\r_\s(X),\qquad X\in L^\infty(\F_\t),\ \s\leq\t\text{ in }\Theta,
    \end{equation}
    coincides with $\r$ at deterministic times, while $u_{\s,\t}:=-\hat\r_{\s,\t}$ is a dynamic concave utility satisfying assumptions (a)--(h) of \cite[Section~3]{Delbaen+Peng+RosazzaGianin_2010_Representation_penalty_term_dynamic_concave_utilities}.
\end{lemma}

\begin{proof}[\textbf{Proof of Proposition~\ref{prop: charact-accept-f}}]
    All the properties of $\r^\A$ are deduced from Lemmas~\ref{LEM:from_A_to_rho}~(a),(c) and~\ref{LEM:from_rho_to_A}.
    Moreover, $c^\A_{t,T}(\P)=0$ by Lemma~\ref{LEM:penalty}, i.e., $\r^\A_{t,T}(X)\geq\E[-X|\F_t]$ for all $t\in[0,T]$ and $X\in L^\infty(\F_T)$.
    
    Then, Lemma~\ref{LEM:extension} applies to $\r^\A$, and \cite[Theorem~3.2]{Delbaen+Peng+RosazzaGianin_2010_Representation_penalty_term_dynamic_concave_utilities} yields $f:\Omega\times[0,T]\times\R^d\to[0,+\infty]$ such that, for all $0\leq s\leq t\leq T$ and $\Q\in\mathcal Q_{s,t}$ (extending $\Q$ to $\F_T$ through its density on $\F_t$)
    \[
        c^\A_{s,t}(\Q)=\esssup_{\xi\in L^\infty(\F_t)}\big\{\E_\Q[-\xi|\F_s]-\r^\A_{s,t}(\xi)\big\}=\E_\Q\left[\int_s^tf(u,q_u)\d u\,\bigg|\,\F_s\right],
    \]
    where the first identity is \eqref{EQ:c^A_dual}; this is \eqref{EQ:c^A_f}.
    
    By \cite[Proposition~3.6\,(ii),(iii),(v)]{Delbaen+Peng+RosazzaGianin_2010_Representation_penalty_term_dynamic_concave_utilities}, $f=\inf_n f_n$ with $f_n(\o,t,q)=\sup_{z\in\mathbb Q^d}\{q\cdot z-g_n(\o,t,z)\}$ outside a $(\P\otimes\ell)$-null set, where $g_n(\cdot,\cdot,z)$ is predictable; hence $f$ admits a $\mathcal P\otimes\mathcal B(\R^d)$-measurable version.
    Redefining $f(\o,t,\cdot):=0$ on a predictable $(\P\otimes\ell)$-null set, we may assume that $f(\o,t,\cdot)$ is proper, convex and lower semicontinuous with $f(\o,t,0)=0$ for every $(\o,t)$; by Fubini's theorem this does not affect $\int_s^tf(u,q_u)\d u$, $\Q$-a.s.
    
    \eqref{eq: rho-A-f} and the minimality of $c^\A$ on $\mathcal Q_{s,t}$ follow from \eqref{EQ:c^A_f} and Lemma~\ref{LEM:from_A_to_rho}\,\ref{item:dual_eq}, which applies since $c^\A_{s,t}(\P)<+\infty$ by Lemma~\ref{LEM:penalty}, and \eqref{eq: A-f} follows from $\A=\A^{\r^\A}$ and \eqref{eq: rho-A-f}.
\end{proof}
 
Proposition~\ref{prop: charact-accept-f} does not characterize BSDE-induced risk measures, since $f$ need not be the convex conjugate of a driver.
It is the conjugate of a driver when $\A$ is induced by a $\mu$-Lipschitz or a quadratic driver; these two cases are discussed in the next two subsections.

\subsection{Lipschitz BSDEs}
\label{SUBSEC:CA-Lipschitz}
 
Throughout this subsection, 
we fix $p\in(1,+\infty]$.
For $\mu>0$, we use the reference $\mu$-Lipschitz monetary driver $g^\mu(z):=\mu|z|$, $z\in\R^d$, with induced risk measure $\r^\mu:=\r^{g^\mu}$ and induced monetary acceptance family $\A^\mu:=\A^{g^\mu}$; note that  $\r^\mu\leq\r^{\tinybullet,\mu}$ by \itemref{IT:B-comparison}.
The cash-additive counterpart of \itemref{IT:N-SL} is \itemref{IT:S-SL}: there exists $\mu>0$ such that $\A_{s,t}+\A^\mu_{s,t}\subseteq\A_{s,t}$ for all $0\leq s\leq t\leq T$; we then say that \itemref{IT:S-SL} holds with constant $\mu$.
Here $\A^\mu_{s,t}$ collects the positions with nonnegative conditional expectation under every model with Girsanov kernel bounded by $\mu$, and \itemref{IT:S-SL} says that adding such a position to an acceptable one preserves acceptability.

\begin{remark}
\label{REM:SL_implies_S1}
    For a monetary acceptance family $\A$ in $L^p$, the following implications hold:
    \begin{multicols}{2}
        \begin{itemize}[nosep, leftmargin=*, label=\scriptsize$\bullet$]
        \item \itemref{IT:S-SL} $\Rightarrow$ \itemref{IT:S-monotone}
        \item \itemref{IT:S-SL} $+$ \itemref{IT:S-normalized} $\Rightarrow$ \itemref{IT:S-upperbound} $+$ \itemref{IT:S-lowerbound}
    \end{itemize}    
    \end{multicols}
    \begin{itemize}[nosep, leftmargin=*, label=\scriptsize$\bullet$]
        \item \itemref{IT:S-SL} with constant $\mu$ $\Rightarrow$ $\A^m_{s,t}:=\A_{s,t}-m$ satisfies \itemref{IT:N-SL} with constant $\mu$.
    \end{itemize}
    
    Indeed, $\r^\mu_{s,t}(U)\leq\r^\mu_{s,t}(0)=0$ for $U\in L^p_+(\F_t)$ by \itemref{IT:B-comparison}, i.e., $L^p_+(\F_t)\subseteq\A^\mu_{s,t}$, and \itemref{IT:S-SL} gives the first implication.
    For the second, let $X\in L^p(\F_t)$ and choose $a\in\A_{s,t}\cap L^p(\F_s)$ (Remark~\ref{REM:not_empty}); since $X+\r^\mu_{s,t}(X)\in\A^\mu_{s,t}$ by \itemref{IT:B-stability} and \itemref{IT:B-cash}, \itemref{IT:S-SL} gives $X+a+\r^\mu_{s,t}(X)\in\A_{s,t}$, with $a+\r^\mu_{s,t}(X)\in L^p(\F_s)$.
    For \itemref{IT:S-lowerbound}, if $X+m\in\A_{s,t}$ with $m\in L^p(\F_s)$, then $-X+\r^\mu_{s,t}(-X)\in\A^\mu_{s,t}$, and \itemref{IT:S-SL} gives $m+\r^\mu_{s,t}(-X)\in\A_{s,t}\cap L^p(\F_s)$, hence $m\geq-\r^\mu_{s,t}(-X)\in L^p(\F_s)$ by \itemref{IT:S-normalized}.
    For the third, if $X\in\A^m_{s,t}$ and $Y\in\A^{n,\mu}_{s,t}$, then $\r^\mu_{s,t}(Y)\leq\r^{\tinybullet,\mu}_{s,t}(Y)\leq n$, so that $Y+n\in\A^\mu_{s,t}$, and \itemref{IT:S-SL} gives $(X+m)+(Y+n)\in\A_{s,t}$, i.e., $X+Y\in\A^{m+n}_{s,t}$.
\end{remark}
 
We now characterize the acceptance families induced by Lipschitz BSDEs in the cash-additive case.
The result is derived from Theorem~\ref{THM:Lp-nonadditive} through the embedding $\A^m_{s,t}:=\A_{s,t}-m$. As expected, the corresponding driver is then monetary, that is, independent of $y$.

\begin{theorem}[Lipschitz BSDEs, cash-additive case]
\label{THM:Lp-additive}
    Let $p\in(1,+\infty]$ and $\mu>0$.
    \begin{enumerate}[label=(\roman*)]
        \item
            If a monetary acceptance family $\A$ in $L^p$ is regular and satisfies Lipschitz domination \itemref{IT:S-SL} with constant $\mu$, then it is induced by a $\mu$-Lipschitz monetary driver $g$.
            Conversely, if $g$ is a $\mu$-Lipschitz monetary driver and $p<\infty$ or $g$ is convex, then $\A^g$ is regular and satisfies \itemref{IT:S-SL} with constant $\mu$.
        \item
            In this case, $g$ is unique, $\r^\A=\r^g$ on $L^p$, and $\A$ also satisfies \itemref{IT:S-upperbound}, \itemref{IT:S-lowerbound} and \itemref{IT:S-monotone}.
            Moreover:
            \begin{itemize}[nosep, leftmargin=*, label=\scriptsize$\bullet$]
                \item
                    if $p<\infty$, then, for $z\in\R^d$, $q\in[1,2)$ and $\ell$-a.e.\ $t\in[0,T)$, with $\Delta_\eps W_t:=W_{t+\eps}-W_t$,
                    \begin{equation}
                    \label{eq: g from A-1}
                        g(t,z)
                        =L^q\text{-}\lim_{\eps\downarrow 0}\frac{
                            \r^\A_{t,t+\eps}(-z\cdot\Delta_\eps W_t)}\eps;
                    \end{equation}
                \item 
                    if $p=\infty$ and $\A$ also satisfies \itemref{IT:S-convex} and \itemref{IT:S-expectation}, then $g=f^\ast$ and $f=g^\ast$ $\P\otimes\ell$-a.e., where $f$ is the function of Proposition~\ref{prop: charact-accept-f}.
            \end{itemize}
        \item
            If $g$ is a $\mu$-Lipschitz monetary driver, then $\A^g$ satisfies
            \begin{multicols}{3}
            \begin{itemize}[nosep, leftmargin=*, label=\scriptsize$\bullet$]
                \item \itemref{IT:S-convex} $\Leftrightarrow$ $g$ convex,
                \item \itemref{IT:S-star} $\Leftrightarrow$ $g$ star-shaped,
                \item \itemref{IT:S-expectation} $\Leftrightarrow$ $g\geq0$.
            \end{itemize}
            \end{multicols}
    \end{enumerate}
\end{theorem}

\begin{proof}
\noindent\textbf{\textit{(i) first statement, (ii).}}
    Let $\A$ satisfy the axioms in (i), and let $\A^m_{s,t}:=\A_{s,t}-m$, so that $\r^{\A^\tinybullet}=\r^\A$.
    By Remark~\ref{REM:SL_implies_S1}, $\A^\tinybullet$ satisfies \itemref{IT:N-SL} with constant $\mu$; by Remark~\ref{REM:P-embedding} and Section~\ref{SUBSEC:N-TC}, it also satisfies \itemref{IT:L-rightcont}, \itemref{IT:N-normalized}, \itemref{IT:N-decomposition} and \itemref{IT:N-pasting}.
    By Theorem~\ref{THM:Lp-nonadditive}~(i),(ii), $\A^\tinybullet=\A^{\tinybullet,\hat g}$ and $\r^\A=\r^{\hat g}$ on $L^p$ for some $\mu$-Lipschitz driver $\hat g$.
    
    Since $\r^{\hat g}=\r^\A$ is cash-additive, $\A^{\tinybullet,\hat g}=\A^\tinybullet$ satisfies \itemref{IT:N-cashadd} by Lemma~\ref{LEM:N-from-rho-to-A}, so that $\hat g=:g$ is monetary by Theorem~\ref{THM:Lp-nonadditive}~(iii), and $\A=\A^0=\A^{0,g}=\A^g$.
    If $g'$ is another $\mu$-Lipschitz monetary driver with $\A^{g'}=\A$, then $\A^{\tinybullet,g'}=\A^{g'}-m=\A^\tinybullet$ by \itemref{IT:B-cash}, and $g'=g$ $\P\otimes\ell$-a.e.\ by the uniqueness in Theorem~\ref{THM:Lp-nonadditive}~(ii).
    For $p<\infty$, \eqref{eq: g from A-1} is \eqref{EQ:N-g-from-A} with $y=0$, given by Theorem~\ref{THM:Lp-nonadditive}~(ii).

    Let now $p=+\infty$ and assume also \itemref{IT:S-convex} and \itemref{IT:S-expectation}.
    Since $\A=\A^g$, (iii) shows that $g$ is convex, so we may assume that $g$ is a $\mu$-Lipschitz convex monetary driver.
    Moreover, $\A$ satisfies \itemref{IT:S-monotone} by (ii), so that Proposition~\ref{prop: charact-accept-f} applies, and we recall from \cite[Proposition~3.6]{Delbaen+Peng+RosazzaGianin_2010_Representation_penalty_term_dynamic_concave_utilities} that its $f$ satisfies $f=\inf_n f_n$, where $f_n=g_n^\ast$.
    Since $g$ is convex and $\mu$-Lipschitz, $g^\ast(\cdot,q)=+\infty$ for $|q|>\mu$, and $\r^g_{s,t}(X)=\esssup_{\Q\in\mathcal Q_{s,t}}\E_\Q\big[-X-\int_s^tg^\ast(u,q_u)\d u\,\big|\,\F_s\big]$, where only kernels with $|q|\leq\mu$ contribute; see \cite[Theorem~7.4]{Barrieu+ElKaroui_2009_Pricing_hedging_optimally_designing_derivatives_minimization_risk_measures}.
    By the minimality of $c^\A$ (Proposition~\ref{prop: charact-accept-f}), $c^\A_{s,t}(\Q)\leq\E_\Q\big[\int_s^tg^\ast(u,q_u)\d u\,\big|\,\F_s\big]$ for all $\Q\in\mathcal Q_{s,t}$.
    Hence, for $n\geq\mu$, restricting the dual representation of $\r^\A=\r^g$ to kernels bounded by $n$, as in \cite[Proposition~3.6]{Delbaen+Peng+RosazzaGianin_2010_Representation_penalty_term_dynamic_concave_utilities}, gives a risk measure that is at most $\r^\A$ and at least $\r^g$, i.e., $\r^{g_n}=\r^g$, and the uniqueness of the driver, applied with constant $n$, yields $g_n=g$.
    Consequently, $f=g^\ast$, and, since $g(\o,t,\cdot)$ is finite, convex and continuous, the Fenchel--Moreau theorem gives $g=f^\ast$.
    \medskip

\noindent\textbf{\textit{(i) converse, axioms in (ii), (iii).}}
    Let $g$ be a $\mu$-Lipschitz monetary driver.
    By Theorem~\ref{THM:Lp-nonadditive}~(i),(ii), $\A^{\tinybullet,g}$ satisfies \itemref{IT:N-normalized}, \itemref{IT:N-decomposition} and \itemref{IT:N-pasting}, and also \itemref{IT:N-closed} if $p<\infty$ or $g$ is convex.
    Since $\A^{\tinybullet,g}=\A^g-m$ by \itemref{IT:B-cash}, Remark~\ref{REM:P-embedding}, Section~\ref{SUBSEC:N-TC} and the definition of \itemref{IT:S-closed} show that $\A^g$ satisfies \itemref{IT:S-normalized}, \itemref{IT:S-decomposition} and \itemref{IT:S-pasting}, and also \itemref{IT:S-closed} in the same cases, so that $\A^g$ is regular if $p<\infty$ or $g$ is convex.
    Moreover, by \itemref{IT:B-stability}, \itemref{IT:B-comparison} and \itemref{IT:B-cash}, $\r^g$ is a cash-additive dynamic risk measure on $L^p$ and $\r^{\A^g}=\r^g$, so that $\A^g$ satisfies \itemref{IT:S-upperbound}, \itemref{IT:S-lowerbound} and \itemref{IT:S-monotone} by Lemma~\ref{LEM:from_rho_to_A}.

    For \itemref{IT:S-SL}, let $X^1,X^2\in L^p(\F_t)$ and let $(Y^i,Z^i)$ be the solution of \eqref{EQ:N-BSDE} with terminal condition $-X^i$, $i=1,2$.
    Then $(Y^1-Y^2,Z^1-Z^2)$ is the solution of \eqref{EQ:N-BSDE} with terminal condition $-(X^1-X^2)$ and monetary driver
    \[
        h(r,z):=g(r,Z^2_r+z)-g(r,Z^2_r)\leq g^\mu(z),
    \]
    which is $\mu$-Lipschitz with $h(\cdot,0)=0$.
    Hence \itemref{IT:B-comparison} gives $\r^g_{s,t}(X^1)-\r^g_{s,t}(X^2)=\r^h_{s,t}(X^1-X^2)\leq\r^\mu_{s,t}(X^1-X^2)$.
    In particular, if $X^2\in\A^g_{s,t}$ and $X^1-X^2\in\A^\mu_{s,t}$, then $\r^g_{s,t}(X^1)\leq0$, i.e., $X^1\in\A^g_{s,t}$; hence $\A^g_{s,t}+\A^\mu_{s,t}\subseteq\A^g_{s,t}$.

    By definition, $\A^g$ satisfies \itemref{IT:S-convex} if and only if $\A^g-m=\A^{\tinybullet,g}$ satisfies \itemref{IT:N-convex}, i.e., by Theorem~\ref{THM:Lp-nonadditive}~(iii), if and only if $g$ is convex.
    Similarly, \itemref{IT:S-star} for $\A^g$ is \itemref{IT:N-star} for $\A^{\tinybullet,g}=\A^g-m$, so that it is equivalent to star-shapedness of $g$ by Theorem~\ref{THM:Lp-nonadditive}~(iii).
    By Theorem~\ref{THM:Lp-nonadditive}~(iii), $g\geq0$ if and only if $\A^{\tinybullet,g}$ satisfies \itemref{IT:N-expectation}, i.e., $\A^g\subseteq\A^\E$, which implies \itemref{IT:S-expectation}.
    Conversely, assume \itemref{IT:S-expectation} and let $X\in\A^g_{s,t}$.
    Since $\r^g$ is time-consistent and conditionally constant by \itemref{IT:B-flow}, $\r^g_{s,T}(X)=\r^g_{s,t}\big(-\r^g_{t,T}(X)\big)=\r^g_{s,t}(X)\leq0$, i.e., $X\in\A^g_{s,T}$, and \itemref{IT:S-expectation} at time $s$ gives $\E[X|\F_s]\geq0$.
    Hence $\A^g\subseteq\A^\E$.
\end{proof}

Closedness, which is part of Definition~\ref{DEF:regular} of a regular family, is only needed in the previous theorem to identify the sets.
In the next corollary we show that, without \itemref{IT:S-closed}, we still obtain $\r^\A=\r^g$, but not $\A=\A^g$.

\begin{corollary}
\label{COR:no-closedness}
    Let $p\in(1,+\infty]$, $\mu>0$, and let $\A$ be a monetary acceptance family in $L^p$ satisfying \itemref{IT:S-normalized}, \itemref{IT:S-pasting}, \itemref{IT:S-decomposition} and \itemref{IT:S-SL} with constant $\mu$.
    Then $\r^\A=\r^g$ on $L^p$ for a unique $\mu$-Lipschitz monetary driver $g$, and $\A\subseteq\A^g$, with equality if $\A$ satisfies \itemref{IT:S-closed}.
    If $p<\infty$, $\A^g$ is the $L^p$-closure of $\A$.
\end{corollary}
\begin{proof}
    The leveled family $\A^m_{s,t}:=\A_{s,t}-m$ satisfies \itemref{IT:N-normalized} and \itemref{IT:N-pasting} by Remark~\ref{REM:P-embedding}, \itemref{IT:N-decomposition} by Section~\ref{SUBSEC:N-TC}, and \itemref{IT:N-SL} with constant $\mu$ by Remark~\ref{REM:SL_implies_S1}.
    By Corollary~\ref{COR:N-no-lrc}, $\r^\A=\r^{\A^\tinybullet}=\r^{\hat g}$ for a unique $\mu$-Lipschitz driver $\hat g$, and, since $\r^{\hat g}=\r^\A$ is cash-additive, $\A^{\tinybullet,\hat g}$ satisfies \itemref{IT:N-cashadd} by Lemma~\ref{LEM:N-from-rho-to-A}, so that $g:=\hat g$ is monetary by Theorem~\ref{THM:Lp-nonadditive}~(iii).
    Then $\A=\A^0\subseteq\A^{0,g}=\A^g$ by Corollary~\ref{COR:N-no-lrc}, with equality if $\A^\tinybullet$ satisfies \itemref{IT:L-rightcont}, which holds under \itemref{IT:S-closed} by Remark~\ref{REM:P-embedding}.
    Let now $p<\infty$.
    Since $\r^g_{s,t}$ is $L^p$-continuous by \itemref{IT:B-stability}, $\A^g_{s,t}=\{X\in L^p(\F_t):\r^g_{s,t}(X)\leq0\}$ is closed and contains $\A_{s,t}$.
    Conversely, let $X\in\A^g_{s,t}$, i.e., $\r^\A_{s,t}(X)\leq0$.
    By Lemma~\ref{LEM:N-directedness} applied to $\A^\tinybullet$, there are $m_k\in L^p(\F_s)$ with $X+m_k\in\A_{s,t}$ and $m_k\downarrow\r^\A_{s,t}(X)$ $\P$-a.s.
    Since $-\r^\A_{s,t}(X)\geq0$ and $\A$ satisfies \itemref{IT:S-monotone} (Remark~\ref{REM:SL_implies_S1}), $X_k:=X+m_k-\r^\A_{s,t}(X)\in\A_{s,t}$, and $X_k\to X$ in $L^p$ by dominated convergence, since $0\leq m_k-\r^\A_{s,t}(X)\leq m_1-\r^\A_{s,t}(X)\in L^p$.
\end{proof}
 
\subsection{Quadratic BSDEs}
\label{SUBSEC:CA-quadratic}
 
Throughout this subsection, $g$ denotes a quadratic convex driver in the sense of Section~\ref{SUBSEC:P-BSDE}, so that Lemma~\ref{LEM:quad_BSDE} applies.
Representation theorems for quadratic filtration-consistent nonlinear expectations are given in \cite{Hu+Ma+Peng+Yao_2008_Representation_theorems_quadratic_F-consistent_nonlinear_expectations} under several domination conditions, and in \cite[Theorem~4.6]{Zheng_2024_g-expectations_filtration-consistent_nonlinear_expectations} under a locally Lipschitz domination condition; both assume independent increments, i.e., that $\r_{s,t}\big(z\cdot(W_t-W_s)\big)$ is deterministic for all $z\in\R^d$ and $0\leq s\leq t\leq T$, which forces the driver to be deterministic.
Here, instead, the driver may be random, and the characterization is obtained from the penalty representation of Proposition~\ref{prop: charact-accept-f}.

The conditional relative entropy of $\Q\in\mathcal Q_{s,t}$ with respect to $\P$ is defined as
\begin{equation}
\label{EQ:entropy}
    H_s(\Q|\P)
    :=\E_\Q\bigg[\log \frac{\d \Q}{\d\P}\bigg|\F_s\bigg]
    =\frac1{2}\,\E_\Q\left[\int_s^t|q_u|^2\d u\,\bigg|\,\F_s\right]\in[0,+\infty],
\end{equation}
where $q$ is the Girsanov kernel of $\Q$; see \cite[Theorems~2.1 and 2.3]{Leonard_2012_Girsanov_theory_finite_entropy_condition}.
 
For $\g>0$, $0\leq s\leq t\leq T$ and $X\in L^\infty(\F_t)$, we consider the dynamic entropic risk measure, with its acceptance set and minimal penalty:
\begin{align}
    \mathfrak e^\g_{s,t}(X)&:=\frac1\g\log\E\big[e^{-\g X}\big|\F_s\big],\\
    \A^\g _{s,t}&=\big\{X\in L^\infty(\F_t)\ :\ \E[e^{-\g X}|\F_s]\leq 1\big\},\\
    c^\g _{s,t}(\Q)
    &=\frac1\g H_s(\Q|\P),
    \qquad \Q\in\mathcal Q_{s,t};
    \label{EQ:support_entropic}
\end{align}
see \cite[Example~4.34]{Follmer+Schied_2016_Stochastic_finance} for the static case and \cite{Detlefsen+Scandolo_2005_Conditional_dynamic_convex_risk_measures} for its conditional version.
Moreover, $\mathfrak e^\g$ and $\A^\g$ are known to be induced by the quadratic convex driver $g^\g:\R^d\to\R$
\[
    g^\g(z):=\frac\g2|z|^2,\qquad z\in\R^d.
\]

For a monetary acceptance family $\A$ in $L^\infty$, the entropic bound \itemref{IT:S-SQ} requires the existence of $\g>0$ such that $\A^\g_{0,T}\subseteq\A_{0,T}$; we then say that \itemref{IT:S-SQ} holds with constant $\g$.
It says that, over the whole horizon, the rule is not more conservative than an agent with exponential utility and risk aversion $\g$: every $X$ with $\E[e^{-\g X}]\leq1$ is acceptable.
 
The next lemma translates \itemref{IT:S-SQ} into a condition on the function $f$ of Proposition~\ref{prop: charact-accept-f}.
 
\begin{lemma}
\label{LEM:SQ_f}
    Let $\A$ be a regular monetary acceptance family in $L^\infty$ satisfying \itemref{IT:S-monotone}, \itemref{IT:S-convex} and \itemref{IT:S-expectation}, let $f$ be as in Proposition~\ref{prop: charact-accept-f}, and let $\g>0$.
    The following are equivalent:
    \begin{enumerate}[label=(\roman*)]
        \item \label{SQ:f}
            there is a $\P\otimes\ell$-null set $N\in\mathcal P$ such that $f(\o,u,q)\geq
                \frac1{2\g}|q|^2$, for $(\o,u)\notin N$, $q\in\R^d$;
        \item \label{SQ:Ast}
            $\A^\g_{s,t}\subseteq\A_{s,t}$ for all $0\leq s\leq t\leq T$;
        \item \label{SQ:A0T}
            \itemref{IT:S-SQ} holds with constant $\g$.
    \end{enumerate}
\end{lemma}
\begin{proof}
    \ref{SQ:f}$\Rightarrow$\ref{SQ:Ast}.
    Let $X\in\A^\g _{s,t}$ and $\Q\in\mathcal Q_{s,t}$.
    Since $N$ is $\P\otimes\ell$-null and $\Q\sim\P$, Fubini's theorem implies that, for $\Q$-a.e.\ $\o$, $(\o,u)\notin N$ for a.e.\ $u\in[s,t]$.
    By \eqref{EQ:dual_quad} for $g^\g$, for which $(g^\g)^\ast(q)=|q|^2/(2\g)$, and since $\mathfrak e^\g_{s,t}(X)\leq0$,
    \[
        \E_\Q[-X|\F_s]
        \leq\frac1{2\g}\,\E_\Q\bigg[\int_s^t|q_u|^2\d u\,\bigg|\,\F_s\bigg]
        \leq\E_\Q\bigg[\int_s^tf(u,q_u)\d u\,\bigg|\,\F_s\bigg].
    \]
    By \eqref{eq: A-f}, $X\in\A_{s,t}$.
    \smallskip
    
    \ref{SQ:Ast}$\Rightarrow$\ref{SQ:A0T} is trivial.
    \smallskip
 
    \ref{SQ:A0T}$\Rightarrow$\ref{SQ:f}.
    Let $\Q\in\mathcal Q_{0,T}$ have a bounded Girsanov kernel $q$.
    By Proposition~\ref{prop: charact-accept-f}, the definition of $c^\A_{0,T}$, \itemref{IT:S-SQ}, and \eqref{EQ:support_entropic},
    \begin{equation}
    \label{EQ:SQ_bounded}
        \E_\Q\left[\int_0^Tf(u,q_u)\d u\right]
        =c^\A_{0,T}(\Q)
        \geq\sup_{X\in\A^\g _{0,T}}\E_\Q[-X]
        =\frac1{2\g}\E_\Q\left[\int_0^T|q_u|^2\d u\right].
    \end{equation}
    Consider
    \[
        \Gamma:=\big\{(\o,u,q)\in\Omega\times[0,T]\times\R^d\ :\ f(\o,u,q)<|q|^2/(2\g)\big\}\in\mathcal P\otimes\mathcal B(\R^d),
    \]
    and let $N'$ be its projection on $\Omega\times[0,T]$.
    By the measurable selection theorem (see \cite{Aumann_1969_Measurable_utility_measurable_choice_theorem} and \cite[Theorems~18.25 and~18.26]{Aliprantis+Border_1999_Infinite_dimensional_analysis}), applied on the complete finite measure space $(\Omega\times[0,T],\overline{\mathcal P},\P\otimes\ell)$ as in the proof of \cite[Proposition~3.6(v)]{Delbaen+Peng+RosazzaGianin_2010_Representation_penalty_term_dynamic_concave_utilities}, we have $N'\in\overline{\mathcal P}$ and there exists a $\overline{\mathcal P}$-measurable map $\sigma:\Omega\times[0,T]\to\R^d$ such that $(\o,u,\sigma(\o,u))\in\Gamma$ for $\P\otimes\ell$-a.e.\ $(\o,u)\in N'$.
    Let $\tilde q$ be a predictable process with $\tilde q=\sigma$ $\P\otimes\ell$-a.e., and set
    \[
        B_n:=\big\{(\o,u)\ :\ |\tilde q(\o,u)|\leq n,\ f\big(\o,u,\tilde q(\o,u)\big)<|\tilde q(\o,u)|^2/(2\g)\big\}\in\mathcal P,\qquad n\in\N,
    \]
    so that $N'\subseteq\bigcup_{n}B_n$ up to a $\P\otimes\ell$-null set.
    Fix $n\in\N$ and let $q^n:=\tilde q\1_{B_n}$, a predictable process bounded by $n$.
    By Novikov's criterion, $\d \Q^n/\d\P:=\mathcal E_T(\int_0^\cdot q^n\cdot\d W)$ defines $\Q^n\in\mathcal Q_{0,T}$ with Girsanov kernel $q^n$.
    Since $f(\cdot,0)=0$,
    \[
        f(u,q^n_u)-\frac{|q^n_u|^2}{2\g}=\1_{B_n}(u)\bigg(f(u,\tilde q_u)-\frac{|\tilde q_u|^2}{2\g}\bigg)\in\bigg[-\frac{n^2}{2\g},0\bigg],
    \]
    and this quantity is strictly negative on $B_n$.
    In particular, both sides of \eqref{EQ:SQ_bounded} with $\Q=\Q^n$ are finite, and
    \[
        0\leq\E_{\Q^n}\left[\int_0^T\1_{B_n}(u)\bigg(f(u,\tilde q_u)-\frac{|\tilde q_u|^2}{2\g}\bigg)\d u\right]\leq0.
    \]
    Therefore $(\Q^n\otimes\ell)(B_n)=0$ and, since $\Q^n\sim\P$, $(\P\otimes\ell)(B_n)=0$.
    Hence $N'$ is $\P\otimes\ell$-null and is contained in a null set $N\in\mathcal P$, which proves \ref{SQ:f}.
\end{proof}
 
We can now state the main result of this subsection.
 
\begin{theorem}[Quadratic BSDEs]
\label{THM:BSDE_quad}
    Let $\g>0$.
    A monetary acceptance family $\A$ in $L^\infty$ is induced by a quadratic convex driver $g$ with constant $\g$ if and only if it is regular and satisfies \itemref{IT:S-monotone}, \itemref{IT:S-convex}, \itemref{IT:S-expectation} and \itemref{IT:S-SQ} with constant $\g$.
    In this case, $\A$ also satisfies \itemref{IT:S-upperbound} and \itemref{IT:S-lowerbound}, $\r^\A=\r^g$ on $L^\infty$, and $g$ can be chosen such that $g=f^\ast$ and $f=g^\ast$ $\P\otimes\ell$-a.e., where $f$ is the function of Proposition~\ref{prop: charact-accept-f}.
\end{theorem}

\begin{proof}
\textbf{\textit{``If''.}}
    We recall from \cite[Proposition 3.6(iii)–(v)]{Delbaen+Peng+RosazzaGianin_2010_Representation_penalty_term_dynamic_concave_utilities} that $f=\inf_{n\in\N}f_n$, where
    \[
        f_n(\o,u,q):=\sup_{z\in\mathbb Q^d}\big\{q\cdot z-g_n(\o,u,z)\big\}
    \]
    for some functions $g_n:\Omega\times[0,T]\times\R^d\to\R$ such that $g_n(\cdot,z)$ is predictable for every $z\in\R^d$.
    Being a countable infimum of countable suprema of $\mathcal P\otimes\mathcal B(\R^d)$-measurable functions, $f$ is $\mathcal P\otimes\mathcal B(\R^d)$-measurable; in particular, $u\mapsto f(u,q_u)$ is predictable for every predictable $q$.
    Moreover, by \cite{Delbaen+Peng+RosazzaGianin_2010_Representation_penalty_term_dynamic_concave_utilities}, Proposition~\ref{prop: charact-accept-f} and Lemma~\ref{LEM:SQ_f}, there is a $\P\otimes\ell$-null set $N_0\in\mathcal P$ such that, for all $(\o,u)\notin N_0$ and $n\in\N$:
    \begin{enumerate}[label=(\alph*)]
        \item $g_n(\o,u,\cdot)$ is convex and $n$-Lipschitz, and $0\leq g_n(\o,u,\cdot)\leq g_{n+1}(\o,u,\cdot)$;
        \item $f(\o,u,\cdot)$ is proper, convex and lower semicontinuous, with $f(\o,u,0)=0$;
        \item $f(\o,u,q)\geq |q|^2/(2\g)$ for all $q\in\R^d$.
    \end{enumerate}
    Fix $(\o,u)\notin N_0$ and omit it from the notation.
    Since $g_n$ is continuous, $f_n=g_n^\ast$; since $g_n$ is moreover finite and convex, the Fenchel--Moreau theorem gives $f_n^\ast=g_n^{\ast\ast}=g_n$.
    Hence, for every $z\in\R^d$,
    \begin{equation}
    \label{EQ:fstar}
        f^\ast(z)
        =\sup_{q\in\R^d}\sup_{n\in\N}\big\{q\cdot z-f_n(q)\big\}
        =\sup_{n\in\N}\sup_{q\in\R^d}\big\{q\cdot z-f_n(q)\big\}
        =\sup_{n\in\N}f_n^\ast(z)
        =\sup_{n\in\N}g_n(z).
    \end{equation}
    Moreover, $f^\ast(z)\geq0\cdot z-f(0)=0$ by (b), and $f^\ast\leq g^\g$ by (c).
 
    We now define $g^\A(\o,u,z):=\1_{N^c_0}(\o,u)\sup_ng_n(\o,u,z)$, $z\in\R^d$.
    Then $g^\A$ is $\mathcal P\otimes\mathcal B(\R^d)$-measurable.
    By \eqref{EQ:fstar}, $g^\A=f^\ast$ on $N_0^c\times\R^d$; hence $g^\A(\o,u,\cdot)$ is convex and $0\leq g^\A\leq g^\g$ everywhere.
 
    Outside $N_0$, the Fenchel--Moreau theorem and (b) give $(g^\A)^\ast=f^{\ast\ast}=f$.
    Hence, by Fubini's theorem $\int_s^t(g^\A)^\ast(u,q_u)\d u=\int_s^tf(u,q_u)\d u$ $\Q$-a.s.\ for every $\Q\in\mathcal Q_{s,t}$.
    Fix $t\in[0,T]$ and $X\in L^\infty(\F_t)$.
    By Lemma~\ref{LEM:quad_BSDE} applied to $g^\A$ and by \eqref{eq: rho-A-f}, the unique solution with bounded first component of the BSDE with parameters $(g^\A,t,X)$ has first component
    \[
        Y_s
        =\esssup_{\Q\in\mathcal Q_{s,t}}\E_\Q\left[-X-\int_s^tf(u,q_u)\d u\,\bigg|\,\F_s\right]
        =\r^\A_{s,t}(X),\qquad s\in[0,t].
    \]
    Hence $\r^\A=\r^{g^\A}$ and, by Lemma~\ref{LEM:from_A_to_rho}~(a), $\A=\A^{\r^\A}=\A^{g^\A}$.
 
\textbf{\textit{``Only if'' and the last statement.}}
    By Lemma~\ref{LEM:quad_BSDE}, $\r^g$ is well defined on $L^\infty$, with $|\r^g_{s,t}(X)|\leq\|X\|_{L^\infty}$.
    By uniqueness of the solution with bounded first component and $g(\cdot,0)=0$, $\r^g$ is cash-additive, normalized, time-consistent and local, i.e., $\1_B\r^g_{s,t}(X)=\r^g_{s,t}(\1_BX)$ for $B\in\F_s$, since $\1_Bg(\cdot,z)=g(\cdot,\1_Bz)$, and it is monotone by \eqref{EQ:dual_quad}.
    Hence, by Lemma~\ref{LEM:from_rho_to_A}, $\A^g$ satisfies \itemref{IT:S-upperbound} and \itemref{IT:S-lowerbound}, $\r^{\A^g}=\r^g$, so that $\A^g=\A^{\r^{\A^g}}$, and \itemref{IT:S-normalized}, \itemref{IT:S-decomposition} and \itemref{IT:S-pasting} hold.
    By \eqref{EQ:dual_quad},
    \[
        \A^g_{s,t}=\bigcap_{\Q\in\mathcal Q_{s,t}}\bigcap_{B\in\F_s}\bigg\{X\in L^\infty(\F_t)\ :\ \E_\Q\big[\1_B(-X)\big]\leq\E_\Q\Big[\1_B\int_s^tg^\ast(u,q_u)\d u\Big]\bigg\},
    \]
    an intersection of $\sigma(L^\infty,L^1)$-closed half-spaces; this gives \itemref{IT:S-monotone}, \itemref{IT:S-convex} and \itemref{IT:S-closed}.
    Choosing $\Q=\P$ in \eqref{EQ:dual_quad} and using $g^\ast(\cdot,0)=0$, we get $\r^g_{s,t}(X)\geq\E[-X|\F_s]$, which gives \itemref{IT:S-expectation}.
    Finally, $g\leq g^\g$ implies $g^\ast(\cdot,q)\geq|q|^2/(2\g)$, so \eqref{EQ:dual_quad} for $g$ and $g^\g$ yields $\r^g_{0,T}\leq\mathfrak e^\g_{0,T}$, which gives \itemref{IT:S-SQ}.
\end{proof}

\begin{remark}[Characterization of leveled families through quadratic BSDEs]
\label{REM:quad-nonadditive}
    A natural extension of Theorem~\ref{THM:BSDE_quad} to leveled acceptance families would allow drivers $g(u,y,z)$, convex in $(y,z)$ and nonincreasing in $y$, with $|g|\leq L|y|+\frac\g2|z|^2$ for some $L\geq 0$, $\g>0$.
    The direction ``driver $\Rightarrow$ axioms'' still holds, by \cite[Assumption~(A.2) and Corollary~6]{Briand+Hu_2008_Quadratic_BSDEs_convex_generators_unbounded_terminal_conditions} and the dual representation of \cite[Theorem~7.5]{ElKaroui+Ravanelli_2009_Cash_sub-additive_risk_measures_interest_rate_ambiguity}, where the dual variables are pairs $(\b,q)$ of a discount rate $\b\in[0,L]$ and a Girsanov kernel $q$.
    The converse is more delicate.
    The proof of the ``if'' part of Theorem~\ref{THM:BSDE_quad} recovers the driver from the penalty integrand of \cite[Proposition~3.6]{Delbaen+Peng+RosazzaGianin_2010_Representation_penalty_term_dynamic_concave_utilities}, and this works because $\Q$ determines its kernel $q$.
    Here, instead, a sub-probability density $Z=e^{-\int\b\d u}\,\mathcal E(\int q\cdot\d W)$ is generated by many pairs $(\b,q)$.
    A characterization would require
    showing that the minimal penalty equals an infimum over all paths $(\b,q)$ generating $Z$, i.e., a continuous-time version of the infimal cocycle property of \cite{Ferrari+Laeven+RosazzaGianin+Zullino_2026_Time-consistency_cash-subadditive_infimal_cocycle}, based on factorizations that are simultaneously optimal at all intermediate dates $u\in[s,t]$.
\end{remark}

\section{Applications and examples}
\label{SEC:applications}

The first subsections contain the applications: the dual characterization of \itemref{IT:N-decomposition} through the infimal cocycle property (Section~\ref{SUBSEC:IC}), the resilience rate (Section~\ref{SUBSEC:resilience}), law-invariance (Section~\ref{SUBSEC:LI}), and VaR and CVaR (Section~\ref{SEC:VaR_CVaR}).
Section~\ref{SUBSEC:examples} applies the characterizations of Sections~\ref{SEC:N} and~\ref{SEC:CA} to two acceptance rules.

\subsection{Level decomposition and infimal cocycle property}
\label{SUBSEC:IC}

For $0\leq s\leq t\leq T$, let $\mathcal Z_{s,t}:=\{Z\in L^1(\F_t)\ :\ Z\geq0,\ \E[Z|\F_s]\leq1\}$ be the set of $\F_s$-conditional sub-probability densities on $\F_t$, and define the penalty of a leveled acceptance family $\A^\tinybullet$ in $L^p$ by
\[
    c^{\A^\tinybullet}_{s,t}(Z):=\esssup\big\{\E[-ZX|\F_s]-m\ :\ m\in L^p(\F_s),\ X\in\A^m_{s,t}\cap L^\infty(\F_t)\big\},
    \qquad Z\in\mathcal Z_{s,t}.
\]
For $u\in[s,t]$ and $Z\in\mathcal Z_{s,t}$, let $\Phi_u(Z):=\{(Z^1,Z^2)\in\mathcal Z_{s,u}\times\mathcal Z_{u,t}\ :\ Z=Z^1Z^2\}$.
Following \cite[Definition~3.8]{Ferrari+Laeven+RosazzaGianin+Zullino_2026_Time-consistency_cash-subadditive_infimal_cocycle}, we consider the infimal cocycle property:
\begin{itemize}[leftmargin=4em, labelwidth=3.4em, labelsep=0.6em, align=left]
    \item[(IC)] \proplabel{IC}{RM:IC}
    $\ds c^{\A^\tinybullet}_{s,t}(Z)=\essinf_{(Z^1,Z^2)\in\Phi_u(Z)}\Big\{c^{\A^\tinybullet}_{s,u}(Z^1)+\E\big[Z^1c^{\A^\tinybullet}_{u,t}(Z^2)\big|\F_s\big]\Big\}, \qquad Z\in\mathcal Z_{s,t}$.
\end{itemize}
 
\begin{corollary}
\label{COR:N-dec-IC}
    Let $p\in[1,+\infty)$ and let $\A^\tinybullet$ be a leveled acceptance family in $L^p$ satisfying \itemref{IT:L-monotone}, \itemref{IT:L-rightcont}, \itemref{IT:N-cashsub}, \itemref{IT:N-wellposed}, \itemref{IT:N-monotone}, \itemref{IT:N-convex}, \itemref{IT:N-normalized} and \itemref{IT:N-pasting}.
    Then the following are equivalent:
    \begin{multicols}{3}
    \begin{itemize}[nosep, leftmargin=*, label=\scriptsize$\bullet$]
        \item $\A^\tinybullet$ satisfies \itemref{IT:N-decomposition},
        \item $\r^{\A^\tinybullet}$ is time-consistent,
        \item $c^{\A^\tinybullet}$ satisfies \itemref{RM:IC}.
    \end{itemize}
    \end{multicols}
    \noindent In this case, the essential infimum in \itemref{RM:IC} is attained.
\end{corollary}
 
\begin{proof}
    Let $\r:=\r^{\A^\tinybullet}$.
    By Lemma~\ref{LEM:N-from-A-to-rho}, \itemref{IT:N-lowerbound} holds and $\r$ is a local, convex, cash-subadditive and normalized dynamic risk measure on $L^p$ with $\A^m_{s,t}=\A^{m,\r}_{s,t}$.
    Hence, for each $X\in L^\infty(\F_t)$, the essential supremum of $\E[-ZX|\F_s]-m$ over all $m\in L^p(\F_s)$ with $X\in\A^m_{s,t}$ is attained at $m=\r_{s,t}(X)$, so that
    \[
        c^{\A^\tinybullet}_{s,t}(Z)=\esssup_{X\in L^\infty(\F_t)}\big\{\E[-ZX|\F_s]-\r_{s,t}(X)\big\},
        \qquad Z\in\mathcal Z_{s,t},
    \]
    which is the penalty of $\r$ defined in \cite[Section~3.1]{Ferrari+Laeven+RosazzaGianin+Zullino_2026_Time-consistency_cash-subadditive_infimal_cocycle}.
    By Proposition~\ref{prop: suff cond-A3}, \itemref{IT:N-decomposition} holds if and only if $\r$ is time-consistent.
    The rest follows from \cite[Theorem~3.14(b)]{Ferrari+Laeven+RosazzaGianin+Zullino_2026_Time-consistency_cash-subadditive_infimal_cocycle}.
\end{proof}

\subsection{Resilience}
\label{SUBSEC:resilience}

The resilience rate of \cite{Laeven+Ferrari+Rosazza+Zullino_2025_measuringfinancialresilienceusing},
\[
    \bm\dot\r_t(X):=\lim_{\eps\downarrow0}\frac1\eps\,\E\big[\r_{t+\eps,T}(X)-\r_{t,T}(X)\big],\qquad t\in[0,T),
\]
measures how fast the risk of a position is expected to recover over time.
Its existence is established in \cite{Laeven+Ferrari+Rosazza+Zullino_2025_measuringfinancialresilienceusing} for risk measures induced by BSDEs, under conditions on the driver.
Our characterizations turn these into conditions on the acceptance family, which can be checked on the set of acceptable positions, without knowing the driver or even the risk measure.
Namely, if $\A^\tinybullet$ satisfies the axioms of Theorem~\ref{THM:Lp-nonadditive}~(i), or $\A$ those of Theorem~\ref{THM:Lp-additive}~(i) or Theorem~\ref{THM:BSDE_quad}, then $\bm\dot\r_t(X)$ exists and equals $-\E[g(t,\r_{t,T}(X),Z_t)]$ for almost every $t$, where $g$ is the driver recovered from the acceptance family and $(\r_{\cdot,T}(X),Z)$ is the solution of the BSDE with parameters $(g,T,X)$, by \cite[Theorem~11~(i)]{Laeven+Ferrari+Rosazza+Zullino_2025_measuringfinancialresilienceusing}.
The sign of the rate can be read off the acceptance family: if, in addition, \itemref{IT:N-expectation} (resp.\ \itemref{IT:S-expectation}) holds, then $g\geq0$ by Theorem~\ref{THM:Lp-nonadditive}~(iii) (resp.\ Theorem~\ref{THM:Lp-additive}~(iii)), so that $\bm\dot\r_t(X)\leq0$, i.e., the position is resilient or resilience neutral in the sense of \cite{Laeven+Ferrari+Rosazza+Zullino_2025_measuringfinancialresilienceusing}.
In the same way, the resilience evaluation of \cite{Ferrari+Laeven+Rosazza+Zullino_2026_financialresilienceevaluation}, which replaces the expectation by a dynamic risk measure, is well defined by \cite[Theorem~3.11]{Ferrari+Laeven+Rosazza+Zullino_2026_financialresilienceevaluation}, under the conditions stated there, for the convex families of Theorem~\ref{THM:Lp-additive} and for those of Theorem~\ref{THM:BSDE_quad}.

For a level $a\in\R$ and a time $t\in[0,T)$, the \emph{resilience-acceptance set} of \cite[Definition~27]{Laeven+Ferrari+Rosazza+Zullino_2025_measuringfinancialresilienceusing} is, with our sign convention for positions,
\[
    \mathcal R^a_t:=\big\{X\in L^p(\F_T)\ :\ \bm\dot\r_t(X)\leq a\big\},
\]
the set of positions whose risk is expected to increase at rate at most $a$ right after time $t$.
It plays for the rate the role that the acceptance set $\A^m_{t,T}$ (which is $\A_{t,T}-m$ in the monetary case) plays for the level: under the axioms above, $\r_{t,T}(X)=\min\{m:X\in\A^m_{t,T}\}$, and $\bm\dot\r_t(X)=\min\{a:X\in\mathcal R^a_t\}$ whenever the rate exists.
Both families are described by the driver $g$ recovered from the acceptance family: if $(Y,Z)$ is the solution of the BSDE with parameters $(g,T,X)$, then
\[
    X\in\A^m_{t,T}\ \Longleftrightarrow\ Y_t\leq m,
    \qquad\qquad
    X\in\mathcal R^a_t\ \Longleftrightarrow\ \E\big[g(t,Y_t,Z_t)\big]\geq-a,
\]
the second equivalence holding for almost every $t$; cf.\ \cite[Proposition~28]{Laeven+Ferrari+Rosazza+Zullino_2025_measuringfinancialresilienceusing}.
Thus acceptability is a constraint on the level $Y_t$ of the solution, and resilience-acceptability on the expected driver along it.

Structural properties of the resilience-acceptance sets can also be read off the acceptance family; for $a\in\R$, $\l>0$ and $0\leq t<u<T$:
\begin{itemize}[nosep, leftmargin=*, label=\scriptsize$\bullet$]
    \item \itemref{IT:N-expectation} (resp.\ \itemref{IT:S-expectation}) $\Rightarrow$ every $X\in L^p(\F_T)$ belongs to $\mathcal R^0_t$ for almost every $t$;
    \item \itemref{IT:N-conic} (resp.\ \itemref{IT:S-conic}) $\Rightarrow$ $\l\mathcal R^a_t=\mathcal R^{\l a}_t$;
    \item $\A$ monetary $\Rightarrow$ $\mathcal R^a_t+L^p(\F_t)=\mathcal R^a_t$ and $\mathcal R^a_t\subseteq\big(\mathcal R^a_t\cap L^p(\F_u)\big)+\A_{u,T}$.
\end{itemize}
The first item follows from $g\geq0$.
The others follow from positive homogeneity, cash-additivity and time-consistency of $\r$, as in \cite[Propositions~19, 29 and~31]{Laeven+Ferrari+Rosazza+Zullino_2025_measuringfinancialresilienceusing}: in the monetary case, $\r_{r,T}(X)=\r_{r,T}\big(-\r_{u,T}(X)\big)$ for $r\leq u$, so that $X\in\mathcal R^a_t$ implies $-\r_{u,T}(X)\in\mathcal R^a_t\cap L^p(\F_u)$, while $X+\r_{u,T}(X)\in\A_{u,T}$.
The last inclusion is the counterpart of the decomposition \itemref{IT:S-decomposition}: a resilience-acceptable position is the sum of an $\F_u$-measurable resilience-acceptable position and of a position that is acceptable over $[u,T]$.

\subsection{Law-invariant acceptance families}
\label{SUBSEC:LI}

Law-invariant $g$-expectations were introduced in \cite{Xu+Xu+Zhou_2022_g-Expectation_distributions}; they were later characterized in \cite{Bensaid+Dumitrescu+Matoussi+Sabbagh_2026_Law-invariant_BSDEs_dynamic_risk_measures}, which also extends to continuous time the result of \cite{Kupper+Schachermayer_2009_Representation_results_law_invariant_time_consistent_functions} that law-invariant time-consistent risk measures are entropic.
We apply the results of \cite{Bensaid+Dumitrescu+Matoussi+Sabbagh_2026_Law-invariant_BSDEs_dynamic_risk_measures} to characterize law-invariance in terms of acceptance families.
For $0\leq s\leq t\leq T$ and $X,Y\in L^0(\F_t)$, we write $\mathrm{Law}(X|\F_s)=\mathrm{Law}(Y|\F_s)$ if $\P(X\leq x\,|\,\F_s)=\P(Y\leq x\,|\,\F_s)$, $\P$-a.s., for rational $x$;
equivalently, $\E[\varphi(X)|\F_s]=\E[\varphi(Y)|\F_s]$ $\P$-a.s.\ for every bounded Borel $\varphi:\R\to\R$; 
see \cite[Remark~3.1]{Bensaid+Dumitrescu+Matoussi+Sabbagh_2026_Law-invariant_BSDEs_dynamic_risk_measures}.
For $s=0$, this reduces to $X$ and $Y$ having the same distribution.
A set $\mathcal C\subseteq L^p(\F_t)$ is \emph{$\F_s$-conditionally law-invariant} if $X\in\mathcal C$, $Y\in L^p(\F_t)$ and $\mathrm{Law}(X|\F_s)=\mathrm{Law}(Y|\F_s)$ imply $Y\in\mathcal C$; for $s=0$ we simply say that $\mathcal C$ is \emph{law-invariant}.
 
Since $\F_0$ is trivial, 
if $\A_{0,T}$ is law-invariant, so is $\r^\A_{0,T}$, i.e., $\r^\A_{0,T}(X)=\r^\A_{0,T}(Y)$ whenever $X$ and $Y$ are identically distributed; the converse holds whenever $\A_{0,T}=\A^{\r^\A}_{0,T}$ (see Lemma~\ref{LEM:from_A_to_rho}~(a)).
Law-invariance of $\r_{0,T}$ is the property (LI) of \cite[Definition~3.1]{Bensaid+Dumitrescu+Matoussi+Sabbagh_2026_Law-invariant_BSDEs_dynamic_risk_measures}.
The following corollary of Theorems~\ref{THM:Lp-additive},~\ref{THM:BSDE_quad}, and \cite[Theorem~3.2]{Bensaid+Dumitrescu+Matoussi+Sabbagh_2026_Law-invariant_BSDEs_dynamic_risk_measures} shows that, in our BSDE setting, law-invariance only occurs in two explicit cases: the conditional expectation of Example~\ref{EX:P-AE} and the entropic family.

\begin{corollary}
\label{COR:LI}
    \begin{enumerate}[label=(\roman*)]
        \item
            Let $p\in(1,+\infty]$, and let $\A$ be a regular monetary acceptance family in $L^p$ satisfying \itemref{IT:S-SL}.
            Then the following are equivalent:
            \begin{enumerate}[label=(\alph*), nosep]
                \item $\A_{0,T}$ is law-invariant;
                \item $\A_{s,t}$ is $\F_s$-conditionally law-invariant for all $0\leq s\leq t\leq T$;
                \item $\A=\A^{\E}$.
            \end{enumerate}
        \item
            Let $\A$ be a monetary acceptance family in $L^\infty$ satisfying the axioms of Theorem~\ref{THM:BSDE_quad} with constant $\g$.
            Then (a) and (b) are equivalent to
            \begin{enumerate}[label=(c$'$), nosep]
                \item either $\A=\A^{\E}$, or $\A=\A^{\g'}$ for some $\g'\in(0,\g]$.
            \end{enumerate}
    \end{enumerate}
\end{corollary}
\begin{proof}
    In both cases, (c)/(c$'$)$\Rightarrow$(b) holds because $\E[X|\F_s]$ and $\E[e^{-\g'X}|\F_s]$ depend on $X$ only through $\mathrm{Law}(X|\F_s)$, and (b)$\Rightarrow$(a) is the case $s=0$.

    (a)$\Rightarrow$(c)/(c$'$).
    By Theorem~\ref{THM:Lp-additive} (resp.\ Theorem~\ref{THM:BSDE_quad}), $\r^\A=\r^g$ for a $\mu$-Lipschitz monetary driver (resp.\ a quadratic convex driver with constant $\g$) $g$, which satisfies $g(\cdot,0)=0$.
    By (a), $\r^g_{0,T}=\r^\A_{0,T}$ is law-invariant on $L^\infty(\F_T)$.
    We apply \cite[Theorem~3.2]{Bensaid+Dumitrescu+Matoussi+Sabbagh_2026_Law-invariant_BSDEs_dynamic_risk_measures} in the form of \cite[Remark~3.2]{Bensaid+Dumitrescu+Matoussi+Sabbagh_2026_Law-invariant_BSDEs_dynamic_risk_measures}, which only requires the growth and continuity conditions (A1)--(A2) of \cite{Bensaid+Dumitrescu+Matoussi+Sabbagh_2026_Law-invariant_BSDEs_dynamic_risk_measures} together with strict monotonicity, the local property and constant preservation.
    Conditions (A1)--(A2) are immediate in the Lipschitz case; in the quadratic case they follow from $0\leq g\leq\frac\g2|z|^2$ and convexity, which give $|g(u,z)-g(u,z')|\leq\frac\g2(R+1)^2|z-z'|$ for $|z|,|z'|\leq R$, uniformly in $(\o,u)$.
    The local property follows from \itemref{IT:S-pasting} by Lemma~\ref{LEM:from_A_to_rho}~(b), and constant preservation follows from cash-additivity and normalization.
    Strict monotonicity follows from the strict comparison theorem for Lipschitz BSDEs \cite[Theorem~2.2]{ElKaroui+Peng+Quenez_1997_Backward_stochastic_differential_equations_finance} in case (i); in case (ii), if $X\geq X'$ and $\Q\in\mathcal Q_{s,t}$ attains the essential supremum in \eqref{EQ:dual_quad} for $X$ (Lemma~\ref{LEM:quad_BSDE}), then $\r^g_{s,t}(X')\geq\E_\Q\big[-X'-\int_s^tg^\ast(u,q_u)\d u\,\big|\,\F_s\big]=\r^g_{s,t}(X)+\E_\Q[X-X'|\F_s]$, and $\E_\Q[X-X'|\F_s]\geq 0$ is not identically zero unless $X=X'$, since $\Q\sim\P$.
    
    Hence $g=k|z|^2$ $\P\otimes\ell$-a.e.\ for a constant $k\in\R$, since $g$ does not depend on $y$.
    In case (i), the Lipschitz continuity of $g$ forces $k=0$; in case (ii), $0\leq g\leq\frac\g2|z|^2$ gives $0\leq k\leq\g/2$.
    If $k=0$, then $\r^g_{s,t}(X)=\E[-X|\F_s]$, hence $\A^g=\A^\E$; if $k>0$, then $\r^g=\mathfrak e^{2k}$, hence $\A^g=\A^{2k}$.
    Since $\A=\A^g$ by Theorem~\ref{THM:Lp-additive}~(i) (resp.\ Theorem~\ref{THM:BSDE_quad}), (c) (resp.\ (c$'$) with $\g':=2k$) follows.
\end{proof}
 
\begin{remark}
\label{REM:LI}
    Corollary~\ref{COR:LI}~(ii) is a continuous-time version of the result of \cite{Kupper+Schachermayer_2009_Representation_results_law_invariant_time_consistent_functions}, formulated in terms of acceptance sets.
    Compared with \cite[Theorem~4.1]{Bensaid+Dumitrescu+Matoussi+Sabbagh_2026_Law-invariant_BSDEs_dynamic_risk_measures}, the dynamic risk measure is not given a priori but defined from $\A$, the condition $\a^m_{0,T}(\P)=0$ is replaced by \itemref{IT:S-expectation} (cf.\ Lemma~\ref{LEM:penalty}), the quadratic growth of the driver follows from \itemref{IT:S-SQ} by Lemma~\ref{LEM:SQ_f}, and neither strict monotonicity nor attainment of the dual representation needs to be assumed.
\end{remark}
 
\subsection{VaR and CVaR}
\label{SEC:VaR_CVaR}
 
As a complement to Section~\ref{SEC:CA}, we determine which axioms are satisfied by the acceptance sets of VaR and CVaR, and we then ask whether these risk measures admit a time-consistent majorant.

For $\a\in(0,1)$ and
$0\leq s \leq t \leq T$, the dynamic versions of $\mathrm{VaR}$ and $\mathrm{CVaR}$ are defined as follows, see \cite[Example~11.7, Definition~11.8]{Follmer+Schied_2016_Stochastic_finance}:
\begin{align}
    \mathrm{VaR}_{s,t}^{\alpha}(X)
    &= \essinf\{m_s \in L^{\infty}(\F_s)\ :\  \P(m_s+X <0 | \F_s) \leq \alpha\},
    \quad X \in L^{\infty}(\F_t),\\
    \mathrm{CVaR}_{s,t}^{\alpha}(X)
    &= \esssup_{\Q \in \mathcal Q_{s,t}^{\alpha}} \E_\Q[-X \vert \mathcal{F}_s],
    \quad X \in L^{\infty}(\F_t),
\end{align}
where $\mathcal{Q}_{s,t}^{\alpha}:=\big\{\Q\in\mathcal Q^{\mathrm{ac}}_{s,t}\ :\ \d \Q/\d\P\leq1/\a\big\}$.
The corresponding acceptance sets are
\begin{align}
\label{eq: A-var-dyn}
    \mathcal{A}^{\mathrm{VaR}^{\alpha}}_{s,t}
    &= \left\{X \in L^{\infty}(\F_t): \P(X <0 \vert \F_s) \leq \alpha \right\},\\
    \mathcal{A}^{\mathrm{CVaR}^{\alpha}}_{s,t}
    &= \left\{X \in L^{\infty}(\F_t): \E_\Q[-X \vert \F_s] \leq 0 \mbox{ for any } \Q \in \mathcal Q^{\alpha}_{s,t} \right\}.
\label{eq: A-cvar-dyn}
\end{align}
 
We now collect the properties satisfied or violated by these acceptance sets.
\begin{proposition}
\label{CVaR}
Let $\a\in(0,1)$.
\begin{enumerate}[label=(\roman*)]
    \item The family $\mathcal{A}^{\mathrm{VaR}^{\alpha}}_{s,t}$ satisfies \itemref{IT:S-monotone}, \itemref{IT:S-conic}, \itemref{IT:S-normalized} and \itemref{IT:S-pasting}, while it violates \itemref{IT:S-convex}, \itemref{IT:S-closed}, \itemref{IT:S-decomposition}, and \itemref{IT:S-expectation}.
 
    \item The family $\mathcal{A}^{\mathrm{CVaR}^{\alpha}}_{s,t}$ satisfies \itemref{IT:S-monotone}, \itemref{IT:S-convex}, \itemref{IT:S-conic}, \itemref{IT:S-normalized}, \itemref{IT:S-closed}, \itemref{IT:S-pasting} and \itemref{IT:S-expectation}, while it violates \itemref{IT:S-decomposition}.
\end{enumerate}
\end{proposition}
 
\begin{proof}
    Both $\mathrm{VaR}^\a$ and $\mathrm{CVaR}^\a$ are local cash-additive dynamic risk measures on $L^\infty$ that are not time-consistent; see \cite[Example~3.1 and Remark~3.2]{Cheridito+Stadje_2009_Time-inconsistency_VaR_time-consistent_alternatives} for $\mathrm{VaR}^\a$, and \cite[Theorem~1.10 and Remark~1.13]{Kupper+Schachermayer_2009_Representation_results_law_invariant_time_consistent_functions} and \cite[Theorems~9 and~12]{Delbaen_2006_Structure_m-stable_sets_particular_set_risk_neutral_measures} for $\mathrm{CVaR}^\a$.
    Hence \itemref{IT:S-decomposition} fails for both families by Lemma~\ref{LEM:from_rho_to_A}.
    Throughout, $B\in\F_s$, $\l>0$, $X,Y$ are $\F_t$-measurable and bounded, and $m\in L^\infty(\F_s)$.
    \smallskip

\noindent\textbf{\textit{(i)}}
    Since $\P(\l X<0|\F_s)=\P(X<0|\F_s)$, $\P(m<0|\F_s)=\1_{\{m<0\}}$ and
    $\P(\1_BX+\1_{B^c}Y<0|\F_s)=\1_B\P(X<0|\F_s)+\1_{B^c}\P(Y<0|\F_s)$, the axioms \itemref{IT:S-monotone}, \itemref{IT:S-conic}, \itemref{IT:S-normalized} and \itemref{IT:S-pasting} follow from \eqref{eq: A-var-dyn}.

    For the violations, let $t\in(0,T]$ and $s=0$.
    Take disjoint $C,D\in\F_t$ with $\P(C)=\a$ and $0<\P(D)\leq\a\wedge(1-\a)$.
    Then $-\1_C,-\1_D\in\A^{\mathrm{VaR}^\a}_{0,t}$, whereas $\l(-\1_C)+(1-\l)(-\1_D)<0$ on $C\cup D$ for $\l\in(0,1)$, and $\P(C\cup D)>\a$; hence \itemref{IT:S-convex} fails.
    For $t=T$, $-\1_C\in\A^{\mathrm{VaR}^\a}_{0,T}$ but $\E[-\1_C]=-\a<0$, so \itemref{IT:S-expectation} fails.
    For \itemref{IT:S-closed}, take independent events $B_n\in\F_t$ with $\P(B_n)=\a$, e.g., built from Brownian increments on disjoint subintervals of $(0,t]$.
    Since $(\1_{B_n}-\a)_n$ is orthogonal and bounded in $L^2$, it converges weakly to $0$ in $L^2$ by Bessel's inequality, hence $\1_{B_n}\to\a$ in $\sigma(L^\infty,L^1)$ by uniform boundedness and density of $L^2$ in $L^1$.
    Thus $-\1_{B_n}\in\A^{\mathrm{VaR}^\a}_{0,t}$ converges to $-\a\notin\A^{\mathrm{VaR}^\a}_{0,t}$.
    \smallskip

\noindent\textbf{\textit{(ii)}}
    By \eqref{eq: A-cvar-dyn}, $\A^{\mathrm{CVaR}^\a}_{s,t}$ is the intersection over $\Q\in\mathcal Q^\a_{s,t}$ of the sets
    $\{X\ :\ \E_\Q[-X|\F_s]\leq0\}=\bigcap_{D\in\F_s}\{X\ :\ \E_\Q[-X\1_D]\leq0\}$.
    Each of them is a monotone, convex, $\sigma(L^\infty,L^1)$-closed cone, is stable under pasting with $B\in\F_s$, and contains $m$ if and only if $m\geq0$.
    Hence \itemref{IT:S-monotone}, \itemref{IT:S-convex}, \itemref{IT:S-conic}, \itemref{IT:S-closed}, \itemref{IT:S-pasting} and \itemref{IT:S-normalized} hold, and \itemref{IT:S-expectation} follows from $\P\in\mathcal Q^\a_{s,T}$.
\end{proof}

In particular, \itemref{IT:S-decomposition} fails for both families, so that neither $\mathrm{VaR}$ nor $\mathrm{CVaR}$ is induced by a BSDE, as is well known.
It is therefore natural to ask whether they admit a time-consistent majorant.
The next result shows that the only such majorant is the trivial one, namely the worst-case risk measure.
 
\begin{theorem}
\label{TH:VaR_majorant}
    Let $\a\in(0,1)$.
    If $\r$ is a normalized, time-consistent, cash-additive dynamic risk measure on $L^\infty$ such that
    \begin{equation}
    \label{EQ:VaR_dom}
        \r_{s,T}(X)\geq\mathrm{VaR}^\a_{s,T}(X),\qquad s\in[0,T],\ X\in L^\infty(\F_T),
    \end{equation}
    then $\r$ is the worst-case risk measure, i.e., for $0\leq s\leq t\leq T$
    \begin{equation}
    \label{EQ:worst}
        \r_{s,t}(X)=\esssup(-X|\F_s):= \essinf\big\{Y\in L^\infty(\F_s)\ :\ Y\geq -X\big\},\qquad X\in L^\infty(\F_t).
    \end{equation}
    In particular, this holds if $\mathrm{VaR}^\a$ is replaced by $\mathrm{CVaR}^\a$ in \eqref{EQ:VaR_dom}.
\end{theorem}

\begin{remark}[Idea of the proof]
\label{REM:VaR_majorant_idea}
    The worst-case risk measure is characterized by the fact that a loss of size $b>0$ on any non-null event $A$ is charged in full, however small $\P(A)$ is; 
    it therefore suffices to show that $\r_{0,T}(-b\1_A)=b$ for every $A\in\F_T$ with $\P(A)>0$. 
    Let $M_u:=\P(A|\F_u)$.
    The proof propagates this ``full charge'' from times at which $A$ is 
    likely enough to trigger $\mathrm{VaR}^\a$  back to earlier times at which $A$ is less likely.
    This is done by induction on a threshold $x\in(0,1]$ for $M$, which is lowered at each step until it falls below $\P(A)=M_0$.
    The claim at threshold $x$ is that $\r_{u,T}(-b\1_A)=b$ on $\{M_u\geq x\}$ at every time $u\in[0,T]$:
    \begin{itemize}[nosep, label=\scriptsize$\bullet$]
        \item \emph{Base case.} 
            By \eqref{EQ:VaR_dom}, $\r_{u,T}(-b\1_A)=b$ wherever $M_u>\a$, since $\mathrm{VaR}^\a$ ignores no event of conditional probability above $\a$; hence the claim holds for any starting threshold $x_0\in(\a,1)$.
        \item \emph{Inductive step.} 
            Assume the claim at threshold $x$, fix a date $s$, and let $G\supseteq A$ be the event that $M$ reaches $x$ before $T$.
            If $G$ occurs, the loss is already charged in full at the first date at which $M$ reaches $x$, so that, by time-consistency, it suffices that $\mathrm{VaR}^\a$ sees the larger event $G$ at date $s$.
            Since $M$ is a continuous martingale, it reaches $x$ with conditional probability $M_s/x$, which exceeds $\a$ wherever $M_s\geq\b x$, for a fixed $\b\in(\a,1)$.
            Hence the claim holds at the lower threshold $\b x$.
        \item \emph{Conclusion.} 
            The thresholds $\b^kx_0$ go to $0$, so the claim holds at every threshold $x>0$, i.e., $\r_{u,T}(-b\1_A)=b$ wherever $M_u>0$; in particular, $\r_{0,T}(-b\1_A)=b$.
    \end{itemize}
    Each application of $\mathrm{VaR}^\a$ only covers events of probability above $\a$, but time-consistency lets it be applied again at every intermediate date, and continuity of the martingales ensures that the event ``$M$ reaches the previous threshold'' still has probability above $\a$.
\end{remark}
 
\begin{proof}[\textbf{Proof of Theorem~\ref{TH:VaR_majorant}}]
    The last statement follows from $\mathrm{CVaR}^\a_{s,T}\geq\mathrm{VaR}^\a_{s,T}$; see \cite[Proposition~11.9]{Follmer+Schied_2016_Stochastic_finance}.
 
    Throughout, $0\leq\r_{u,T}(-b\1_B)\leq b$ for all $b>0$, $B\in\F_T$ and $u\in[0,T]$, by monotonicity, normalization and cash-additivity.
    Moreover, for $B\in\F_T$, $b>0$ and $s\in[0,T]$,
    \begin{equation}
    \label{EQ:VaR_indicator}
        \mathrm{VaR}^\a_{s,T}(-b\1_B)\geq b\quad\text{a.s.\ on }\{\P(B|\F_s)>\a\}.
    \end{equation}
    Indeed, let $m\in L^\infty(\F_s)$ be such that $\P(m-b\1_B<0|\F_s)\leq\a$.
    Since $\{m<b\}\in\F_s$ and $\{m<b\}\cap B\subseteq\{m-b\1_B<0\}$, we get $\a\geq\P(m-b\1_B<0|\F_s)\geq\1_{\{m<b\}}\P(B|\F_s)$.
    Hence $m\geq b$ a.s.\ on $\{\P(B|\F_s)>\a\}$, and taking the essential infimum over all such $m$ yields \eqref{EQ:VaR_indicator}.\par
    \smallskip
 
    \emph{Step 1: reduction.}
    By cash-additivity and normalization, $\r_{t,T}(X)=-X$ for $X\in L^\infty(\F_t)$; hence, by time-consistency and monotonicity, $\r_{s,t}(X)\leq0$ implies $\r_{0,T}(X)=\r_{0,t}(X)=\r_{0,s}(-\r_{s,t}(X))\leq\r_{0,s}(0)=0$, so $L^\infty_+(\F_t)\subseteq\A^\r_{s,t}\subseteq\A^\r_{0,T}$.
    Hence it suffices to show $\A^\r_{0,T}\subseteq L^\infty_+(\F_T)$: then $\A^\r_{s,t}=L^\infty_+(\F_t)$, and \eqref{EQ:worst} holds by \eqref{eq: rho given A},  
    cf.\ \cite[Example~4.39]{Follmer+Schied_2016_Stochastic_finance}.
    Let $X\in\A^\r_{0,T}$ and suppose, by contradiction, that $A:=\{X<-\eps\}$ has positive probability for some $\eps>0$.
    With $K:=\|X\|_{L^\infty}$, we have $X\leq K-(K+\eps)\1_A$, so monotonicity and cash-additivity yield $0\geq\r_{0,T}(X)\geq\r_{0,T}\big(-(K+\eps)\1_A\big)-K$.
    A contradiction is therefore obtained once we show that
    \begin{equation}
    \label{EQ:indicator_wc}
        \r_{0,T}(-b\1_A)=b,\qquad b>0,\ A\in\F_T\text{ with }\P(A)>0.
    \end{equation}
 
    \emph{Step 2: induction.}
    Fix $b>0$ and $A\in\F_T$ with $\P(A)>0$.
    Set $R_u:=\r_{u,T}(-b\1_A)$ and $M_u:=\P(A|\F_u)$, $u\in[0,T]$, where $M$ is chosen continuous, so that $M_T=\1_A$.
    For $x\in(0,1]$, consider the statement
    \begin{itemize}
        \item[($L_x$)] for every $s\in[0,T]$, $R_s=b$ on $\{M_s\geq x\}$.
    \end{itemize}
    If we manage to prove that ($L_x$) holds for all $x\in(0,1]$, then ($L_{\P(A)}$) with $s=0$ gives \eqref{EQ:indicator_wc}.
    By \eqref{EQ:VaR_dom} and \eqref{EQ:VaR_indicator}, $R_s\geq b$ on $\{M_s>\a\}$, hence ($L_x$) holds for all $x\in(\a,1]$.
    Set $\b:=(1+\a)/2\in(\a,1)$.
    In Step 3, we show that ($L_x$) implies ($L_{\b x}$) for every $x\in(0,1)$.
    Fixing $x_0\in(\a,1)$, induction then yields ($L_{\b^kx_0}$) for all $k\in\N$.
    Since ($L_z$) implies ($L_x$) for every $x\in[z,1]$, this yields ($L_x$) for all $x\in(0,1]$.
    \smallskip
 
    \emph{Step 3: ($L_x$)$\,\Rightarrow$($L_{\b x}$).}
    Fix $x\in(0,1)$ such that ($L_x$) holds, set $x':=\b x$, and fix $s\in[0,T)$; the case $s=T$ is trivial because $R_T=b\1_A$ and $\{M_T\geq x'\}=A$.
    For $n\in\N$, let $\pi_n:=\{s+k(T-s)2^{-n}\ :\ k=1,\dots,2^n\}$, define the stopping time $\t_n:=\min\{u\in\pi_n\ :\ M_u\geq x\}$, with $\min\varnothing:=T$, and set $G_n:=\{M_{\t_n}\geq x\}=\{\max_{u\in\pi_n}M_u\geq x\}$.
    For $u\in\pi_n$, we have $\{\t_n=u\}\cap G_n\subseteq\{M_u\geq x\}$, where $R_u=b$ by ($L_x$); since $R\geq0$, it follows that $R_{\t_n}=\r_{\t_n,T}(-b\1_A)\geq b\1_{G_n}$, with the notation \eqref{EQ:rho_theta}.
    Lemma~\ref{LEM:optional_sampling}, monotonicity, \eqref{EQ:VaR_dom} and \eqref{EQ:VaR_indicator} then yield
    \begin{equation}
    \label{EQ:Rs_geq_b}
        R_s=\r_{s,T}(-R_{\t_n})\geq\r_{s,T}(-b\1_{G_n})\geq\mathrm{VaR}^\a_{s,T}(-b\1_{G_n})\geq b\quad\text{a.s.\ on }\big\{\P(G_n|\F_s)>\a\big\}.
    \end{equation}
    The sets $G_n$ increase to $G_\infty:=\bigcup_nG_n$ and, by continuity of $M$ and density of $\bigcup_n\pi_n$ in $[s,T]$, $G_\infty\supseteq\{\sup_{u\in[s,T]}M_u>x\}$.
    Hence $\P(G_n|\F_s)\uparrow\P(G_\infty|\F_s)\geq\P(\sup_{u\in[s,T]}M_u>x|\F_s)$ a.s.
    We claim that
    \begin{equation}
    \label{EQ:upcrossing}
        \P\Big(\sup_{u\in[s,T]}M_u>x\,\Big|\,\F_s\Big)>\a\quad\text{a.s.\ on }\{M_s\geq x'\}.
    \end{equation}
    Given \eqref{EQ:upcrossing}, a.s.\ on $\{M_s\geq x'\}$ we have $\P(G_n|\F_s)>\a$ for $n$ large enough, so that $R_s\geq b$ by \eqref{EQ:Rs_geq_b}; together with $R_s\leq b$, this proves ($L_{x'}$).
 
    To prove \eqref{EQ:upcrossing}, note first that, by continuity, $\{M_s>x\}\subseteq\{\sup_{u\in[s,T]}M_u>x\}$, so the conditional probability in \eqref{EQ:upcrossing} equals $1$ on $\{M_s>x\}$.
    Next, since $x'/\a=\b x/\a>x$, we can choose $y\in(x,1\wedge(x'/\a))$.
    Let $\s:=\inf\{u\geq s\ :\ M_u\geq y\}\wedge T$ and $H:=\{\sup_{u\in[s,T]}M_u\geq y\}$.
    On $\{M_s<y\}$, continuity gives $M_\s=y$ on $H$, while $M_\s=M_T=\1_A=0$ on $H^c$.
    By optional sampling, $M_s=\E[M_\s|\F_s]=y\,\P(H|\F_s)$ on $\{M_s<y\}\in\F_s$.
    Since $H\subseteq\{\sup_{u\in[s,T]}M_u>x\}$, we obtain on $\{x'\leq M_s\leq x\}$
    \[
        \P\Big(\sup_{u\in[s,T]}M_u>x\,\Big|\,\F_s\Big)\geq\P(H|\F_s)=\frac{M_s}{y}\geq\frac{x'}{y}>\a.
    \]
    This proves \eqref{EQ:upcrossing} and concludes the proof.
\end{proof}

Note that the proof only uses that $\F_0$ is trivial and that all $\mathbb F$-martingales are continuous, as in \cite[Theorem~9]{Delbaen_2006_Structure_m-stable_sets_particular_set_risk_neutral_measures}.
    
\begin{remark}
\label{REM:VaR_majorant_lit}
    In discrete time, composing the one-step conditional versions of $\mathrm{VaR}^\a$ or $\mathrm{CVaR}^\a$ produces time-consistent risk measures that differ from the worst-case risk measure; see \cite[Exercise~11.2.3]{Follmer+Schied_2016_Stochastic_finance}, \cite[Sections~4 and~5]{Cheridito+Stadje_2009_Time-inconsistency_VaR_time-consistent_alternatives} and \cite[Section~2.3.1]{Cheridito+Kupper_2011_Composition_time-consistent_dynamic_monetary_risk_measures_discrete_time}.
    Theorem~\ref{TH:VaR_majorant} shows that this cannot happen when all martingales are continuous: with jumps, $M$ may skip the threshold, and the mechanism of Remark~\ref{REM:VaR_majorant_idea} fails.
    A nontrivial continuous-time limit of these compositions, given by a BSDE-induced risk measure, is obtained in \cite{Stadje_2010_Extending_dynamic_convex_risk_measures_discrete_continuous_time} by rescaling the level $\a$ with the time step.
\end{remark}

Since the worst-case risk measure \eqref{EQ:worst} is not induced by a BSDE (for $A:=\{z\cdot W_T>0\}$ with $z\neq0$, $\r_{t,T}(-\1_A)$ equals $1$ for $t<T$ and $\1_A$ for $t=T$, so it has no continuous modification), $\mathrm{VaR}^\a$ and $\mathrm{CVaR}^\a$ are not even dominated, in the sense of \eqref{EQ:VaR_dom}, by a normalized cash-additive BSDE-induced risk measure.

\subsection{From acceptance rules to BSDEs}
\label{SUBSEC:examples}

We now provide two examples that show how the characterizations are used: an acceptance rule is described only through the positions it accepts, with no risk measure given a priori, the axioms are checked on the sets, and Theorem~\ref{THM:Lp-nonadditive} and Corollary~\ref{COR:no-closedness} show that the risk measure it defines is induced by a BSDE.
In particular, it is time-consistent, and its properties are read off the driver.
When the driver can be computed, it reveals features of the rule that are not visible in its definition.

\begin{example}[Stress tests with an interest-rate corridor]
\label{EX:corridor}
    When capital is posted at time $s$ and the position is settled at time $t$, acceptability depends on the interest rate between the two dates, and cash-additivity is lost if this rate is uncertain \cite{ElKaroui+Ravanelli_2009_Cash_sub-additive_risk_measures_interest_rate_ambiguity}.
    A supervisor who does not want to commit to a model may prescribe a stress test: the capital must cover the expected discounted loss in every scenario in which the short rate stays in a corridor and the probability is not too far from $\P$.
    The rule does not say which rate is applied when.
 
    \smallskip
    \noindent\emph{The rule.}
    Let $p\in(1,+\infty)$, $\kappa>0$ and $r\leq R$ in $\R$.
    A \emph{scenario} is a pair $(\beta,q):\Omega\times[0,T]\to\R\times\R^d$ of predictable processes: the short rate $\beta$, with values in $[r,R]$, and the Girsanov kernel $q$, with $|q|\leq\kappa$.
    We denote by $\mathcal S$ the set of scenarios and, for $0\leq s\leq t\leq T$, we let
    \[
        \Gamma^{\beta,q}_{s,t}:=\exp\Big(-\int_s^t\beta_u\d u+\int_s^tq_u\cdot\d W_u-\frac12\int_s^t|q_u|^2\d u\Big),
    \]
    so that $\E[\Gamma^{\beta,q}_{s,t}X|\F_s]=\E_\Q\big[e^{-\int_s^t\beta_u\d u}X\big|\F_s\big]$ is the expected value of $X$, discounted at the rate $\beta$, under the model $\Q\in\mathcal Q_{s,t}$ with Girsanov kernel $q\1_{(s,t]}$.
    A position is acceptable with capital $m$ if, in every scenario, the capital covers the expected discounted loss:
    \[
        \A^m_{s,t}:=\big\{X\in L^p(\F_t)\ :\ m+\E[\Gamma^{\beta,q}_{s,t}X|\F_s]\geq0\ \text{ for all }(\beta,q)\in\mathcal S\big\}, \qquad m\in L^p(\F_s).
    \]
    The induced capital requirement $\r:=\r^{\A^\tinybullet}$ is the worst-case expected discounted loss, $\r_{s,t}(X)=\esssup_{(\beta,q)\in\mathcal S}\E[-\Gamma^{\beta,q}_{s,t}X|\F_s]$.
    The worst rate is $r$ for a liability $X\leq0$ and $R$ for a claim $X\geq0$; for a position with gains and losses, it is not known in advance.
 
    \smallskip
    \noindent\emph{The axioms.}
    They are checked on the sets, with $\mu:=\max\{|r|,|R|,\kappa\}$:
    \begin{itemize}[nosep, leftmargin=*, label=\scriptsize$\bullet$]
        \item \itemref{IT:N-normalized}
            follows from $0\in\A^m_{s,t}$ if and only if $m\geq0$;
        \item \itemref{IT:N-pasting}, \itemref{IT:L-rightcont} follow by direct inspection;
        \item \itemref{IT:N-decomposition}: 
            see Appendix~\ref{APP:examples};
        \item \itemref{IT:N-SL}: 
            for $(\beta,q)\in\mathcal S$, $g^{\beta,q}(u,y,z):=-\beta_uy+q_u\cdot z$ is a $\mu$-Lipschitz driver with $g^{\beta,q}\leq g^{\tinybullet,\mu}$.
            Hence, for $X\in L^p(\F_t)$, 
            \begin{equation}
            \label{EQ:corridor-linear}
                \E[-\Gamma^{\beta,q}_{s,t}X|\F_s]=\r^{g^{\beta,q}}_{s,t}(X)\leq\r^{\tinybullet,\mu}_{s,t}(X)
            \end{equation}
            by the formula for linear BSDEs \cite[Proposition~2.2]{ElKaroui+Peng+Quenez_1997_Backward_stochastic_differential_equations_finance} and by \itemref{IT:B-comparison}; if $X\in\A^m_{s,t}$ and $Y\in\A^{n,\mu}_{s,t}$, then $m+n+\E[\Gamma^{\beta,q}_{s,t}(X+Y)|\F_s]\geq n-\r^{\tinybullet,\mu}_{s,t}(Y)\geq0$.
    \end{itemize}
    \smallskip
    \noindent\emph{The driver.}
    Hence $\A^\tinybullet$ is regular and satisfies \itemref{IT:N-SL}.
    By Theorem~\ref{THM:Lp-nonadditive}, it is induced by a unique $\mu$-Lipschitz driver $g$, and \eqref{EQ:N-g-from-A} gives (see Appendix~\ref{APP:examples})
    \begin{equation}
    \label{EQ:g-corridor}
        g(t,y,z)=R\,y^--r\,y^++\kappa|z|=\max\big\{-\beta y+q\cdot z\ :\ \beta\in[r,R],\ |q|\leq\kappa\big\}.
    \end{equation}
    The driver reveals what the rule leaves implicit: 
    the capital requirement $Y_u=\r_{u,t}(X)$ accrues at the lowest rate $r$ when it is positive and at the highest rate $R$ when it is negative, as for an agent who deposits at the rate $r$ and borrows at the rate $R$, and $\kappa|Z_u|$ is the charge for model uncertainty.
    By Theorem~\ref{THM:Lp-nonadditive}~(iii), the properties of the rule are read off the driver. For instance, \itemref{IT:N-convex} holds since $g$ is convex, while \itemref{IT:N-cashsub} is satisfied if and only if $r\geq0$: the rule is cash-subadditive exactly when rates are nonnegative.
\end{example}

\begin{example}[Capital requirements relative to approved benchmarks]
\label{EX:benchmarks}
    Acceptability is sometimes prescribed in relative terms: a supervisor approves a list of reference portfolios, and a position is acceptable if it performs at least as well as one of them, up to a tolerance.
 
    \smallskip
    \noindent\emph{The rule.}
    Let $p\in(1,+\infty]$ and $\mu>0$.
    The \emph{benchmarks} are the elements of a set $\mathcal G$ of adapted processes $G:\Omega\times[0,T]\to\R$ such that $G_t-G_s\in L^p(\F_t)$ is the P\&L over $[s,t]$.
    We only assume that $0\in\mathcal G$, i.e., doing nothing is approved, and that benchmarks do not lose on average, i.e., $\E[G_t-G_s|\F_s]\geq0$ for $s\leq t$.
    For instance, if $p<\infty$, $\mathcal G:=(G^i)_{i=0}^n$ may consist of the null process $G^0\equiv 0$ and $n$ It\^o processes $G^i=\int_0^{\cdot}a^i_u\d u +\int_0^{\cdot} b^i_u\cdot \d W_u$ with bounded predictable drifts $a^i\geq 0$ and volatilities $b^i$, $i=1,\dots,n$.
    The agent may revise the choice of the benchmark at finitely many dates, on the basis of the available information.
    Accordingly, for $0\leq s\leq t\leq T$, we let $\mathcal H_{s,t}$ be the set of \textit{switching strategies}
    \[
        H=\sum_{k=1}^N\big(G^k_{t_k}-G^k_{t_{k-1}}\big)\in L^p(\F_t),
        \qquad s=t_0\leq\dots\leq t_N=t,
    \]
    where the benchmark $G^k$ followed on $[t_{k-1},t_k]$ is chosen on the basis of $\F_{t_{k-1}}$.\footnote{Rigorously, $G^k=\sum_{j=1}^J\1_{B_{k,j}}G^{k,j}$ for some $G^{k,j}\in\mathcal G$ and some $\F_{t_{k-1}}$-measurable partition $(B_{k,j})_{j=1}^J$ of $\Omega$.}
    A position $X\in L^p(\F_t)$ is acceptable at time $s$ if its P\&L follows a switching strategy of benchmarks up to an error that is acceptable for the reference risk measure $\r^\mu$:
    \[
        \A_{s,t}:=\mathcal H_{s,t}+\A^\mu_{s,t}
        =\big\{X\in L^p(\F_t)\ :\ \r^\mu_{s,t}(X-H)\leq0\ \text{for some }H\in\mathcal H_{s,t}\big\}.
    \]
    The induced capital requirement $\r^\A_{s,t}(X)=\essinf_{H\in\mathcal H_{s,t}}\r^\mu_{s,t}(X-H)$ is the smallest amount of cash that makes $X$ comparable to a switching strategy.
 
    \smallskip
    \noindent\emph{The axioms} 
    \itemref{IT:S-decomposition}, \itemref{IT:S-pasting}, \itemref{IT:S-SL}, \itemref{IT:S-normalized} and \itemref{IT:S-expectation}
    are checked using the definition of $\mathcal H_{s,t}$ and the fact that $\A^\mu$ is regular, satisfies \itemref{IT:S-SL} and is contained in $\A^\E$.
 
    \smallskip
    \noindent\emph{The driver.}
    By Corollary~\ref{COR:no-closedness}, $\r^\A=\r^g$ for a unique $\mu$-Lipschitz monetary driver $g\geq 0$, since $\A\subseteq\A^g\subseteq \A^\E$ by Theorem~\ref{THM:Lp-additive}~(iii).
    Therefore, the capital requirement $\r^\A_{s,t}(X)$, which is defined as an infimum over switching strategies, is time-consistent and solves a single BSDE with driver $g$.
    In the example above with finite It\^o benchmarks, \eqref{eq: g from A-1} also gives  (see Appendix~\ref{APP:examples})
    \begin{equation}
    \label{EQ:g-benchmarks}
        g(u,z)=\min_{0\leq i\leq n}\big\{a^i_u+\mu\,|z+b^i_u|\big\},\qquad a^0=0,\ b^0=0.
    \end{equation}
    In particular, the driver is the pointwise minimum of the individual drivers: although no finite switching strategy is optimal in general, the infimum is approached by following, at each time, the benchmark that minimizes $a^i_u+\mu|Z_u+b^i_u|$.
    Here, $Z_u$ is the Brownian exposure of the position, $a^i_u$ acts as a hurdle rate, and $|Z_u+b^i_u|$ as a tracking error.
    Moreover,  if the benchmarks are martingales, $g(u,z)=\mu\,\mathrm{dist}\big(z,\{-b^0_u,\dots,-b^n_u\}\big)$: replicating a benchmark is free.
    Finally, $g$ is in general not convex, so that mixing two acceptable positions may require capital; by Theorem~\ref{THM:Lp-additive}~(iii), convexity is restored when mixtures of benchmarks are approved, i.e., when $\mathcal G$ is convex.

    Since $g\geq0$, every position is resilient or resilience neutral (Section~\ref{SUBSEC:resilience}); for It\^o benchmarks, $\bm\dot\r_t(X)=-\E\big[\min_i\{a^i_t+\mu|Z_t+b^i_t|\}\big]$, so risk recovers at the hurdle rate plus the tracking error of the best benchmark, and the position is resilience neutral at $t$ exactly when, almost surely, it replicates a martingale benchmark, i.e., $Z_t=-b^i_t$ and $a^i_t=0$ for some $i$, which may depend on the state.
\end{example}

\appendix

\section{Proofs of the correspondence results}
\label{APP:correspondence_proofs}
 
This appendix contains the proofs of the correspondence results of Sections~\ref{SEC:N} and~\ref{SEC:CA}.
They are standard or technical, and they are included for completeness and for ease of reference.
 
\subsection{Leveled acceptance families}
\label{APP:N}
 
Throughout, $\A^\tinybullet$ is a leveled acceptance family in $L^p$, $p\in[1,+\infty]$, and $\r:=\r^{\A^\tinybullet}$.
 
\begin{proof}[\textbf{Proof of Lemma~\ref{LEM:N-from-A-to-rho}}]
    Fix $0\leq s\leq t\leq T$, $m\in L^p(\F_s)$ and $X\in L^p(\F_t)$.
    We only prove the statements that do not follow directly from \eqref{eq: rho given A-m} or \eqref{eq: A-m given rho}.
    \smallskip
 
    \emph{(a).}
    For the equality $\A^m_{s,t}=\A^{m,\r}_{s,t}$, assume \itemref{IT:L-monotone}, \itemref{IT:L-rightcont} and \itemref{IT:N-pasting}, and let $X\in\A^{m,\r}_{s,t}$.
    Since $m$ is finite, $\r_{s,t}(X)<+\infty$ $\P$-a.s., hence Lemma~\ref{LEM:N-directedness} yields $(m_k)_{k\in\N}\subset L^p(\F_s)$ such that $X\in\A^{m_k}_{s,t}$ for every $k\in\N$ and $m_k\downarrow\r_{s,t}(X)$ $\P$-a.s.
    Set $n_k:=m_k\vee m$, $k\in\N$.
    Then $n_k\in L^p(\F_s)$, because $|n_k|\leq|m_k|+|m|$, and $n_k\geq m_k$, so that \itemref{IT:L-monotone} gives $X\in\A^{m_k}_{s,t}\subseteq\A^{n_k}_{s,t}$ for all $k\in\N$.
    Moreover, $n_k\downarrow\r_{s,t}(X)\vee m=m$ $\P$-a.s., and \itemref{IT:L-rightcont} yields $X\in\A^m_{s,t}$.
 
    For well-posedness, \itemref{IT:N-wellposed} and \itemref{IT:N-lowerbound} say exactly that $k\leq\r_{s,t}(X)\leq m$ for some $k,m\in L^p(\F_s)$.
    
    For the item on \itemref{IT:N-lowerbound}, by convexity (see (b)) and $\r_{s,t}(0)=0$, we get $0\leq\frac12\r_{s,t}(X)+\frac12\r_{s,t}(-X)$, and \itemref{IT:N-wellposed} gives $m_1\in L^p(\F_s)$ with $\r_{s,t}(-X)\leq m_1$; hence $\r_{s,t}(X)\geq-m_1$, i.e., \itemref{IT:N-lowerbound} holds with $k:=-m_1$.
    
    For cash-additivity, \itemref{IT:N-constancy} implies \itemref{IT:N-normalized}, so that $\r$ is well posed in $L^p$, convex and conditionally constant by the items above and (b).
    For $m\in L^p(\F_s)$ and $\l\in[1/2,1)$, convexity and conditional constancy give $\r_{s,t}(X+m)\leq\l\r_{s,t}(X/\l)-m$, and convexity on the segment between $X$ and $2X$ gives $\r_{s,t}(X/\l)\leq\r_{s,t}(X)+(1/\l-1)\big(\r_{s,t}(2X)-\r_{s,t}(X)\big)$.
    Letting $\l\uparrow1$ along a sequence gives $\r_{s,t}(X+m)\leq\r_{s,t}(X)-m$, and replacing $(X,m)$ by $(X+m,-m)$ gives the reverse inequality; see also \cite{Cheridito+Kupper_2009_Recursiveness_indifference_prices_translation-invariant_preferences} and \cite[Proposition~2 and footnote~25]{Drapeau+Kupper_2013_Risk_preferences_robust_representation}.
    \smallskip

    \emph{(b).}
    For locality, let $X,Y\in L^p(\F_t)$, $A\in\F_s$ and $Z:=\1_AX+\1_{A^c}Y$; by \itemref{IT:N-wellposed}, $\r_{s,t}(X),\r_{s,t}(Y)<+\infty$.
    If $X\in\A^{m}_{s,t}$ and $Y\in\A^{n}_{s,t}$, then $Z\in\A^{\1_Am+\1_{A^c}n}_{s,t}$ by \itemref{IT:N-pasting}; choosing $m=m_k\downarrow\r_{s,t}(X)$ and $n=n_k\downarrow\r_{s,t}(Y)$ as in Lemma~\ref{LEM:N-directedness} gives $\r_{s,t}(Z)\leq\1_A\r_{s,t}(X)+\1_{A^c}\r_{s,t}(Y)$.
    Applied to $X=\1_AZ+\1_{A^c}X$ and to $Y=\1_{A^c}Z+\1_AY$, this gives $\1_A\r_{s,t}(X)\leq\1_A\r_{s,t}(Z)$ and $\1_{A^c}\r_{s,t}(Y)\leq\1_{A^c}\r_{s,t}(Z)$.
    \smallskip
 
    \emph{(c).}
    Let $c:=\|X\|_{L^\infty}$.
    By \itemref{IT:N-constancy} and \itemref{IT:N-monotone}, $X\geq-c\in\A^c_{s,t}$ gives $X\in\A^c_{s,t}$, hence $\r_{s,t}(X)\leq c$.
    If $X\in\A^m_{s,t}$, then $c\geq X$ gives $c\in\A^m_{s,t}\cap L^\infty(\F_s)$, hence $m\geq-c$ by \itemref{IT:N-constancy}, so $\r_{s,t}(X)\geq -c$.
\end{proof}
 
\begin{proof}[\textbf{Proof of Lemma~\ref{LEM:N-from-rho-to-A}}]
    Fix $0\leq s\leq t\leq T$, $m\in L^p(\F_s)$ and $X\in L^p(\F_t)$.
    \itemref{IT:L-monotone} is immediate from \eqref{eq: A-m given rho}, and \itemref{IT:L-rightcont} holds because $\r_{s,t}(X)\leq n_k$ for all $k\in\N$ and $n_k\downarrow n$ imply $\r_{s,t}(X)\leq n$.
    Since $\r_{s,t}(X)\in L^p(\F_s)$, we have $X\in\A^{\r_{s,t}(X),\r}_{s,t}$, and $X\in\A^{m,\r}_{s,t}$ if and only if $m\geq\r_{s,t}(X)$; hence \itemref{IT:N-wellposed} holds, \itemref{IT:N-lowerbound} holds with $k:=\r_{s,t}(X)$, and $\r^{\A^{\tinybullet,\r}}_{s,t}(X)=\r_{s,t}(X)$.
    By $\r^{\A^{\tinybullet,\r}}=\r$, the implications in Lemma~\ref{LEM:N-from-A-to-rho}~(b) hold for $\A^{\tinybullet,\r}$.
    Conversely, each property of $\r$ gives the corresponding axiom by \eqref{eq: A-m given rho}; e.g., if $\r$ is local, $\r_{s,t}(X)\leq m$, $\r_{s,t}(Y)\leq n$ and $A\in\F_s$, then $\r_{s,t}(\1_AX+\1_{A^c}Y)=\1_A\r_{s,t}(X)+\1_{A^c}\r_{s,t}(Y)\leq\1_Am+\1_{A^c}n$, and, if $\r$ is cash-additive, then $\A^{m,\r}_{s,t}+m=\{Y\in L^p(\F_t)\ :\ \r_{s,t}(Y-m)\leq m\}=\A^{0,\r}_{s,t}$.
    The other cases are analogous.
\end{proof}
 
\subsection{Monetary acceptance families}
\label{APP:CA}

 Throughout, $\A$ is a monetary acceptance family in $L^p$, $p\in[1,+\infty]$, and $\r:=\r^\A$.
 
\begin{proof}[\textbf{Proof of Lemma~\ref{LEM:penalty}}]
    For $X\in L^p(\F_t)$, define
    \[
        M_{s,t}^X:=\{m\in L^p(\F_s) \ : \ X+m\in\A_{s,t}\},
    \]
    so that $\r^\A_{s,t}(X)=\essinf_{m\in M_{s,t}^X}m$.
    For every $\Q\in\mathcal Q^{\mathrm{ac}}_{s,t}$ such that $\d \Q / \d \P$ is $q$-integrable,
    \begin{align}
        \E_\Q[-X|\F_s]-\r^\A_{s,t}(X)
        =
        \esssup_{m\in M_{s,t}^X}\{\E_\Q[-X|\F_s]-m\}
        =
        \esssup_{m\in M_{s,t}^X}\E_\Q[-(X+m)|\F_s]
        \leq
        c^\A_{s,t}(\Q),
    \end{align}
    where the inequality follows from  $X+m\in \A_{s,t}$.
    Take the essential supremum over $X\in L^p(\F_t)$ to get
    \[
        \esssup_{X\in L^p(\F_t)}
        \big\{\E_\Q[-X|\F_s]-\r^\A_{s,t}(X)\big\}
        \leq c^\A_{s,t}(\Q).
    \]
    The reverse inequality easily follows from $\A_{s,t}\subseteq\A^{\r^\A}_{s,t}\subseteq L^p(\F_t)$.
    Hence the sought representation for $c^\A_{s,t}$; cf.\ \cite[Remark~9]{Detlefsen+Scandolo_2005_Conditional_dynamic_convex_risk_measures} for $p=+\infty$.
 
    By definition of $c^\A_{s,t}$ and \itemref{IT:S-normalized}, we have for all $\Q\in\mathcal Q^{\mathrm{ac}}_{s,t}$ with $\d \Q/\d\P\in L^q(\F_t)$:
    \begin{align}
        c^\A_{s,t}(\Q)
        \geq
        \esssup_{X\in \A_{s,t}\cap L^p(\F_s)}\E_\Q[-X|\F_s]
        =
        -\essinf_{X\in \A_{s,t}\cap L^p(\F_s)}X
        =0,
    \end{align}
    hence $c^\A_{s,t}(\Q)\in [0,+\infty]$.
 
    \itemref{IT:S-expectation} gives $\E[-X|\F_s]\leq 0$ for all $X\in \A_{s,T}$.
    Taking the essential supremum and recalling that $c_{s,T}^\A\geq 0$ from \itemref{IT:S-normalized}, we get $c^\A_{s,T}(\P)=0$.
 
    Assume \itemref{IT:S-normalized}, \itemref{IT:S-decomposition} and \itemref{IT:S-expectation}.
    Let $Y\in \A_{t,T}\cap L^p(\F_t)$, which exists by \itemref{IT:S-normalized}, and $X\in \A_{s,t}$; see Remark \ref{REM:not_empty}.
    By \itemref{IT:S-decomposition}, $X+Y\in \A_{s,T}$, hence, by \itemref{IT:S-expectation}, $\E[-X| \F_s] \leq \E[Y| \F_s]$.
    Taking the essential supremum over $X\in \A_{s,t}$ gives $c^\A_{s,t}(\P)\leq \E[Y|\F_s]\in L^p(\F_s)$, and $c^\A_{s,t}(\P)\geq 0$ by the first item.
\end{proof}
 
\begin{proof}[\textbf{Proof of Lemma~\ref{LEM:from_A_to_rho}}]
    Let $\A^m_{s,t}:=\A_{s,t}-m$, so that $\r^{\A^\tinybullet}=\r^\A$ (Section~\ref{SUBSEC:P-acceptance}).
    By Remark~\ref{REM:P-embedding} and Section~\ref{SUBSEC:N-TC}, each axiom on $\A$ in (a) and (b) is, or implies, the corresponding axiom on $\A^\tinybullet$.
    Hence 
    (a) and (b) follow from Lemma~\ref{LEM:N-from-A-to-rho} with $m=0$.
    For the last item of (a), $\A\subseteq\A^\E$ is \itemref{IT:N-expectation}, which gives $\r^\A_{s,t}(X)\geq\E[-X|\F_s]\in L^p(\F_s)$.
    \smallskip
 
    \emph{(c).}
    By (a) and (b), $\r^\A$ is monotone, cash-additive and normalized, hence $1$-Lipschitz on $L^\infty$ with $\|\r^\A_{s,t}(X)\|_{L^\infty}\leq\|X\|_{L^\infty}$ (see \cite[Proposition 3.3]{Cheridito+Delbaen+Kupper_2006_Dynamic_monetary_risk_measures_bounded_discrete-time_processes} and \cite[Lemma~4.3]{Follmer+Schied_2016_Stochastic_finance} for the static case), and local \cite[Remark~2.2]{Bion-Nadal_2008_Dynamic_risk_measures_time_consistency__risk_measures_BMO_martingales}.
    Conditional convexity and conditional positive homogeneity follow from convexity and positive homogeneity (see (b)), locality and the Lipschitz property by approximating $\Lambda$ with simple random variables; see \cite[Section~2, Remark~2)]{Kloppel+Schweizer_2007_Dynamic_Indifference_Valuation_Convex_Risk_Measures} and \cite[Remark~7]{Detlefsen+Scandolo_2005_Conditional_dynamic_convex_risk_measures}.
 
    For \ref{item:cont_above}, let $X_n\downarrow X$ in $L^\infty(\F_t)$.
    By monotonicity, $\r^\A_{s,t}(X_n)\uparrow R\leq\r^\A_{s,t}(X)$ for some $R\in L^\infty(\F_s)$.
    By cash-additivity and $\A=\A^{\r^\A}$ (see (a)), $X_n+R\in\A_{s,t}$, and Remark~\ref{REM:P-embedding} gives $X+R\in\A_{s,t}$, i.e., $\r^\A_{s,t}(X)\leq R$.
 
    For \ref{item:dual_eq}, \eqref{EQ:c^A_dual} holds for all $\Q\in\mathcal Q^{\mathrm{ac}}_{s,t}$ by Lemma~\ref{LEM:penalty}, since $q=1$.
    By (c)(i) and \ref{item:cont_above}, \cite[Theorem~1]{Detlefsen+Scandolo_2005_Conditional_dynamic_convex_risk_measures} represents $\r^\A_{s,t}$ as an essential supremum over $\mathcal Q^{\mathrm{ac}}_{s,t}$ with minimal penalty $\Q\mapsto\esssup_{X}\{\E_\Q[-X|\F_s]-\r^\A_{s,t}(X)\}$, which is $c^\A_{s,t}$ by \eqref{EQ:c^A_dual}.
    Since $c^\A_{s,t}(\P)<+\infty$, \cite[Lemma~3.5]{Follmer+Penner_2006_Convex_risk_measures_dynamics_their_penalty_functions}, applied with $\P^*=\P$ and $\F_s$, $\F_t$ in place of $\F_t$, $\F_T$, allows one to replace $\mathcal Q^{\mathrm{ac}}_{s,t}$ by $\mathcal Q_{s,t}$, which gives \eqref{EQ:dual_ac}; minimality follows from \eqref{EQ:c^A_dual}.
\end{proof}
 
\begin{proof}[\textbf{Proof of Lemma~\ref{LEM:from_rho_to_A}}]
    Fix $0\leq s\leq t\leq T$ and $X\in L^p(\F_t)$.
    By cash-additivity, $X+m\in\A^\r_{s,t}$ if and only if $m\geq\r_{s,t}(X)$, for $m\in L^p(\F_s)$; since $\r_{s,t}(X)\in L^p(\F_s)$, this gives $\r^{\A^\r}_{s,t}(X)=\r_{s,t}(X)$, and \itemref{IT:S-upperbound} and \itemref{IT:S-lowerbound} follow from Lemma~\ref{LEM:from_A_to_rho}~(a).
    By $\r^{\A^\r}=\r$, the implications in Lemma~\ref{LEM:from_A_to_rho}~(b) hold for $\A^\r$.
    Conversely, each property of $\r$ gives the corresponding axiom by \eqref{eq: A given rho}; e.g., if $\r$ is local, $X,Y\in\A^\r_{s,t}$ and $A\in\F_s$, then $\r_{s,t}(\1_AX+\1_{A^c}Y)=\1_A\r_{s,t}(X)+\1_{A^c}\r_{s,t}(Y)\leq0$.
    The other cases are analogous.
 
    The characterization of time-consistency is well known; see, e.g., \cite[Theorem~2.5]{Bion-Nadal_2008_Dynamic_risk_measures_time_consistency__risk_measures_BMO_martingales}, \cite[Theorem~4.6]{Cheridito+Delbaen+Kupper_2006_Dynamic_monetary_risk_measures_bounded_discrete-time_processes} and \cite[Proposition~11.15]{Follmer+Schied_2016_Stochastic_finance}.
    It also follows from Proposition~\ref{prop: suff cond-A3} applied to $\A^{m,\r}_{s,t}=\A^\r_{s,t}-m$.
\end{proof}

\section{Extension to stopping times}
\label{APP:extension}
This appendix provides the extension to stopping times used in the proof of Proposition~\ref{prop: charact-accept-f} and of Theorem~\ref{TH:VaR_majorant}.
The results hold in any filtration satisfying the usual conditions.

We denote by $\Theta^{\mathrm f}$ the set of $\mathbb F$-stopping times taking finitely many values in $[0,T]$.
Given a dynamic risk functional $\r$ on $L^\infty$ and $\theta\in\Theta^{\mathrm f}$ with values in $\{t_1<\dots<t_m\}$, we set, as in \cite{Cheridito+Delbaen+Kupper_2006_Dynamic_monetary_risk_measures_bounded_discrete-time_processes},
\begin{equation}
\label{EQ:rho_theta}
    \r_{\theta,T}(X):=\sum_{j=1}^m\1_{\{\theta=t_j\}}\r_{t_j,T}(X),\qquad X\in L^\infty(\F_T).
\end{equation}
The following nonlinear optional sampling property is essentially a discrete-time result of \cite{Cheridito+Delbaen+Kupper_2006_Dynamic_monetary_risk_measures_bounded_discrete-time_processes}; analogous results are known on $L^2$ under \itemref{RM:Dom}; see \cite[Lemma 5.5]{Coquet+Hu+Memin+Peng_2002_Filtration-consistent_nonlinear_expectations_related_g-expectations}, \cite[Section 10]{Peng_2005_dynamically_consistent_nonlinear_evaluations}.
 
\begin{lemma}
\label{LEM:optional_sampling}
    Let $\r$ be a normalized, time-consistent, cash-additive dynamic risk measure on $L^\infty$, and let $\theta,\theta'\in\Theta^{\mathrm f}$.
    \begin{enumerate}[label=(\alph*)]
        \item
            $\r_{\theta,T}:L^\infty(\F_T)\to L^\infty(\F_\theta)$ is well defined, monotone, normalized, $\F_\theta$-cash-additive and local with respect to $\F_\theta$.
            If each $\r_{t,T}$ is convex, continuous from above, or satisfies $\r_{t,T}(X)\geq\E[-X|\F_t]$, then the same holds for $\r_{\theta,T}$, with $\F_\theta$ in place of $\F_t$.
        \item
            If $\theta\leq\theta'$, then
            \begin{equation}
            \label{EQ:tc_finite_st}
                \r_{\theta,T}\big(-\r_{\theta',T}(X)\big)=\r_{\theta,T}(X),\qquad X\in L^\infty(\F_T).
            \end{equation}
            If moreover $\r_{t,T}(X)\geq\E[-X|\F_t]$ for all $t$, then $\E[\r_{\theta',T}(X)|\F_\theta]\leq\r_{\theta,T}(X)$.
    \end{enumerate}
\end{lemma}
 
\begin{proof}
    Fix a grid $u_1<\dots<u_m=T$ containing the values of $\theta$ and $\theta'$, and consider the discrete-time filtration $(\F_{u_j})_{j=1,\dots,m}$ and the dynamic monetary utility function $\varphi_j(Y):=-\r_{u_j,T}(Y_m)$ on bounded $(\F_{u_j})_j$-adapted processes $Y=(Y_j)_j$, which is concave or continuous for bounded decreasing sequences whenever each $\r_{u_j,T}$ is convex or continuous from above.
    Then $\r_{\theta,T}(X)=-\varphi_\theta(X\1_{[m,\infty)})$, with $\varphi_\theta$ as in \cite[(4.1)]{Cheridito+Delbaen+Kupper_2006_Dynamic_monetary_risk_measures_bounded_discrete-time_processes}, so that (a) follows from the observation after \cite[(4.1)]{Cheridito+Delbaen+Kupper_2006_Dynamic_monetary_risk_measures_bounded_discrete-time_processes} and \cite[Proposition~3.3]{Cheridito+Delbaen+Kupper_2006_Dynamic_monetary_risk_measures_bounded_discrete-time_processes}; the lower bound $\r_{\theta,T}(X)\geq\E[-X|\F_\theta]$ holds on each $\{\theta=u_j\}$.
    By time-consistency of $\r$, $(\varphi_j)_j$ satisfies the one-step condition \cite[(4.6)]{Cheridito+Delbaen+Kupper_2006_Dynamic_monetary_risk_measures_bounded_discrete-time_processes}, hence it is time-consistent by \cite[Proposition~4.5]{Cheridito+Delbaen+Kupper_2006_Dynamic_monetary_risk_measures_bounded_discrete-time_processes}, and \eqref{EQ:tc_finite_st} follows from \cite[Remark~4.3]{Cheridito+Delbaen+Kupper_2006_Dynamic_monetary_risk_measures_bounded_discrete-time_processes}; the case of a deterministic $s$ corresponds to $\theta\equiv s$.
    The last assertion follows from \eqref{EQ:tc_finite_st} and (a) applied to $-\r_{\theta',T}(X)$.
\end{proof}

For $\s\in\Theta$, $\s_n:=2^{-n}T\lceil 2^n\s/T\rceil\in\Theta^{\mathrm f}$ denotes its dyadic approximation, so that $\s_n\downarrow\s$ and $\F_\s\subseteq\F_{\s_n}$.

\begin{proof}[\textbf{Proof of Lemma~\ref{LEM:extension}}]
    We write $\r_t:=\r_{t,T}$ for $t\in[0,T]$ and $\r_\theta:=\r_{\theta,T}$ for $\theta\in\Theta^{\mathrm f}$, see \eqref{EQ:rho_theta}.
    By time-consistency and normalization, $\r_{s,t}=\r_s$ on $L^\infty(\F_t)$, hence $\r_{s,t}(X)\geq\E[-X|\F_s]$, i.e., the minimal penalty of $\r_{s,t}$ vanishes at $\P$.
    For $s\in[0,T]$ and $\Q\in\mathcal Q_{s,T}$, let $c_s(\Q)$ be the minimal penalty of $\r_s$ at $\Q$ and $\phi_s(\Q):=\E[c_s(\Q)]\in[0,+\infty]$; for $\theta\in\Theta$ with $\theta\geq s$, let $\Q^\theta\in\mathcal Q_{s,T}$ be defined by $\d \Q^\theta/\d\P:=Z^\Q_\theta$, where $Z^\Q$ is the càdlàg density process of $\Q$, so that $\Q^\theta=\Q$ on $\F_\theta$.
 
    \emph{Step 1: penalties.}
    By \cite[Theorem~1, Remark~9]{Detlefsen+Scandolo_2005_Conditional_dynamic_convex_risk_measures} and $c_s(\P)=0$ (see the proof of Lemma~\ref{LEM:from_A_to_rho}\,\ref{item:dual_eq}), $\r_s(X)=\esssup_{\Q\in\mathcal Q_{s,T}}\{\E_\Q[-X|\F_s]-c_s(\Q)\}$.
    This family is upward directed and contains $\E[-X|\F_s]$, and $\Q=\P$ on $\F_s$; hence \cite[Lemma~1]{Detlefsen+Scandolo_2005_Conditional_dynamic_convex_risk_measures} and \cite[proof of Proposition~1(i)]{Bion-Nadal_2009_Time_consistent_dynamic_risk_processes} give
    \begin{equation}
    \label{EQ:phi_dual}
        \E[\r_s(X)]=\sup_{\Q\in\mathcal Q_{s,T}}\big\{\E_\Q[-X]-\phi_s(\Q)\big\},
        \qquad
        \phi_s(\Q)=\sup_{Y\in L^\infty(\F_T)}\big\{\E_\Q[-Y]-\E[\r_s(Y)]\big\};
    \end{equation}
    in particular, $\phi_s$ is lower semicontinuous with respect to the $L^1(\P)$-convergence of densities.
    Let now $\theta\in\Theta^{\mathrm f}$ with $\theta\geq s$ and $\Q\in\mathcal Q_{s,T}$ with $\phi_s(\Q)<\infty$.
    By Lemma~\ref{LEM:optional_sampling}, $\r_s=\r_s(-\r_\theta)$, so the cocycle condition \cite[Theorem~1]{Bion-Nadal_2009_Time_consistent_dynamic_risk_processes} for $\F_s\subseteq\F_\theta\subseteq\F_T$ gives $c_s(R)=\a_{s,\theta}(R)+\E_R[\a_{\theta,T}(R)|\F_s]$ for $R\in\mathcal Q_{s,T}$, where $\a_{s,\theta}$ and $\a_{\theta,T}$ are the minimal penalties of $\r_s|_{L^\infty(\F_\theta)}$ and $\r_\theta$.
    Since $\a_{s,\theta}(\Q^\theta)=\a_{s,\theta}(\Q)$ and $\a_{\theta,T}(\Q^\theta)=\a_{\theta,T}(\P)=0$ (Lemma~\ref{LEM:optional_sampling}\,(a)), taking expectations for $R=\Q$ and $R=\Q^\theta$ yields $\phi_s(\Q)=\phi_s(\Q^\theta)+\E_\Q[\a_{\theta,T}(\Q)]$, hence
    \begin{equation}
    \label{EQ:key_ineq}
        \phi_s(\Q^\theta)\leq\phi_s(\Q),
        \qquad
        \E_\Q[\r_\theta(X)]-\phi_s(\Q^\theta)\geq\E_\Q[-X]-\phi_s(\Q).
    \end{equation}
 
    \emph{Step 2: càdlàg modification.}
    $(\r_t(X))_t$ is a supermartingale by Lemma~\ref{LEM:optional_sampling}\,(b).
    If $t_n\downarrow t$ and $\Q\in\mathcal Q_{t,T}$ with $\phi_t(\Q)<\infty$, then \eqref{EQ:key_ineq} and $Z^\Q_{t_n}\to1$ in $L^1(\P)$ yield $\liminf_n\E[\r_{t_n}(X)]\geq\E_\Q[-X]-\phi_t(\Q)$. Hence $t\mapsto\E[\r_t(X)]$ is right-continuous by \eqref{EQ:phi_dual}; see also \cite[Lemma~4]{Bion-Nadal_2009_Time_consistent_dynamic_risk_processes}.
    Hence a càdlàg modification $\bar\r(X)$ exists \cite[Theorem~1.3.13]{Karatzas+Shreve_1991_Brownian_motion_stochastic_calculus}; it is bounded by $\|X\|_{L^\infty}$, $\bar\r(0)\equiv0$, $\bar\r_\theta(X)=\r_\theta(X)$ for $\theta\in\Theta^{\mathrm f}$, and it is a supermartingale also along stopping times \cite[Theorem~1.3.22]{Karatzas+Shreve_1991_Brownian_motion_stochastic_calculus}.
 
    \emph{Step 3: $\r_s(-\bar\r_\t(X))=\r_s(X)$ for $s\in[0,T]$ and $\t\in\Theta$ with $\t\geq s$.}
    By Lemma~\ref{LEM:optional_sampling}, $\r_s(-\bar\r_{\t_n}(X))=\r_s(X)$ for all $n$, and $\bar\r_{\t_n}(X)\to\bar\r_\t(X)$ a.s.
    Since $W_n:=\sup_{k\geq n}(-\bar\r_{\t_k}(X))\downarrow-\bar\r_\t(X)$ and $\r_s(W_n)\leq\r_s(X)$, continuity from above gives ``$\leq$''.
    Conversely, let $\Q\in\mathcal Q_{s,T}$ with $\phi_s(\Q)<\infty$.
    Since $Z^\Q_{\t_n}\to Z^\Q_\t$ in $L^1(\P)$, we have $\E_\Q[\bar\r_{\t_n}(X)]\to\E_{\Q^\t}[\bar\r_\t(X)]$ and $\phi_s(\Q^\t)\leq\liminf_n\phi_s(\Q^{\t_n})$, so \eqref{EQ:key_ineq} with $\theta=\t_n$ and \eqref{EQ:phi_dual} yield
    \[
        \E\big[\r_s(-\bar\r_\t(X))\big]\geq\E_{\Q^\t}[\bar\r_\t(X)]-\phi_s(\Q^\t)\geq\E_\Q[-X]-\phi_s(\Q).
    \]
    Taking the supremum over $\Q$ and using \eqref{EQ:phi_dual} gives $\E[\r_s(-\bar\r_\t(X))]\geq\E[\r_s(X)]$, and the claim follows.
 
    \emph{Step 4: assumptions (a)--(h).}
    At deterministic times, $\hat\r_{s,t}=\r_s=\r_{s,t}$ on $L^\infty(\F_t)$.
    In particular, the minimal penalty of $\hat\r_{t,T}=\r_t$ vanishes at $\P$ for every $t\in[0,T]$, as shown at the beginning of the proof.
    This is assumption (h) of \cite[Section~3]{Delbaen+Peng+RosazzaGianin_2010_Representation_penalty_term_dynamic_concave_utilities}, which, unlike (a)--(g), is only required at deterministic times.
    If $U,V$ are càdlàg with $U_t\leq V_t$ a.s.\ for every $t$, then $U_\s\leq V_\s$ a.s.\ for every $\s\in\Theta$; hence monotonicity, convexity and normalization pass to $\hat\r_{\s,\t}$.
    Cash-additivity and the local property with respect to $\F_\s$ hold at each $\s_n$ by Lemma~\ref{LEM:optional_sampling}\,(a), since $\F_\s\subseteq\F_{\s_n}$, and pass to $\s$ by right-continuity.
    For continuity from above, let $X_k\downarrow X$; then $\bar\r_\s(X_k)\uparrow\Lambda\leq\bar\r_\s(X)$, and $a_{k,n}:=\E[\bar\r_{\s_n}(X_k)]$ is nondecreasing in $n$ (supermartingale property) with limit $\E[\bar\r_\s(X_k)]$, and nondecreasing in $k$ (continuity from above of $\r_{\s_n}$) with limit $\E[\bar\r_{\s_n}(X)]$; exchanging the two suprema gives $\E[\Lambda]=\E[\bar\r_\s(X)]$, hence $\Lambda=\bar\r_\s(X)$.
    For time-consistency, let $\s\leq\t\leq\upsilon$, $X\in L^\infty(\F_\upsilon)$ and $Y:=-\bar\r_\t(X)$.
    For $t\in[0,T]$, Step~3 applied to $\t\vee t$ and locality on $\{t\leq\t\}\in\F_t$ give $\r_t(Y)=\r_t(X)$ a.s.\ on $\{t\leq\t\}$; by right-continuity, $\bar\r_\s(Y)=\bar\r_\s(X)$ on $\{\s<\t\}$, while on $\{\s=\t\}$ cash-additivity gives $\bar\r_\s(Y)=-Y=\bar\r_\s(X)$.
    Hence $\hat\r_{\s,\t}(-\hat\r_{\t,\upsilon}(X))=\hat\r_{\s,\upsilon}(X)$.
\end{proof}

\section{Verifications for the examples}
\label{APP:examples}

This appendix contains the verifications omitted in the examples of Section~\ref{SUBSEC:examples}; we use their notation.

\begin{proof}[\textbf{Verification for Example~\ref{EX:corridor}}]
    We prove \itemref{IT:N-decomposition} and \eqref{EQ:g-corridor}.
    Let $s\leq u\leq t$, $m\in L^p(\F_s)$ and $X\in L^p(\F_t)$.
    For $(\beta,q)\in\mathcal S$,
    \begin{equation}
    \label{EQ:corridor-tower}
        \E[\Gamma^{\beta,q}_{s,t}X|\F_s]=\E\big[\Gamma^{\beta,q}_{s,u}\,\E[\Gamma^{\beta,q}_{u,t}X|\F_u]\,\big|\,\F_s\big].
    \end{equation}
    If $n\in\A^m_{s,u}$ and $X\in\A^{-n}_{u,t}$, then $\E[\Gamma^{\beta,q}_{u,t}X|\F_u]\geq n$, and \eqref{EQ:corridor-tower} gives $m+\E[\Gamma^{\beta,q}_{s,t}X|\F_s]\geq m+\E[\Gamma^{\beta,q}_{s,u}n|\F_s]\geq0$, i.e., $X\in\A^m_{s,t}$.
    Conversely, let $X\in\A^m_{s,t}$ and let $n:=-\r_{u,t}(X)$, so that $X\in\A^{-n}_{u,t}$; 
    by \eqref{EQ:corridor-linear} $-\r^{\tinybullet,\mu}_{u,t}(X)\leq n\leq\r^{\tinybullet,\mu}_{u,t}(-X)$, hence $n\in L^p(\F_u)$ by \itemref{IT:B-stability}.
    The family $\{\E[\Gamma^{\beta,q}_{u,t}X|\F_u]:(\beta,q)\in\mathcal S\}$ is directed downward: for $(\beta^i,q^i)\in\mathcal S$ and $B:=\{\E[\Gamma^{\beta^1,q^1}_{u,t}X|\F_u]\leq\E[\Gamma^{\beta^2,q^2}_{u,t}X|\F_u]\}\in\F_u$, the scenario equal to $(\beta^1,q^1)$ on $B\times(u,T]$ and to $(\beta^2,q^2)$ elsewhere belongs to $\mathcal S$ and attains the minimum. 
    Hence, as in the proof of Lemma~\ref{LEM:N-directedness}, \cite[Theorem~A.37(b)]{Follmer+Schied_2016_Stochastic_finance} gives $n=\lim_k\E[\Gamma^{\beta^k,q^k}_{u,t}X|\F_u]$ for a decreasing sequence with $(\beta^k,q^k)\in\mathcal S$. 
    For $(\beta,q)\in\mathcal S$, the scenario equal to $(\beta,q)$ on $[0,u]$ and to $(\beta^k,q^k)$ on $(u,T]$ belongs to $\mathcal S$.
    Applying \eqref{EQ:corridor-tower} to this scenario and letting $k\to\infty$, monotone convergence gives $m+\E[\Gamma^{\beta,q}_{s,u}n|\F_s]\geq0$, i.e., $n\in\A^m_{s,u}$.
    This proves \itemref{IT:N-decomposition}.

    As for \eqref{EQ:g-corridor}, let $\hat g$ be its right-hand side, which is a $\mu$-Lipschitz driver with $g^{\beta,q}\leq\hat g$, hence $\r^{g^{\beta,q}}\leq\r^{\hat g}$ for $(\beta,q)\in\mathcal S$ by \itemref{IT:B-comparison}.
    Taking the $\esssup$ yields $\r\leq \r^{\hat g}$, hence $g\leq\hat g$ by \eqref{EQ:N-g-from-A}, applied to $\A^\tinybullet$ and to $\A^{\tinybullet,\hat g}$.
    Conversely, for constant $\beta\in[r,R]$ and $|q|\leq\kappa$, and with $\Delta_\eps W_t:=W_{t+\eps}-W_t$,
    \[
        \r_{t,t+\eps}(-y-z\cdot\Delta_\eps W_t)\geq\E[\Gamma^{\beta,q}_{t,t+\eps}(y+z\cdot\Delta_\eps W_t)|\F_t]=e^{-\beta\eps}(y+\eps\,q\cdot z),
    \]
    and \eqref{EQ:N-g-from-A} gives $g(t,y,z)\geq-\beta y+q\cdot z$.
    Taking the supremum yields $g\geq \hat g$, hence $g=\hat g$.
\end{proof}

\begin{proof}[\textbf{Verification for Example~\ref{EX:benchmarks}}]
    We prove \eqref{EQ:g-benchmarks} in the finite It\^o example.
    Let $g_i(t,z):=a^i_t+\mu|z+b^i_t|$ for $i\in\{0,1,\dots,n\}$ and $\hat g:=\min_ig_i$, where $a^0=0$, $b^0=0$.
    Fix $H\in\mathcal H_{s,t}$ and let $I:\Omega\times[0,T]\to\{0,1,\dots,n\}$ be the stochastic index of the benchmark followed by the strategy $H$, namely $H_s:=0$, $\d H_u:=\d G^{I_u}_u$, for $u\in[s,t]$ and $H_t:=H$.
    Then the shift $(Y,Z)\mapsto(Y-H,Z-b^{I})$ shows that $\r^\mu_{s,t}(X-H)=\r^{g_I}_{s,t}(X)$, with $g_I(u,z):=g_{I_u}(u,z)\geq\hat g(u,z)$; by comparison \cite[Theorem~2.2]{ElKaroui+Peng+Quenez_1997_Backward_stochastic_differential_equations_finance}, $\r^\mu_{s,t}(X-H)\geq\r^{\hat g}_{s,t}(X)$, and taking the essential infimum over $H\in\mathcal H_{s,t}$ gives $\r^g=\r^\A\geq\r^{\hat g}$, hence $g\geq\hat g$ by \eqref{eq: g from A-1} and \cite[Lemma~2.1]{Jiang_2008_Convexity_translation_invariance_subadditivity_g-expectations_related_risk_measures}.
    Conversely, choosing the constant strategy $H=G^i_{t+\eps}-G^i_t\in\mathcal H_{t,t+\eps}$ gives $\r^g_{t,t+\eps}(X)=\r^\A_{t,t+\eps}(X)\le\r^\mu_{t,t+\eps}(X-H)=\r^{g_i}_{t,t+\eps}(X)$; evaluating at $X=-z\cdot\Delta_\eps W_t$, dividing by $\eps$ and letting $\eps\downarrow0$ yields $g\leq g_i$ by \eqref{eq: g from A-1} and \cite[Lemma~2.1]{Jiang_2008_Convexity_translation_invariance_subadditivity_g-expectations_related_risk_measures}, hence $g\leq\hat g$.
\end{proof}

\renewcommand{\O}{\char"D8}
\phantomsection\addcontentsline{toc}{section}{References}
\bibliographystyle{alphaurlnames}
\bibliography{Biblio,Biblio_ASMF}

\end{document}